\documentclass[%
 reprint,
 amsmath,amssymb,
 aps,
]{revtex4-2}

\usepackage{graphicx}
\usepackage{dcolumn}
\usepackage{physics}
\usepackage{tikz}
\usetikzlibrary{quantikz}
\usepackage{enumitem}
\usepackage{multirow}
\usepackage{comment}
\usepackage{todonotes}

\usepackage{orcidlink}

\newtheorem{prop}{Proposition}

\usepackage{bm}

\begin{document}

\preprint{APS/123-QED}

\title{Quantum density estimation with density matrices: Application to quantum anomaly detection}


\author{Diego H. Useche \orcidlink{0000-0001-5985-7066}}
\email{diusecher@unal.edu.co}
\author{Oscar A. Bustos-Brinez \orcidlink{0000-0003-0704-9117}}
\author{Joseph A. Gallego-Mejia \orcidlink{0000-0001-8971-4998}}
\author{Fabio A. González \orcidlink{0000-0001-9009-7288}}%
\email{fagonzalezo@unal.edu.co}
\affiliation{%
 MindLab Research Group, Universidad Nacional de Colombia, 111321, Bogotá, Colombia
}%

\date{\today}

\begin{abstract}

Density estimation is a central task in statistics and machine learning. This problem aims to determine the underlying probability density function that best aligns with an observed data set. Some of its applications include statistical inference, unsupervised learning, and anomaly detection. Despite its relevance, few works have explored the application of quantum computing to density estimation. In this article, we present a novel quantum-classical density matrix density estimation model, called Q-DEMDE, based on the expected values of density matrices and a novel quantum embedding called quantum Fourier features. The method uses quantum hardware to build probability distributions of training data via mixed quantum states. As a core subroutine, we propose a new algorithm to estimate the expected value of a mixed density matrix from its spectral decomposition on a quantum computer. In addition, we present an application of the method for quantum-classical anomaly detection. We evaluated the density estimation model with quantum random and quantum adaptive Fourier features on different data sets on a quantum simulator and a real quantum computer. An important result of this work is to show that it is possible to perform density estimation and anomaly detection with high performance on present-day quantum computers.

\end{abstract}

\maketitle



\section{Introduction}

Density estimation (DE) is an important problem in machine learning and statistics, which consists of determining the probability density function $p(\mathbf{x})$ of a set of data samples drawn from an unknown probability distribution. It is central to many areas of machine learning, including generative modeling \cite{bigdeli2020learning}, anomaly detection \cite{PhysRevD.101.075042, hu2018anomaly}, supervised classification \cite{bortoloti2021supervised} and clustering \cite{fraley2002model, anderson2009kernel}. Among its many applications include medical image analysis \cite{gudhe2022area}, epidemic policy \cite{kato2021development}, and urban planning \cite{srikanth2020case}. An important area of application of DE is anomaly detection (AD), which aims to detect anomalous data that deviate from normal data. As an example, it can be applied to credit card fraud detection \cite{VanVlasselaer2015APATE:Extensions}, where fraudulent transactions are seen as anomalies among normal transactions.

Density estimation can be addressed using parametric \cite{papamakarios2017masked, varanasi1989parametric, vapnik1999support} and non-parametric methods \cite{Rosenblatt1956RemarksFunction, Parzen1962OnMode, wang2019nonparametric}. Parametric density estimation aims to adjust the observed data to a prior probability distribution $p(\mathbf{x}| \boldsymbol{\theta})$, for instance, it may search for the parameters $\boldsymbol{\theta}$ that fit the data with a Gaussian distribution.  By contrast, non-parametric methods such as kernel density estimation (KDE) \cite{Rosenblatt1956RemarksFunction, Parzen1962OnMode}, do not assume a particular form of the distribution and they are therefore more flexible by reproducing arbitrary probability distributions. However, they are computationally expensive because they require storing in memory the whole data set.

Methods for density estimation have been widely addressed in classical computers \cite{Rosenblatt1956RemarksFunction, Parzen1962OnMode, wang2019nonparametric, papamakarios2017masked};  however, there are only a handful of works that explore the use of quantum computers for DE, including both parametric \cite{liang2019quantum, guo2022quantum} and non-parametric methods \cite{Useche2021QuantumQudits, vargas2022optimisation}, of which some claim quantum advantage \cite{guo2022quantum}.

This recent enthusiasm for quantum computing has its roots in various theoretical works that illustrate that quantum computers may reduce the computational complexity of some classically-intractable problems \cite{lloyd2014quantum, Shor1994AlgorithmsFactoring}. Nevertheless, today's quantum hardware has not reached its full potential and it is prone to quantum noise \cite{haroche1998entanglement, preskill2018quantum}, hence some works have focused on building quantum machine learning algorithms \cite{lee2021quantum, banchi2022robust} for present-day quantum computers.

In this article, we present a novel non-parametric quantum-classical density matrix density estimation strategy, called Q-DEMDE, suitable for current noisy quantum computers. It combines a proposed quantum algorithm for computing the expectation values of density matrices with a novel quantum representation of data called quantum Fourier features. The proposed method, unlike previous quantum density estimation methods \cite{Useche2021QuantumQudits, vargas2022optimisation}, builds probability distributions of data in the form of mixed quantum states in qubit-based quantum hardware. It is based on the density matrix kernel density estimation (DMKDE) algorithm \cite{gonzalez2022learning} originally developed for classical computers; it fuses density matrices with random Fourier features (RFF) \cite{Rahimi2009RandomMachines} to build a non-parametric density estimator.  We show that the proposed Q-DEMDE method can be implemented in current
IBM quantum computers.

As an important quantum subroutine, we present a quantum algorithm, called Q-DMKDE, which estimates the expected value of a mixed density matrix in a quantum computer; this protocol extends to qubits a previous algorithm that performs the same operation in a high-dimensional quantum computer \cite{Useche2021QuantumQudits}. In contrast to pure-state classical sampling \cite{blank2022compact, nghiem2021unified}, or methods like quantum state tomography \cite{James2005OnQubits}, which build a density matrix in terms of the Pauli matrices, the proposed protocol prepares a density matrix in a quantum computer from its spectral decomposition. In addition, we propose a randomized quantum map called quantum random Fourier features (QRFF) and a quantum variational mapping called quantum adaptive Fourier features (QAFF). These methods are analogous to RFF \cite{Rahimi2009RandomMachines} and adaptive Fourier features (AFF) \cite{li2019learning} respectively, and they are suitable for both classical and quantum computers; they represent classical data as quantum states, where the squared modulus of the inner product in the Hilbert space approximates a Gaussian kernel in the original space. We show that the Q-DEMDE method, which combines the QRFF and QAFF with the Q-DMKDE algorithm, allows us to perform quantum-classical density estimation (QDE) with a low number of qubits in both a quantum simulator and a real noisy quantum computer. Furthermore, we present a quantum-classical anomaly detection (QAD) strategy based on the proposed QDE method. We show that it is possible to perform both quantum density estimation and anomaly detection on current IBM quantum devices. In summary, the contributions of the article are as follows:

\begin{enumerate}[label=(\roman*)]
	\item Two new quantum feature mappings called QRFF and QAFF, which can be viewed as quantum analogues of RFF \cite{Rahimi2009RandomMachines} and AFF \cite{li2019learning}.
	\item An innovative quantum algorithm that estimates the expected value of a density matrix from its eigendecomposition.
	\item A quantum-classical density estimation method for current qubit-based quantum hardware that combines quantum Fourier features and quantum algorithms for expected value estimation.
	\item A quantum-classical anomaly detection strategy built on the proposed quantum-classical density estimation approach.
\end{enumerate}

The structure of the document is as follows: in Sect. \ref{QDE:Related work}, we describe the related work, in Sect. \ref{QDE:Preliminaries}, we present the theoretical background of the DMKDE algorithm and the notation for the description of the quantum circuits, in Sect. \ref{QDE:QDE Method}, we introduce our proposed Q-DEMDE method, outline the proposed Q-DMKDE protocol, and show the proposed QRFF and QAFF, in Sect. \ref{QDE: density estimation results}, we present the results of our proposed method for quantum-classical density estimation, in Sect. \ref{QDE: anomaly detection results} we illustrate the application of the method for quantum-classical anomaly detection, and in Sect. \ref{QDE: conclusions} we establish our conclusions and future outlook.

\section{Related Work}\label{QDE:Related work}

Density estimation is an important challenge in both classical and quantum machine learning. Both parametric and non-parametric methods for DE have been widely explored in classical computers \cite{Rosenblatt1956RemarksFunction, Parzen1962OnMode, gonzalez2022learning, wang2019nonparametric, papamakarios2017masked, varanasi1989parametric, vapnik1999support}. Some early works on quantum density estimation build parametric models based on multivariate Gaussian distributions on quantum devices \cite{liang2019quantum, guo2022quantum}. On the non-parametric side, it is possible to combine the formalism of density matrices and kernels for density estimation as in the classical algorithms of kernel density matrices (KDM) \cite{gonzalez2023quantum} and DMKDE \cite{gonzalez2022learning}; the latter combining density matrices with random Fourier features. Although DMKDE was originally developed for classical computers, it can be implemented in quantum computers with both pure-state and mixed-state approaches; the pure-state approach has been developed in qubit-based quantum computers \cite{vargas2022optimisation}, while the mixed-state version has been implemented in a high-dimensional quantum computer \cite{Useche2021QuantumQudits}. In contrast to these previous works, we present a QDE strategy built on mixed states for present-day qubit-based quantum devices and propose a quantum algorithm for estimating the expected value of a density matrix that extend the mixed-state DMKDE \cite{Useche2021QuantumQudits} to qubit-based quantum hardware.

Other quantum machine learning models based on mixed states include quantum support vector machines \cite{rebentrost2014quantum} and other kernel-based methods \cite{blank2022compact, blank2020quantum, sergioli2019new, nghiem2021unified}. These models require the preparation of mixed quantum states in quantum computers by building a purification of the mixed-state \cite{bennett1996purification, rebentrost2014quantum}, or by classically sampling the pure states that form the ensemble \cite{blank2022compact, nghiem2021unified}. In addition, it is possible to build a density matrix in the Pauli basis, like in the variational quantum eigensolver \cite{peruzzo2014variational}, and quantum state tomography \cite{Nielsen2010QuantumEdition, Cramer2010EfficientTomography, James2005OnQubits}, and using the density matrix exponentiation method \cite{lloyd2014quantum}, which prepares an exponential form of the density matrix by a unitary transformation, in analogy with the hamiltonian as the generator of time evolution. Our method differs from the previous techniques by preparing a density matrix for expected value estimation from its spectral decomposition.

Gonzaléz et. al. \cite{Gonzalez2021ClassificationMeasurements} proposed the use of random Fourier features (RFF) \cite{Rahimi2009RandomMachines} as quantum feature maps. RFF approximates shift-invariant kernels by calculating an explicit mapping to a low-dimensional feature space.  Improving the power of random Fourier features to approximate shift-invariant kernels is an active area of research. Two of the main approaches to the problem include refining the original RFF Montecarlo sampling in either classical \cite{2639-8001_2020_3_309, Liu2019AAlgorithm, Agrawal2018Data-dependentApproximation} or quantum computers \cite{yamasaki2020learning} and using optimization techniques to find the optimal Fourier parameters \cite{xie2019deep, Bazavan2012FourierLearning}. Among the methods that rely on optimization, adaptive Fourier features (AFF) \cite{li2019learning} uses a siamese neural network to look for the optimal Fourier weights which reduce the mean squared distance between the Gaussian kernel and its Fourier feature approximation.

In this document, we present quantum random Fourier features, a randomized quantum feature map based on RFF \cite{Rahimi2009RandomMachines},  which induces a Gaussian kernel between data samples. In contrast to conventional RFF which builds an embedding in the real domain whose inner product approximates shift-invariant kernels, QRFF uses the complex form of RFF to approximate the kernel through the squared modulus of Hilbert inner product of the quantum features. Furthermore, we propose quantum adaptive Fourier features, a quantum variational method analogous to AFF suitable for both classical and quantum hardware. Previous works \cite{schuld2021effect, havlivcek2019supervised} have shown that quantum mappings in the Fourier basis can be used to build models that are universal function approximators. In addition, it is also possible to use coherent states as quantum maps to induce Gaussian kernels \cite{chatterjee2016generalized}. The proposed QAFF method learns the frequencies of the QRFF using a strategy based on optimization, which in contrast to most quantum feature maps \cite{schuld2021quantum, stoudenmire2016supervised, havlivcek2019supervised}, allows to learn the optimal quantum embedding.  

\section{Preliminaries}\label{QDE:Preliminaries}

\subsection{Kernel density estimation}

Kernel density estimation, also called the Parzen-Rosenblat window  \cite{Rosenblatt1956RemarksFunction, Parzen1962OnMode}, is a non-parametric method for density estimation. Given a training data set, $\{\mathbf{x_j}\} \in \mathbb{R}^D$ and a kernel, the method builds a density estimator of a new test sample $\mathbf{x^*} \in \mathbb{R}^D$ by computing the kernel between all elements in the training data set and the test sample. Considering a Gaussian kernel $k(\mathbf{x}_j, \mathbf{x^*}) = e^{-\gamma\norm{\mathbf{x}_j-\mathbf{x^*}}^2}$, where $\gamma$ (gamma) is a parameter associated to the bandwidth, the probability density of the test sample $\mathbf{x^*}$ is given by,
\begin{equation}
    p_\gamma(\mathbf{x^*}) = \frac{1}{N M_\gamma}\sum_{j=0}^{N-1} k(\mathbf{x}_j, \mathbf{x^*}) = \frac{1}{N M_\gamma} \sum_{j=0}^{N-1} e^{-\gamma \norm{\mathbf{x}_j-\mathbf{x^*}}^2}, \label{QDE: Eq KDE estimator}
\end{equation}
where $M_\gamma = (\pi/\gamma)^{\frac{D}{2}}$ is a normalization constant. This method scales with the size of the training data set and requires storing the whole data set in memory. 

\subsection{Random and Adaptive Fourier features}

Random Fourier features \cite{Rahimi2009RandomMachines}, is a method that maps the space of characteristics $\mathbf{x} \in \mathbb{R}^D$ to a feature space $\mathbf{z}(\mathbf{x}) \in \mathbb{R}^d$ whose inner product approximates a shift-invariant kernel in the original space, i.e., $\mathbf{z}(\mathbf{x})^T\mathbf{z}(\mathbf{y}) \approx k(\mathbf{x}- \mathbf{y})$ for all $\mathbf{x}, \mathbf{y} \in \mathbb{R}^D$. For the case of the Gaussian kernel $k(\mathbf{x}, \mathbf{y})=e^{-\gamma{\norm{\mathbf{x}-\mathbf{y}}}^2}$, it builds a $d$-dimensional vector with parameter $\gamma$ whose $i^\text{th}$ component is given by, 
\begin{eqnarray}
    \mathbf{z}_{\gamma, i}(\mathbf{x}) = \sqrt{\frac{2}{d}}\cos{(\sqrt{2\gamma}\mathbf{w}_i\cdot\mathbf{x}+b_i)}
\end{eqnarray}
by sampling i.i.d  the weights $\{\mathbf{w}_i\} \in \mathbb{R}^D$ from $\mathcal{N}(\mathbf{0}, \mathbf{I})$, and i.i.d. biases $\{b_i\} \in \mathbb{R}$ from $\text{Uniform}[0, 2\pi]$, leading to $\mathbf{z}_\gamma(\mathbf{x})^T\mathbf{z}_\gamma(\mathbf{y}) \approx e^{-\gamma{\norm{\mathbf{x}-\mathbf{y}}}^2}$. This method follows from Bochner theorem \cite{rudin2017fourier}, which states that a shift-invariant kernel is the Fourier transform of a probability measure, e.g., the Gaussian kernel with parameter $\gamma$ is the Fourier transform of a normal distribution with variance $2\gamma$. In addition, the kernel approximation can be improved by means of adaptive Fourier features \cite{li2019learning}, which learns the optimal RFF weights and biases $\{(\mathbf{w}_i, b_i)\}$ by minimizing the loss function, $\mathcal{L}_{\{(\mathbf{w}_i, b_i)\}} = \sum_{l, m}(k(\mathbf{x}_l, \mathbf{x}_m) - \mathbf{z}_\gamma(\mathbf{x_l})^T\mathbf{z}_\gamma(\mathbf{x}_m))^2$, that reduces the distance between the kernel values of the samples in the data set and their Fourier feature approximation. 

Furthermore, Ref. \cite{gonzalez2022learning} showed that it is possible to build a quantum mapping by normalizing the RFF in the form, 
\begin{equation}
    \ket{\bar{\psi}(\mathbf{x})} = \frac{1}{\mathcal{N}}\sum_{i=0}^{d-1}\mathbf{z}_{\gamma/2, i}(\mathbf{x})\ket{i},\label{QDE: Eq RFF quantum map}
\end{equation}
where, $\mathcal{N}=\norm{\mathbf{z}_{\gamma/2}(\mathbf{x})}$, and the Fourier components constructed with parameter $\gamma/2$; it approximates the Gaussian kernel via $\langle\bar{\psi}(\mathbf{x})\vert\bar{\psi}(\mathbf{y})\rangle^2 \approx e^{-\gamma{\norm{\mathbf{x}-\mathbf{y}}}^2}$ where $\ket{\bar{\psi}(\mathbf{x})} \in \mathbb{R}^d$. 

In this article, we propose the quantum random and quantum adaptive Fourier features, which map the space of the characteristics $\mathbf{x} \in \mathbb{R}^D$, to a complex Hilbert space $\ket{\psi(\mathbf{x})}\in \mathbb{C}^d$, where the square of the modulus approximates the Gaussian kernel $\abs{\langle{\psi}(\mathbf{x})\vert{\psi}(\mathbf{y})\rangle}^2 \approx e^{-\gamma{\norm{\mathbf{x}-\mathbf{y}}}^2}$, see Sect. \ref{QDE:Quantum Adp Fourier features}. For notational purposes, we use a bar to distinguish the quantum map based on RFF in the real domain $\ket{\bar{\psi}(\mathbf{x})}$ from the quantum random and quantum adaptive Fourier features $\ket{\psi(\mathbf{x})}$.

\subsection{Density matrix kernel density estimation}\label{QDE: preliminaries DMKDE}

The machine learning algorithm density matrix kernel density estimation (DMKDE) \cite{gonzalez2022learning} is a non-parametric method for density estimation which combines density matrices and random Fourier features \cite{Rahimi2009RandomMachines}, it does not require optimization and it works as an efficient approximation of kernel density estimation \cite{Rosenblatt1956RemarksFunction, Parzen1962OnMode}.

The DMKDE method \cite{gonzalez2022learning} starts by applying a quantum feature map based on RFF, see Eq. \ref{QDE: Eq RFF quantum map}, to each sample in the train data set $\{\mathbf{x}_j\} \rightarrow \{\ket{\bar{\psi}(\mathbf{x}_j)}\}$, and to the test sample whose density we aim to estimate $\mathbf{x^*} \rightarrow \ket{\bar{\psi}(\mathbf{x^*})}$, where $\mathbf{x} \in \mathbb{R}^D$ and $\ket{\bar{\psi}(\mathbf{x})} \in \mathbb{R}^d$. 

Once the Fourier feature map is established, the method builds a training density matrix. For the training states $\{\ket{\bar{\psi}(\mathbf{x}_j)}\}_{j = 0, \cdots, N-1}$, the training mixed state is constructed by,
\begin{equation}
    \bar{\rho}_{\text{train}} = \frac{1}{N}\sum_{j=0}^{N-1}\ket{\bar{\psi}(\mathbf{x}_j)}\bra{\bar{\psi}(\mathbf{x}_j)},\label{QDE: rho train}
\end{equation}
and the probability density of the testing sample $\mathbf{x^*} \rightarrow \ket{\bar{\psi}(\mathbf{x^*})}$ is computed by,
\begin{equation}
    \hat{p}_\gamma(\mathbf{x^*}) = \frac{1}{M_\gamma}\bra{\bar{\psi}(\mathbf{x^*})}\bar{\rho}_{\text{train}}\ket{\bar{\psi}(\mathbf{x^*})}.\label{QDE: density estimator}
\end{equation}
The complexity of the prediction phase of the DMKDE algorithm is $O(d^2)$, which is independent of the size of the training data. Furthermore, by replacing Eq. \ref{QDE: rho train} on Eq. \ref{QDE: density estimator}, we obtain,
\begin{align}
    \hat{p}_\gamma(\mathbf{x^*}) &= \frac{1}{M_\gamma}\bra{\bar{\psi}(\mathbf{x^*})}\Big(\frac{1}{N}\sum_{j=0}^{N-1}\ket{\bar{\psi}(\mathbf{x}_j)}\bra{\bar{\psi}(\mathbf{x}_j)}\Big)\ket{\bar{\psi}(\mathbf{x^*})} \notag \\
    &= \frac{1}{N M_\gamma}\sum_{j=0}^{N-1}\abs{\langle\bar{\psi}(\mathbf{x}_j)\vert\bar{\psi}(\mathbf{x^*})\rangle}^2 \notag \\
    & \approx \frac{1}{N M_\gamma} \sum_{j=0}^{N-1} e^{-\gamma \norm{\mathbf{x}_j-\mathbf{x^*}}^2} = p_\gamma(\mathbf{x^*}).\label{QDE: Eq estimator}
\end{align}

Hence, $\hat{p}_\gamma(\mathbf{x^*})$ as defined above converges to the Gaussian kernel density estimator $p_\gamma(\mathbf{x^*})$. The following proposition from Ref. \cite{gonzalez2022learning} illustrates the precision of the approximation.

\begin{prop}(González et. al. \cite{gonzalez2022learning})
Let $\mathcal{M}$ be a compact subset of $\mathbb{R}^D$ with diameter $diam(\mathcal{M})$, let $X = \{
\mathbf{x}_i\}_{i=0, \cdots, N-1}\subset \mathcal{M}$ a set of i.i.d samples, then $\hat{p}_\gamma$ and $p_\gamma$ (Eq. \ref{QDE: Eq estimator}) satisfy:
\begin{eqnarray}
\Pr\Big[ \sup_{\mathbf{x}\in\mathcal{M}}\abs{\hat{p}_\gamma(\mathbf{x})-p_\gamma(\mathbf{x})} \ge \epsilon \Big] \le \notag \\
2^8\Big(\frac{\sqrt{2D\gamma}diam(\mathcal{M})}{3M_\gamma \epsilon}\Big)\exp\Big(-\frac{d(3M_\gamma \epsilon)^2}{4(D+2)}\Big)\label{QDE: Eq precision of DE} 
\end{eqnarray}
\end{prop}

Thus, the DMKDE algorithm as an approximation of the KDE method \cite{Rosenblatt1956RemarksFunction, Parzen1962OnMode} can be improved by increasing the number of Fourier features $d$. This approximation has the advantage of reducing the computational complexity, since for $M$ samples, the DMKDE algorithm combined with RFF reduces the number of operations of the Parzen-Rosenblatt window from $O(NM)$ to $O(M+N)$, see Sect. \ref{QDE: Complexity analysis}, and its complexity does not grow exponentially with the size $D$ of the data, in contrast to the fast multipole method for KDE \cite{raykar2010fast, yang2004efficient, yang2003improved}.

In addition, Eqs. \ref{QDE: rho train}, \ref{QDE: density estimator}, and \ref{QDE: Eq estimator} can be extended to our proposed Q-DEMDE method, see Sect. \ref{QDE:QDE Method}, which combines a quantum implementation of the DMKDE with quantum random and quantum adaptive Fourier features for density estimation; the precision of the density estimation method (Eq. \ref{QDE: Eq precision of DE}) also applies to the Q-DEMDE method with QRFF.

\subsection{Notation for the quantum circuits}\label{QAD:DMKDE circuit pure}

We now introduce the notation of the quantum circuits used in the article, which is the same notation as in Ref. \cite{Iten2016QuantumIsometries}. Each state of the canonical basis of an $n$-qubit state can be written as $\ket{b_0b_1\cdots b_{n-1}}$ with $\{b_i\}  \in \{0, 1\}$, hence, we may write any state in the canonical basis as a string in its binary form. This notation can be simplified by writing such state as $\ket{\sum_{i=0}^{n-1}b_i2^i}_n$, namely, the bit string is written in decimal form and the subindex indicates the number of qubits of the state. For example, the 4-qubit state $\ket{1010}$ can be written as $\ket{5}_4$. We use of the same subindex to emphasize the number of qubits that make up a quantum state preparation or a unitary transformation. For example, a quantum state of size $2^n$ and a unitary matrix of size ${2^n\times2^n}$ can be built on $n$ qubits and written as $\ket{\psi}_n$ and $U_n$, respectively.

\subsection{Amplitude encoding, unitaries and isometries}\label{QDE: Explanation Isometries}

Two key ingredients of the quantum algorithms presented in this article are quantum state preparation and isometries. Several quantum protocols have been proposed for quantum state preparation, also called amplitude encoding \cite{shende2006synthesis, mottonen2005transformation}. The problem consists on the construction of an arbitrary quantum state $\ket{\psi} \in \mathbb{C}^d$ in a quantum computer. The protocol requires at least $n = \lceil\log{(d)}\rceil$ qubits and it allows to prepare the complex components of the state in the amplitudes of the basis of binary bit strings, in the form,
\begin{equation}
    \ket{\psi}_n = \sum_{i=0}^{d-1}a_i\ket{i}_n,
\end{equation}
where, $\{a_i\}$ are the complex amplitudes of the quantum state $\ket{\psi}$ and $\{\ket{i}_n\}$ is the binary basis for an $n$ qubit state written in decimal form. The depth (number of CNOTs) of the circuit required to build an arbitrary quantum state scales exponentially \cite{shende2006synthesis, mottonen2005transformation} with the number of qubits $O(2^n) = O(d)$.

The application of an arbitrary unitary transformation $U \in \mathbb{C}^{d \times d}$ on $n =  \lceil\log{(d)}\rceil$ qubits is an important quantum subroutine, whose depth is of the order $O(d^2)$. Unitary transformations are a subset of a more general transformation called isometries \cite{Iten2016QuantumIsometries}, which correspond to transformations from two Hilbert spaces that may have different dimensions, say $\mathbb{C}^r \rightarrow \mathbb{C}^d$. An isometry from $m = \log{(r)}$ to $n = \log{(d)}$ qubits can be regarded as a $\mathbb{C}^{d\times r}$ transformation built from the first $r$ columns of a $\mathbb{C}^{d\times d}$ unitary matrix; amplitude encoding can also be seen as an isometry with $r=1$. Isometries are of particular interest because the circuit depth of isometries scales in the form $O(dr)$, which has lower complexity than arbitrary unitary transformations. Table \ref{table:1} summarizes the number of qubits and circuit depth of amplitude encoding, unitary matrices, and isometries.

\begin{table}[h!]
\centering
\begin{tabular}{ | c | c | c | c |}
\hline
Subroutine & Dimension & Num. qubits & Circuit depth\\
\hline
State preparation & $d$ & $\log{(d)}$ & $d$\\
\hline
Unitary & $d\times d$ & $\log{(d)}$ & $d^2$ \\
\hline
Isometry & $d\times r$ & $\log{(d)}$ & $dr$ \\
\hline
\end{tabular}
\caption{Scale of the number of qubits and circuit depth (number of CNOTS) to build arbitrary quantum states, unitaries, and isometries.}
\label{table:1}
\end{table}

\section{Quantum-classical density matrix density estimation (Q-DEMDE)}\label{QDE:QDE Method}
\begin{figure*}
	\includegraphics[scale = 0.55]{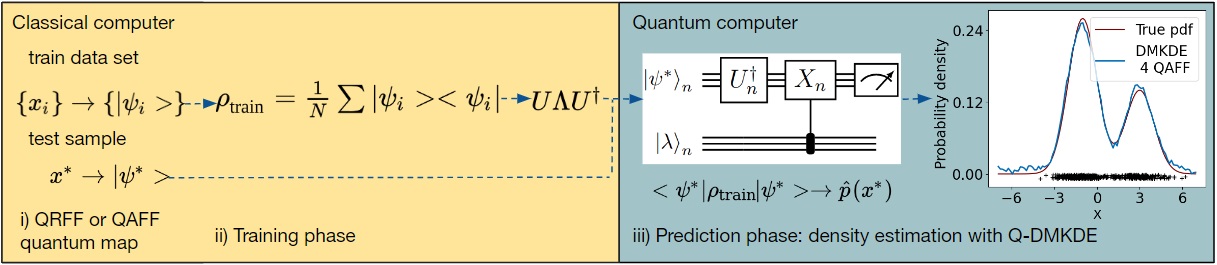}
	\caption{Q-DEMDE method for quantum-classical density estimation with density matrices and quantum Fourier features. In step i) we perform the feature map based on QRFF or QAFF in a classical computer, in step ii) we construct the training density matrix with the training data and find its spectral decomposition using a classical computer, and finally in step iii) we find the density estimate of the test sample with the training density matrix in a quantum computer using the proposed Q-DMKDE quantum algorithm.}
	\label{fig:QDE DensityMethod}
\end{figure*}

We propose a quantum-classical density matrix-based density estimation model called Q-DEMDE based on a novel quantum implementation of the DMKDE algorithm \cite{gonzalez2022learning}, see Sect. \ref{QDE: preliminaries DMKDE}, and the novel quantum random and quantum adaptive Fourier features. Fig. \ref{fig:QDE DensityMethod} shows the main steps of the Q-DEMDE method: (i) quantum feature map, (ii) training phase, and (iii) density estimation of new samples. Steps (i) and (ii) are calculated on a classical computer, while step (iii) is computed on a quantum computer. These steps are described in more detail below.

\begin{enumerate}[label=(\roman*)]
	\item Quantum feature map: Apply a feature map based on QRFF or QAFF, see Sects. \ref{QDE:Quantum Adp Fourier features} and \ref{sec: QAFF method}, in a classical computer, to the training data set $\{\mathbf{x}_j\} \rightarrow \{\ket{\psi(\mathbf{x}_j)}\}$, and to the test sample $\mathbf{x^*} \rightarrow \ket{\psi(\mathbf{x^*})}$, whose density with respect to the training data we want to estimate.
    
	\item Training phase: Use a classical computer to build the training density matrix
	$\rho_{\text{train}} = (1/N)\sum\ket{\psi(\mathbf{x}_j)}\bra{\psi(\mathbf{x}_j)}$ from the training features, as in Eq. \ref{QDE: rho train}, and calculate its spectral decomposition, $\rho_{\text{train}} = V \Lambda V^\dag$. The training mixed quantum state encodes a probability distribution of the training data set.
    
	\item Prediction phase: From the test quantum state and the spectral decomposition of training density matrix, use a quantum computer to estimate the probability density of the test sample by calculating the expectation value of the test state with the training density matrix, $M_\gamma^{-1}\bra{\psi(\mathbf{x^*})}\rho_{\text{train}}\ket{\psi(\mathbf{x^*})} = M_\gamma^{-1}\bra{\psi(\mathbf{x^*})}V \Lambda V^\dag\ket{\psi(\mathbf{x^*})}$, as in Eq. \ref{QDE: density estimator}, using the proposed Q-DMKDE quantum protocol shown in Fig. \ref{fig:QDE DMKDE circuit}. See Sect. \ref{QAD:DMKDE circuit} for the mathematical details of this quantum algorithm.

\end{enumerate}

In the following sections, we present the theoretical details of the new quantum Fourier features and the proposed Q-DMKDE quantum protocol, which estimates the expected value of an eigen-decomposed density matrix.

\subsection{Quantum random Fourier features}\label{QDE:Quantum Adp Fourier features}

To build the quantum features of the quantum density estimation method, we propose a novel embedding called quantum random Fourier features (QRFF), which correspond to RFF \cite{Rahimi2009RandomMachines} in the complex domain. QRFF build a quantum feature map that approximates a Gaussian kernel by mapping classical data samples $\mathbf{x} \in \mathbb{R}^D$ to quantum state-like representations $ \ket{\psi(\mathbf{x})} \in \mathbb{C}^d$, such that, $\hat{k}(\mathbf{x}, \mathbf{y}) = \abs{\langle\psi(\mathbf{x})\vert\psi(\mathbf{y})\rangle}^2$ approximates $k(\mathbf{x}, \mathbf{y})= e^{-\gamma{\norm{\mathbf{x}-\mathbf{y}}}^2}$. The QRFF is distinct from conventional RFF \cite{Rahimi2009RandomMachines} by approximating the kernel via $\abs{\langle\psi(\mathbf{x})\vert\psi(\mathbf{y})\rangle}^2$ instead of a dot product $\mathbf{z}_\gamma(\mathbf{x})^T\mathbf{z}_\gamma(\mathbf{y})$.

The proposed QRFF builds the following quantum mapping, which approximates a Gaussian kernel,

\begin{equation}
    \ket{\psi(\mathbf{x})} = \sqrt{\frac{1}{d}}\begin{bmatrix}e^{i\sqrt{\gamma}\mathbf{w}_0\cdot\mathbf{x}}, e^{i\sqrt{\gamma}\mathbf{w}_1\cdot\mathbf{x}}, \cdots , e^{i\sqrt{\gamma}\mathbf{w}_{d-1}\cdot\mathbf{x}}  \end{bmatrix}'\label{QDE: QAFF map}
\end{equation}

by sampling the i.i.d. weights $\{\mathbf{w}_j\} \in \mathbb{R}^D$ from $\mathcal{N}(\mathbf{0}, \mathbf{I}_D)$,  where $\mathbf{I}_D$ is the $D$-dimensional identity matrix. This method finds justification from Bochner theorem \cite{rudin2017fourier}, which states that a continuous kernel $k(\mathbf{x}-\mathbf{y})$ on $\mathbb{R}^D$ is positive definite if it is the Fourier transform of a non-negative measure. Extending the argument of Ref. \cite{Rahimi2009RandomMachines} to the quantum domain, this theorem can be applied to the Fourier relation between position $\psi(\Delta)$ and momentum $\phi(\mathbf{p})$ wavefunctions, with $\Delta = \mathbf{x} - \mathbf{y}$, i.e., $\psi(\mathbf{x}-\mathbf{y}) = (2\pi)^{-(D/2)}\int_{\mathbb{R}^D}\phi(\mathbf{p})e^{i\mathbf{p}\cdot(\mathbf{x}-\mathbf{y})}d \mathbf{p}$, the position wavefunction results from the inverse Fourier transform of the momentum wavefunction. Considering a Gaussian wavepacket $\phi(\mathbf{p}) = (\pi\gamma)^{-(D/4)} e^{-\norm{\mathbf{p}}^2/(2\gamma)}$, we have that,
\begin{eqnarray}
   (\pi/\gamma)^{D/4}\psi(\mathbf{x}-\mathbf{y}) = \notag \\ \int_{\mathbb{R}^D}(2\pi\gamma)^{-(D/2)}e^{-\norm{\mathbf{p}}^2/(2\gamma)}e^{i\mathbf{p}\cdot(\mathbf{x}-\mathbf{y})}d \mathbf{p}, \label{QDE: QAFF map integral form}
\end{eqnarray}
by writing the last integral as a sum, we have that, $E[e^{i\mathbf{p}\cdot\mathbf{x}}e^{-i\mathbf{p}\cdot\mathbf{y}}]  \approx \langle\psi(\mathbf{y})\vert\psi(\mathbf{x})\rangle$ with $\ket{\psi(\mathbf{x})} = \sqrt{1/d}\sum_{j=0}^{d-1} e^{i\mathbf{p}_j\cdot\mathbf{x}}\ket{j}$ is an unbiased estimate of $(\pi/\gamma)^{D/4}\psi(\mathbf{x}-\mathbf{y})$, when $\{\mathbf{p}_j\}$ is sampled i.i.d. from $\mathcal{N}(\mathbf{0}, \gamma\mathbf{I}_D)$, since the term that complements the complex exponential in Eq. \ref{QDE: QAFF map integral form} is a multivariate Gaussian with mean $\mathbf{0}$ and covariance matrix $\gamma\mathbf{I}_D$. Therefore, we may built the Gaussian kernel by $k(\mathbf{x}-\mathbf{y}) = e^{-\gamma{\norm{\mathbf{x}-\mathbf{y}}}^2} = (\pi/\gamma)^{D/2} \psi(\mathbf{x}-\mathbf{y})^*\psi(\mathbf{x}-\mathbf{y}) \approx \abs{\langle\psi(\mathbf{x})\vert\psi(\mathbf{y})\rangle}^2$, considering that, $\psi(\mathbf{x}-\mathbf{y}) = (\gamma/\pi)^{D/4} e^{-\gamma\norm{\mathbf{x}-\mathbf{y}}^2/2}$, for the Gaussian wavepacket in the position representation.

\subsection{Quantum adaptive Fourier features}\label{sec: QAFF method}

In analogy with adaptive Fourier features \cite{li2019learning}, we also propose quantum adaptive Fourier features (QAFF) which improves the Gaussian kernel approximation of QRFF. It uses optimization techniques to find the optimal parameters $\{\mathbf{w}_j\}$, from Eq. \ref{QDE: QAFF map}, by minimizing a loss function that reduces the distance between the actual kernel and its Fourier feature representation such as the mean square error loss function,
\begin{equation}
\mathcal{L}_{\{\mathbf{w}_j\}} = \frac{1}{T}\sum_{l=1}^{\mathcal{M}}\sum_{m=l}^{\mathcal{M}}(k(\boldsymbol{x}_l, \boldsymbol{x}_m) - \hat{k}_{\{\mathbf{w}_j\}}(\boldsymbol{x}_l, \boldsymbol{x}_m))^2, \label{QDE: QAFF loss}
\end{equation}

where $\{\boldsymbol{x}_l\}_{1\cdots \mathcal{M}}\in \mathbb{R}^D$ is refered to as kernel training data set, which is written in italics to distinguish it from the training data set of the density estimation task $\{\mathbf{x}_l\}_{1\cdots N}\in \mathbb{R}^D$. Also, $\{(\boldsymbol{x}_l, \boldsymbol{x}_m)\}_{1 \cdots T}$, is called the set of all the data pairs of the kernel training data set, clearly, $T = \mathcal{M}(\mathcal{M}+1)/2$.

To optimize the weights $\{\mathbf{w}_j\}$ of QAFF, we initialize randomly these parameters (we may use, for instance, the QRFF weights, however, its initialization is not restricted to them), and apply the usual gradient descend update rule, 
\begin{equation}
\mathbf{w}_j^{(t+1)} = \mathbf{w}_j^{(t)} - \eta \nabla_{\mathbf{w}_j}\mathcal{L}_{\{\mathbf{w}_k\}}^{(t)},\label{QDE: QAFF map update rule}
\end{equation}

where $\eta$ is the learning rate, and $t$ is the time of the iteration. 

To train the QAFF, we can use a siamese neural network to estimate $\abs{\langle\psi(\boldsymbol{x}_l)\vert\psi(\boldsymbol{x}_m)\rangle}^2$ and to fit that value as close as possible to $e^{-\gamma{\norm{\boldsymbol{x}_l-\boldsymbol{x}_m}}^2}$, as in conventional AFF \cite{li2019learning}, see Fig. \ref{fig:QDE Classical QAFF Method} for its neural architecture, however, in the Appendix. \ref{QDE sec: gradient classical QAFF}, we present an explicit mathematical expression of the gradient $\nabla_{\mathbf{w}_j}\mathcal{L}_{\{\mathbf{w}_k\}}$, which avoids the use of the siamese neural network, also showing that the complexity of training QAFF in a classical computer is $O(TdD+Td^2)$.

Once the appropriate weights ${\{\mathbf{w}_k\}}$ of the QAFF are learned, we apply the quantum Fourier map, see Eq. \ref{QDE: QAFF map}, to the training data set $\{\mathbf{x}_j\} \rightarrow \{\ket{\psi(\mathbf{x}_j)}\}$ and to the test sample $\mathbf{x^*} \rightarrow \ket{\psi(\mathbf{x^*})}$ whose probability density we aim to estimate.

\begin{figure*}
	\centering
	\includegraphics[scale = 0.458]{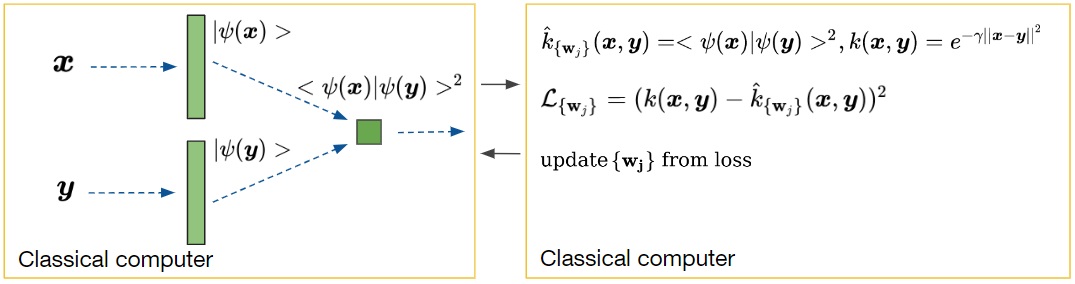} 
	\caption {Quantum adaptive Fourier features method for learning the Gaussian kernel in a classical computer. A Siamese neural network constructs the quantum Fourier map of the data pair $(\boldsymbol{x}, \boldsymbol{y}) \rightarrow (\ket{\psi(\boldsymbol{x})}, \ket{\psi(\boldsymbol{y})})$, the weights of the feature map $\{\mathbf{w}_j\}$ are the same on both feature mappings and they are randomly initialized; these weights are learned by minimizing a loss function $\mathcal{L}_{\{\mathbf{w}_j\}}$ that reduces the distance between the Gaussian kernel and its Fourier feature approximation. To approximate the Gaussian kernel with QAFF, we explore two alternative kernel training data sets: the Gaussian kernel training data set and the ML kernel training data set, see Sect. \ref{QAFF kernel traning data set}.}
	\label{fig:QDE Classical QAFF Method}
\end{figure*}

\subsubsection{The kernel training data set to optimize the QAFF} \label{QAFF kernel traning data set}

One key element in the trainability of the quantum adaptive Fourier features is the kernel training data set $\{\boldsymbol{x}_l\}_{1\cdots \mathcal{M}}$, and its data pairs $\{(\boldsymbol{x}_l, \boldsymbol{x}_m)\}_{1 \cdots T}$. In this section, we present two alternatives as kernel training data sets to train the QAFF.

\textit{The Gaussian kernel training data set}: The main kernel training data set used in this article is the Gaussian data set. To approximate the Gaussian kernel $e^{-\gamma{\norm{\boldsymbol{x}_l-\boldsymbol{x}_m}}^2}$, whose original space is $\boldsymbol{x} \in \mathbb{R}^D$, we build a synthetic data set $\{\boldsymbol{x}_l\}_{1\cdots \mathcal{M}}$ sampled from a multivariate normal distribution $\mathcal{N}(\mathbf{0}, \frac{1}{2\gamma D}\mathbf{I}_D)$. In addition, since the Gaussian kernel is shift-invariant, in this kernel training strategy we set one of the elements of each data pair equal to $\mathbf{0}$, so the new data pairs are $\{(\boldsymbol{x}_l, \mathbf{0})\}_{1 \cdots \mathcal{M}}$, and the loss function to optimize becomes $\mathcal{L}_{\{\mathbf{w}_j\}} = \frac{1}{\mathcal{M}}\sum_l( \abs{\langle\psi(\boldsymbol{x}_l)\vert\psi(\mathbf{0})\rangle}^2 - e^{-\gamma\norm{\boldsymbol{x}_l}^2})^2$. The motivation of this kernel learning strategy is that the Gaussian data resulting from sampling $\mathcal{N}(\mathbf{0}, \frac{1}{2\gamma D}\mathbf{I}_D)$ is equivalent to sampling from the kernel function $e^{-\gamma{\norm{\boldsymbol{x}}^2}}$, so there would be a greater number of kernel training points in the regions with larger values of the kernel function, leading to better approximation of the kernel in those regions. Furthermore, since $\gamma$ acts as a scaling factor in this kernel learning strategy, the learned weights do not depend on $\gamma$, and we can simplify the method by setting $\gamma = 0.5$, nonetheless, when predicting the QAFF mapping on the training data set $\{\mathbf{x}_j\}_{1 \cdots N}$ and testing sample $\mathbf{x^*}$, which are distinct from the kernel training data samples, it is necessary to choose the appropriate $\gamma$ value that approximates the desired kernel as in QRFF.

\textit{The ML problem kernel training data set}: When solving a machine learning (ML) problem with training data points $\{\mathbf{x}_j\}_{1\cdots N} \in \mathbb{R}^D$, one can use a subset of samples from training data set $\{\boldsymbol{x}_l\}_{1\cdots \mathcal{M}} \subseteq \{\mathbf{x}_j\}_{1\cdots N}$ to approximate the Gaussian kernel for the training data set. Namely, in this setting the loss function, see Eq. \ref{QDE: QAFF loss}, is optimized using the data pairs $\{(\boldsymbol{x}_l, \boldsymbol{x}_m)\}_{1 \cdots T}$, which are samples from the original training data. This approach has been explored in conventional adaptive Fourier features \cite{li2019learning}, however, it does not guarantee an appropriate approximation of the kernel for some novel data that deviates from the training data of the ML problem.

\subsubsection{Additional: Quantum variational circuit implementation of QAFF} \label{QAFF Quantum implementation}

\begin{figure*}
	\centering
	\includegraphics[scale = 0.458]{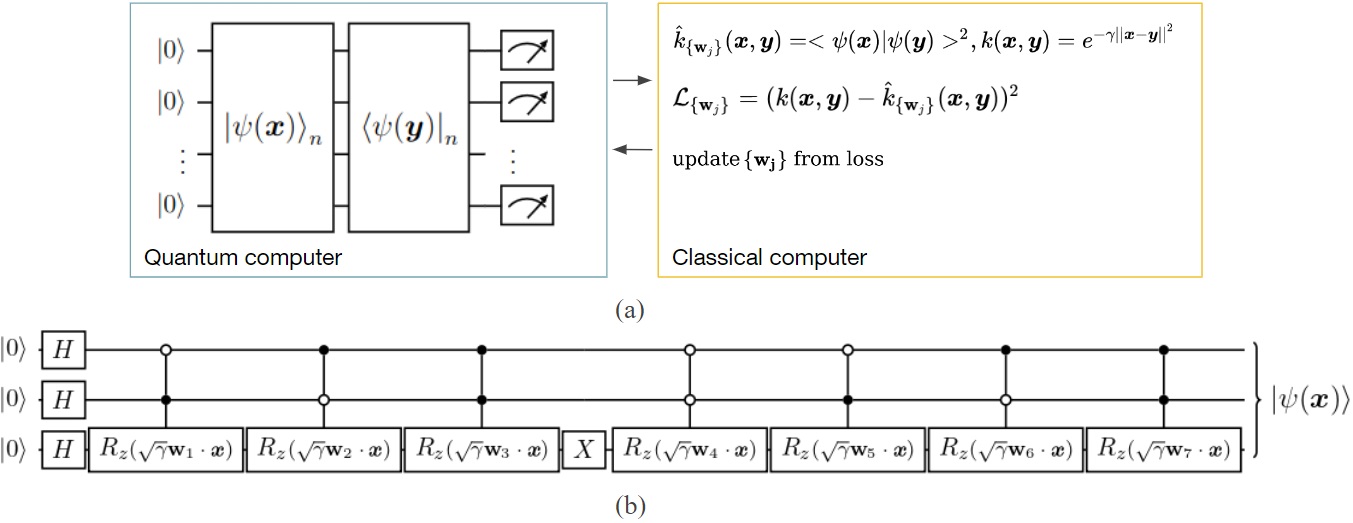} 
	\caption {Althought the Q-DEMDE uses classical QAFF, we present an alternative variational algorithm to train the QAFF in quantum hardware. (a) Hybrid QAFF strategy with quantum kernel estimation protocol \cite{liu2021rigorous}. (b) Example of a quantum circuit to prepare the QAFF with 8 Fourier features and three qubits by means of multicontrolled $R_z$ rotations, however, we may also prepare this state with single $R_z$ rotations and CNOT gates as explained in Appendix \ref{QDE: QAFF 1-2 gates}.}
	\label{fig:QDE QAFF Method}
\end{figure*}

Although the loss function of the proposed classical QAFF method can be trained with classical computers using siamese neural networks as in AFF \cite{li2019learning} or by computing the gradient of the loss function, see Eq. \ref{QDE: eq: gradient loss function}, for completeness, we also present a quantum variational strategy to optimize in a quantum computer the loss function (Eq. \ref{QDE: QAFF loss}) of the proposed QAFF method, see Fig. \ref{fig:QDE QAFF Method}.

We show how to prepare in a quantum computer the QAFF mapping $\ket{\psi(\boldsymbol{x})}=\sqrt{\frac{1}{d}}\begin{bmatrix}1, e^{i\sqrt{\gamma}\mathbf{w}_1\cdot\boldsymbol{x}}, \cdots , e^{i\sqrt{\gamma}\mathbf{w}_{d-1}\cdot \boldsymbol{x}}  \end{bmatrix}'$, which learns variationally the optimal weights of the QRFF (note that the global phase might be ignored by setting $\mathbf{w}_0 = \mathbf{0}$). To prepare this $n = \log{d}$ qubit state, we may use a method based on amplitude encoding \cite{shende2006synthesis}, it consists on applying the unitary $H^{\otimes n}$ on the initial $\ket{0}^{\otimes n}$ state which builds the quantum state $\ket{+}^{\otimes n}$, then it applies a series $d-1$ of uniformly controlled $R_z(\alpha) = \ketbra{0}{0}+e^{i\alpha}\ketbra{1}{1}$ rotations with angles $\{\sqrt{\gamma}\mathbf{w}_j\cdot\boldsymbol{x}\}_{j=1, \cdots, d-1}$, whose controls correspond to the first $n-1$ qubits and its target corresponds to the $n^{\text{th}}$ qubit, each $k^\text{th}$ uniformly controlled rotation works by controlling a quantum state of the first $n-1$ qubits corresponding to the bitstring $\ket{b_0^kb_1^k\cdots b_{n-2}^k}$ with $\{b_i^k\} \in \{0, 1\}$ where the $\circ$ symbol indicates a control on the $\ket{0}$ qubit state and the $\bullet$ symbol indicates a control on the $\ket{1}$ qubit state, as shown in Fig. \ref{fig:QDE QAFF Method}b. By writing the bit string in decimal form $\ket{b_0^kb_1^k\cdots b_{n-2}^k} = \ket{k \ (\text{mod}(d/2))}_{n-1}$, we may define $U_\text{uc}^k$ as the $k^\text{th}$ uniformly controlled $R_z$ rotation with angle $\sqrt{\gamma}\mathbf{w}_k\cdot \boldsymbol{x}$, control the $\ket{k \ (\text{mod}(d/2))}_{n-1}$ state and target the $n^\text{th}$ qubit, this operation performs the following phase trasformation to the $n^\text{th}$ qubit when the control is the $\ket{k \ (\text{mod}(d/2))}_{n-1}$ state, $U_\text{uc}^k(\ket{k \ (\text{mod}(d/2))}_{n-1}\otimes 2^{-1/2}(\ket{0} + \ket{1})) = \ket{k \ (\text{mod}(d/2))}_{n-1}\otimes 2^{-1/2}(\ket{0} + e^{i\sqrt{\gamma}\mathbf{w}_k\cdot \boldsymbol{x}}\ket{1})$, the desired QAFF mapping is obtained by applying the $d-1$ uniformly controlled rotations $\ket{\psi(\boldsymbol{x})} =(\prod_{k = d/2}^{d-1} U_\text{uc}^k) (I^{\otimes n-1}\otimes X) (\prod_{k=1}^{d/2-1} U_\text{uc}^k) \ket{+}^{\otimes n}$; this method requires an $X$ gate in the middle to change the phases of both the $\ket{0}$ and the $\ket{1}$ basis states of the $n^\text{th}$ qubit; Fig \ref{fig:QDE QAFF Method}b shows a particular 3-qubit circuit example to prepare the QAFF. In Appendix \ref{QDE: QAFF 1-2 gates}, we illustrate a proposal to prepare the QAFF with single $R_z$ rotations and CNOTs gates, instead of uniformly controlled rotations, by performing the reparametrization $\{\sqrt{\gamma}\mathbf{w}_j\cdot\boldsymbol{x}\}\rightarrow\{\sqrt{\gamma}\boldsymbol{\theta}_j\cdot\boldsymbol{x}\}$, where the latter are the angles of the single-qubit $R_z$ rotations; we also show that the depth of the algorithm, measured in the number of CNOTs, has order $O(d)$, which is independent of the size $D$ of the original space.

Once $\ket{\psi(\boldsymbol{x})}$ and $\ket{\psi(\boldsymbol{y})}$ are prepared, we can compute $\hat{k}_{\{\mathbf{w}_j\}}(\boldsymbol{x}, \boldsymbol{y}) = \abs{\langle\psi(\boldsymbol{x})\vert\psi(\boldsymbol{y})\rangle}^2$, by the quantum kernel estimation protocol \cite{liu2021rigorous}, see Fig. \ref{fig:QDE QAFF Method}a. Finally, to optimize the loss function, see Eq. \ref{QDE: QAFF loss}, we initialize the components of the variational parameters $\{\mathbf{w}_j\}$ randomly from $U[0, 1]$, and use parameter-shift rule \cite{mitarai2018quantum} to estimate the gradients. 

The total complexity to train the QAFF in quantum computers is $O(TDd + Td^2R)$, where $T$ is, as before, the number of data pairs $\{(\boldsymbol{x}_l, \boldsymbol{x}_m)\}_{1\cdots T}$ and $R$ is the number of shots to estimate the gradients, see Appendix \ref{QDE sec: gradient classical QAFF}. This quantum complexity does not improve the classical complexity of $O(TDd+Td^2)$, since the quantum algorithm requires $R$ measurements of the kernel estimation circuit to get a good estimate of the gradients of the feature mapping. Considering that the classical QAFF has a lower complexity than the quantum QAFF and that the Q-DEMDE method requires an intermediate classical step to build the training density matrix, we choose to use classical QAFF to prepare the quantum mapping, however this quantum implementation may have applications in other quantum algorithms that do not require an intermediate classical preprocessing.

\subsection{Q-DMKDE quantum circuit: estimating expectation values of density matrices}\label{QAD:DMKDE circuit}

\begin{figure}
	\centering
	\includegraphics[scale =0.16]{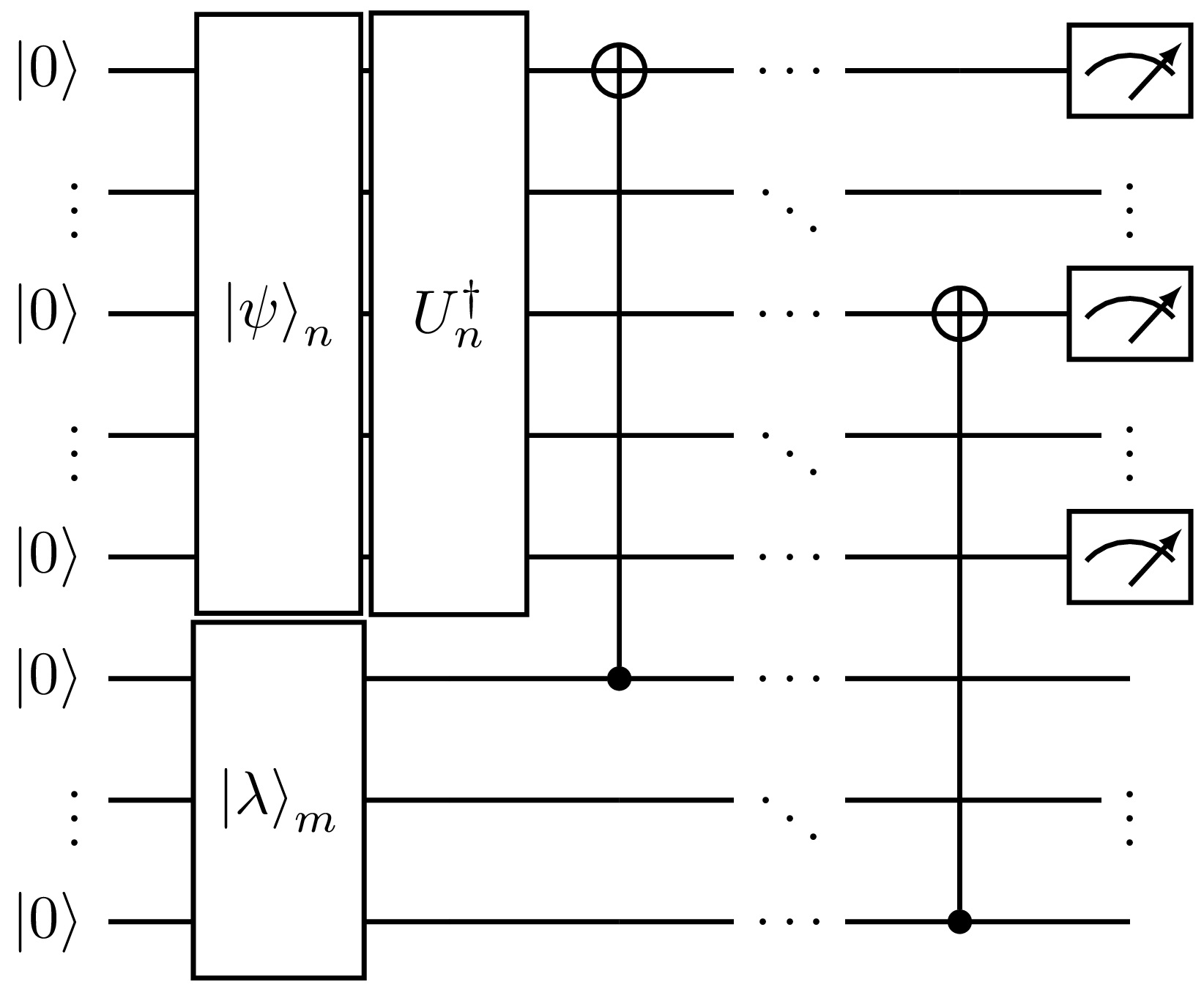}
	\caption{Q-DMKDE quantum circuit to estimate the expected value of a density matrix from its spectral decomposition. The expected value of the density matrix is estimated by $P(\ket{0}_n) = \bra{\psi}\rho\ket{\psi} = \bra{\psi}V\Lambda V^\dag\ket{\psi}$, where $\ket{\psi}$ and $V$ are prepared in the first $n$ qubits, and $\Lambda$ in the last $m$ qubits.}
	\label{fig:QDE DMKDE circuit}
\end{figure}

To implement the prediction phase of the Q-DEMDE, or the classical DMKDE in a quantum computer, see Eq. \ref{QDE: density estimator}, we also propose a novel quantum protocol called Q-DMKDE to estimate the expected value of a mixed state density matrix from its eigendecomposition in a qubit-based quantum computer, see Fig. \ref{fig:QDE DMKDE circuit}. This circuit extends to qubits a previous method for estimating the expected value of a density operator in a high-dimensional quantum computer \cite{Useche2021QuantumQudits}.

We want to compute the expected value of a density matrix $\rho \in \mathbb{C}^{d\times d}$ of rank $r \le d$ with a quantum state $\ket{\psi} \in \mathbb{C}^d$ in a quantum computer. This calculation requires $n + m$ qubits, with $ n = \lceil \log{(d)} \rceil$ and $m = \lceil \log{(r)} \rceil$. The first $n$ qubits prepare the state $\ket{\psi}$ and the unitary matrix $V^\dag$ whose first $r$ rows are the complex conjugate eigenvectors of $\rho$, and the remaining $m$ qubits prepare the eigenvalues of the density matrix.

To begin with, as noted in \cite{Useche2021QuantumQudits}, we have that,
\begin{eqnarray}
 \bra{\psi}\rho\ket{\psi} =   \bra{\psi}V\Big(\sum_{i=0}^{r-1}\lambda_i\ket{i}\bra{i}+\sum_{i=r}^{d-1}0\ket{i}\bra{i}\Big) V^{\dag}\ket{\psi} \notag \\ = \sum_{i=0}^{r-1}\lambda_i\abs{\bra{i}V^\dag\ket{\psi}}^2, \label{QAD:eq8}
\end{eqnarray}
where, $V \in \mathbb{C}^{d\times d}$ is a unitary matrix whose first $r$ columns are the eigenvectors of $\rho$ and $\Lambda = \sum_{i=0}^{r-1}\lambda_i\ket{i}\bra{i}$ is the diagonal matrix of eigenvalues.

The proposed Q-DMKDE quantum circuit starts by initializing the first $n$ qubits with the state $\ket{\psi}_n$, see the notation in Sect. \ref{QAD:DMKDE circuit pure}, and the remaining $m$ qubits with the state $\ket{\lambda}_m=\sum_{j=0}^{r-1}\sqrt{\lambda_j}\ket{j}_m$ that encodes the eigenvalues of $\rho$, see Fig. \ref{fig:QDE DMKDE circuit}. This operation can be carried out thanks to amplitude encoding \cite{shende2006synthesis, mottonen2005transformation}. We then have the complete state
\begin{equation}
\ket{\psi}_n\otimes\ket{\lambda}_m = \ket{\psi}_n \otimes \sum_{j=0}^{r-1}\sqrt{\lambda_j}\ket{j}_m.
\end{equation}
Next, we build the unitary matrix $U^\dag_n$ in the first $n$ qubits, see Eq. \ref{QAD:eq2}, whose first quadrant is composed of the unitary matrix $V^\dag$ and its fourth quadrant with an identity matrix $I$ with rank $2^n-d$,
\begin{equation}
	U_n^\dag =
\left( \begin{array}{c|c}
   V^\dag & 0  \\
   \hline
	0 & I \\
\end{array}\right), \label{QAD:eq2}
\end{equation}
this unitary matrix $U^\dag_n$ is the conjugate transpose of $U_n$. Since the eigenvectors of $\rho$ correspond to the first $r$ columns of $U$, we can build $U_n$ as an isometry from $m$ to $n$ qubits \cite{Iten2016QuantumIsometries}, see Sect. \ref{QDE: Explanation Isometries}. We would have,
\begin{equation}
U^\dag_n \ket{\psi}_n \otimes \sum_{j=0}^{r-1}\sqrt{\lambda_j}\ket{j}_m.
\end{equation}
We can write the state of the first $n$ qubits as, $U^\dag_n \ket{\psi}_n = \sum_{i=0}^{d-1}a_i\ket{i}_n$. Where,
\begin{equation}
	\abs{a_i}^2=\abs{\bra{i}_nU^\dag_n\ket{\psi}_n}^2 = \abs{\bra{i}V^\dag\ket{\psi}}^2, \label{QAD:eq3}
\end{equation}
namely, $\abs{a_i}^2$ is the probability to measure $U^\dag_n \ket{\psi}_n$ in the canonical state $\ket{i}_n$. Therefore, the circuit leads,
\begin{eqnarray}
U^\dag_n \ket{\psi}_n \otimes \ket{\lambda}_m  =  \sum_{i=0}^{d-1}a_i\ket{i}_n \otimes \sum_{j=0}^{r-1}\sqrt{\lambda_j}\ket{j}_m = \notag \\
  \sum_{i=0}^{r-1}a_i\sqrt{\lambda_i}\ket{i}_n \otimes \ket{i}_m + \sum_{\{(i, j): i\ne j\}}a_i\sqrt{\lambda_j}\ket{i}_n  \otimes \ket{j}_m. \label{QDE:Eq temp state}
\end{eqnarray}

Then, as shown in Fig. \ref{fig:QDE DMKDE circuit}, we apply a cascade of $m$ CNOT gates between the first and second halves of the circuit.  The $i^{\text{th}}$ CNOT operates with control the $(i+n)^{\text{th}}$ qubit and target the $i^{\text{th}}$ qubit, with $i \in \{0, \cdots, m-1\}$. The effect of this series of $m$ CNOT gates can be observed by writing,
\begin{equation}
\ket{i}_n  \otimes \ket{j}_m = {\textstyle \ket{\sum_{k=0}^{n-1}b_k^i2^k}_n \otimes \ket{\sum_{l=0}^{m-1}b_l^j2^l}_m},
\end{equation}
where, $\ket{b_0^ib_1^i\cdots b_{n-1}^i}$ and $\ket{b_0^jb_1^j\cdots b_{m-1}^j}$  are the binary representations of $\ket{i}_n$ and $\ket{j}_m$ respectively. The $m$ CNOT gates have the same depth as a single CNOT, as they can be parallelized in a quantum computer \cite{cincio2018learning}.

We represent this series of $m$ CNOT gates with the $(n+m)$-qubit unitary operation $U_{n+m}^{\text{CN}}$, where CN stands for CNOT. The outcome of this transformation is a qubit-wise summation modulo 2 (denoted by $\oplus$) on the first $m$ qubits of the circuit,
\begin{eqnarray}
    U_{n+m}^{\text{CN}}(\ket{i}_n  \otimes \ket{j}_m) = \notag \\ 
    {\textstyle \ket{\sum_{k=0}^{m-1}(b_k^i\oplus b_k^j)2^k + \sum_{k=m}^{n-1}b_k^i2^k}_n \otimes \ket{\sum_{l=0}^{m-1}b_l^j2^l}_m} \notag \\
    = {\textstyle \ket{\sum_{k=0}^{n-1}(b_k^i\oplus b_k^j)2^k}_n \otimes \ket{\sum_{l=0}^{m-1}b_l^j2^l}_m} \notag
\end{eqnarray}

In the last line, we set $b_k^j=0$ for $k \ge m$.  If $i=j$, we have that,
\begin{equation}
U_{n+m}^{\text{CN}}(\ket{i}_n  \otimes \ket{i}_m) = {\textstyle \ket{0}_n \otimes \ket{\sum_{l=0}^{n-1}b_l^i2^l}_m} \notag
\end{equation}
In contrast, if $i \ne j$, the state $\ket{\sum_k(b_k^i\oplus b_k^j)2^k}_n$ would be distinct to the $\ket{0}_n$ state.

Therefore, after applying the series of $m$ CNOT gates to the state of Eq. \ref{QDE:Eq temp state}, the resulting state of the Q-DMKDE quantum circuit is
\begin{eqnarray}
\sum_{i=0}^{r-1}a_i\sqrt{\lambda_i}\ket{0}_n \otimes \ket{i}_m \notag \\+ \sum_{\{(i, j): i\ne j\}}a_i\sqrt{\lambda_j}{\textstyle\ket{\sum_k(b_k^i\oplus b_k^j)2^k}_n}  \otimes \ket{j}_m. \notag 
\end{eqnarray}
By measuring the first $n$ qubits the probability of state $\ket{0}_n$ would be,
\begin{equation}
P(\ket{0}_n) = \sum_{i=0}^{r-1} \abs{a_i}^2 \lambda_i =
\sum_{i=0}^{r-1}\lambda_i\abs{\bra{i}V^\dag\ket{\psi}}^2 = \bra{\psi}\rho\ket{\psi},\notag
\end{equation}
see Eqs. \ref{QAD:eq8} and \ref{QAD:eq3}.

Assuming we have access to the eigenvectors and eigenvalues of the density matrix, the total complexity of the Q-DMKDE quantum algorithm is $O(drS)$ since it takes $O(d)$ to prepare $\ket{\psi}_n$, $O(r)$ to prepare $\ket{\lambda}_m$, and $O(dr)$ to construct the isometry $U_n$, see Table \ref{table:1}. In addition, we need to perform $S$ measurements of the $\ket{0}_n$ bit string. Note that the classical algorithm has complexity $O(dr)$, therefore to achieve quantum advantage it is required to approximate the states $\ket{\psi}_n$, $\ket{\lambda}_m$ and the unitary matrix $U_n$, with quantum circuits of depth $O(\log{d})$. Furthermore, the ansatz used in the Q-DMKDE quantum algorithm can be used to initialize a density matrix from its spectral decomposition starting from the $ \ket{0}_{n}\otimes\ket{0}_m$ state, as illustrated in the SDM (spectral density matrix) ansatz in Appendix \ref{QDE sec:mixedstateinitialization2}. 

\subsection{Computational complexity analysis}\label{QDE: Complexity analysis}

In this section, we present the computational complexity of the proposed quantum-classical density estimation strategy Q-DEMDE using QRFF and QAFF, and we compare it with the complexities of the classical density estimation algorithms KDE \cite{Rosenblatt1956RemarksFunction, Parzen1962OnMode} and DMKDE \cite{gonzalez2022learning}. For this purpose, we set $T$ sample pairs to train classically the QAFF, $N$ training data points which build the training density matrix (assuming $T>N$), and $M$ test samples whose density we want to estimate, recalling that QRFF and QAFF map $D$-dimensional classical data points to a quantum feature space of dimension $d$, and $r$ is the rank of the training density matrix. Furthermore, we set $S$, the number of quantum circuit measurements to perform the density estimation with Q-DMKDE algorithm. Table \ref{table: complexity parameters} summarizes the parameters of the density estimation methods.

\begin{table}[h!]
\centering
\begin{tabular}{ | c | c |}
\hline
Parameter & Description \\
\hline
$N$ & No of training samples \\
\hline
$M$ & No of testing samples \\
\hline
$T$ & No of data pairs to train QAFF \\
\hline
$D$ & Size of original data space \\
\hline
$d$ & Size of the quantum feature space\\
\hline
$r$ & No of eigenvalues of the density matrix\\
\hline
$S$ & No of DE shots with Q-DMKDE\\
\hline
$\gamma$ & Bandwidth of the Gaussian kernel\\
\hline
\end{tabular}
\caption{Summary of the parameters involved in the classical and quantum-classical density estimation algorithms KDE, DMKDE, Q-DEMDE.}
\label{table: complexity parameters}
\end{table}

Table \ref{table:0} summarizes the computational complexity of the training and prediction phase from top to bottom of the methods: kernel density estimation (KDE) \cite{Rosenblatt1956RemarksFunction, Parzen1962OnMode}, DMKDE in classical hardware (C-DMKDE) \cite{gonzalez2022learning}, and the proposed quantum-classical DE method Q-DEMDE. For the C-DMKDE and Q-DEMDE we illustrate the complexity of the algorithms using both QRFF and QAFF and separate the complexity into training and testing stages, where the training phase consists of the construction of the training density matrix in classical or quantum hardware, see Eq. \ref{QDE: rho train}, and the test phase consists of the computation of the density estimation over test samples, see Eq. \ref{QDE: density estimator}. We present the complexity analysis of each method separately and then the comparison between the methods.

\begin{table*}
\centering
\begin{tabular}{ | c | c | c | c |}
\hline
DE Method & Quantum map & Train complexity & Test complexity \\
\hline
\multirow{2}*{KDE \cite{Rosenblatt1956RemarksFunction, Parzen1962OnMode}}  & \multirow{2}*{-} & \multirow{2}*{-} & \multirow{2}*{$O(NMD)$}  \\
&  &  & \\
\hline
\multirow{2}*{C-DMKDE \cite{gonzalez2022learning}} & QRFF & $O(Nd\max{(d, D)})$ & \multirow{2}*{$O(Md\max{(d, D)})$}\\
 & QAFF & $O(Td\max{(d, D)})$ & \\
\hline
\multirow{2}*{Q-DEMDE} & QRFF & $O(Nd\max{(d, D)}+ d^2r)$ & \multirow{2}*{$O(MDd + MdrS)$} \\
& QAFF & $O(Td\max{(d, D)}+ d^2r)$ & \\
\hline
\end{tabular}
\caption{Comparison of the computational complexity of the classical and quantum density estimation methods with QRFF and QAFF in training and prediction phases, with $T$ sample pairs to train the QAFF, $N$ training data points, $M$ test samples, $D$ the dimension of the classical data, $d$ the dimension of the quantum feature map, $r$ the number of eigenvalues of the training density matrix, and $S$ the number of measurements in a quantum computer to find the density estimates with the Q-DMKDE algorithm.}
\label{table:0}
\end{table*}

\subsubsection{Computational complexity of KDE}

The computational complexity for simultaneously training and testing the KDE algorithm \cite{Rosenblatt1956RemarksFunction, Parzen1962OnMode} is $O(NMD)$, since it involves the estimation of the kernel between all the testing samples with all the training samples which leads to $O(NM)$ operations and each kernel calculation has complexity $O(D)$.

\subsubsection{Computational complexity of C-DMKDE}

The complexity of the C-DMKDE \cite{gonzalez2022learning} with QAFF in the training phase is $O(Td\max{(d, D)})$, since it takes $O(T(Dd+d^2))$ to train the QAFF in a classical computer, see Appendix. \ref{QDE sec: gradient classical QAFF}, and $O(Nd^2)$ to prepare the training density matrix, and the complexity of the testing phase is $O(Md\max{(d, D)})$, since it is required $O(MDd)$ to build the quantum features of the testing samples and $O(Md^2)$ to find the density estimates by computing the expectation value between each test quantum state and the training density matrix in a classical computer. The complexity analysis of the C-DMKDE with QRFF is similar to the case of QAFF, except that the training complexity changes from $O(Td\max{(d, D)})$ to $O(Nd\max{(d, D)})$, since it takes $O(NDd)$ to build the QRFF of the training samples.

\subsubsection{Computational complexity and circuit depth of Q-DEMDE}

The training complexity of the proposed quantum-classical density estimation model Q-DEMDE with QAFF is $O(Td\max{(d, D)}+ d^2r)$, since it takes $O(TDd + Td^2)$ to train classically the quantum adaptive Fourier features, see Appendix. \ref{QDE sec: gradient classical QAFF}, and $O(NDd)$ to estimate the QAFF map of the training data set, it also takes $O(Nd^2)$ to classically construct the training density matrix, and $O(d^2r)$ to classically compute its $r$ eigenvalues and eigenvectors. The complexity of the prediction step of the Q-DEMDE with QAFF is $O(MDd + MdrS)$ which can be divided into $O(MDd)$ to build classically the QAFF of the test samples and $O(MdrS)$ to estimate in a quantum computer their density values, because for each test sample, it takes $O(d)$ and $O(dr)$ to prepare the QAFF test state and the training density matrix respectively using the Q-DMKDE quantum algorithm, see Sect. \ref{QAD:DMKDE circuit}, and it is required to measure the $\ket{0}_n$ bit string $S$ times to obtain a good estimate of the density value. As in the C-DMKDE, the complexity of the Q-DEMDE with QRFF resembles the complexity of Q-DEMDE with QAFF, except that in the training phase its complexity changes to $O(Nd\max{(d, D)} + d^2r)$, since it is not required to optimize the quantum features.

\begin{table}[h!]
\centering
\begin{tabular}{ | c | c | c |}
\hline
Q-DEMDE Subroutine & Num. of qubits & Circuit depth\\
\hline
QRFF & $\log{(d)}$ & $d$\\
\hline
QAFF & $\log{(d)}$ & $d$\\
\hline
Q-DMKDE & $\log{(dr)}$ & $dr$ \\
\hline
\end{tabular}
\caption{Number of qubits and approximate depth (number of CNOTS) of the Q-DEMDE quantum-classical subroutines: QAFF, QRFF and Q-DMKDE, in terms of the number of Fourier features $d$ (same size of the training density matrix) and rank $r$ of the eigendecomposition of the training density matrix. Here we assume $d$, $r$ as powers of two}
\label{table:4}
\end{table}

To illustrate the circuit depth of the Q-DEMDE algorithm, we summarize in Table \ref{table:4} the approximate depth (number of CNOTS) and the qubit count of the quantum-classical subroutines QAFF, QRFF and Q-DMKDE. It is worth highlighting that the depth of the QRFF and QAFF does not depend on the size of the classical features $D$, but on the size of the Fourier representation $d$, which allows to represent high dimensional data in a quantum computer.

\subsubsection{Computational complexity comparison}

We summarize the complexity of the classical KDE, C-DMKDE and quantum-classical Q-DEMDE density estimation algorithms in Table \ref{table:0}. This complexity analysis shows that the Q-DEMDE reduces the computational complexity of KDE in the regime where $N$ is larger than $d$, and $ND$ is larger than $drS$, nevertheless, it also illustrates no quantum advantage of using the Q-DEMDE  over the classical DMKDE. Indeed, we observe that the proposed Q-DEMDE requires more resources compared to the classical algorithm since it needs to perform multiple measurements to find good estimates of the probability density values; this complexity drawback is shared by some other quantum machine learning algorithms of the state-of-art \cite{schuld2020circuit, vargas2022optimisation, watkins2023quantum} that rely on estimating expectation values on quantum hardware, so solving this challenge represents an opportunity for the design of future quantum machine learning algorithms \cite{cardoso2021detailed}.

\subsection{Alternative of the spectral decomposition: quantum-classical non-orthogonal density matrix density estimation (QNO-DEMDE)}\label{QDE: QNO-DEMDE}

We present an additional model called QNO-DEMDE, for quantum-classical density estimation with non-orthogonal density matrices. The method is an alternative of the Q-DEMDE method that does not require the spectral decomposition of the training density matrix, however, in Sect. \ref{Sec: Complexity QNO-DMKDE quantum algorithm}, we illustrate that the proposed Q-DEMDE is a more optimal method because the QNO-DEMDE has greater computational complexity. The steps of the QNO-DEMDE method are as follows: i) Use a classical computer to apply a feature map to the training data samples $\{\mathbf{x}_j\} \rightarrow \{\ket{\psi(\mathbf{x}_j)}\}$ and the test sample $\mathbf{x^*} \rightarrow \ket{\psi(\mathbf{x^*})}$ based on quantum random or quantum adaptive Fourier features, see Sect. \ref{QDE:Quantum Adp Fourier features}, ii) Use a quantum computer to prepare the training density matrix $\rho_\text{train}$, see Eq. \ref{QDE: rho train}, and to estimate the probability density of the test sample $\mathbf{x^*}$, see Eq. \ref{QDE: density estimator}, by means of the QNO-DMKDE quantum circuit. 

\subsubsection{QNO-DMKDE quantum algorithm}\label{NO-DMKDE quantum algorithm}

\begin{figure}
	\centering
	\includegraphics[scale =0.14]{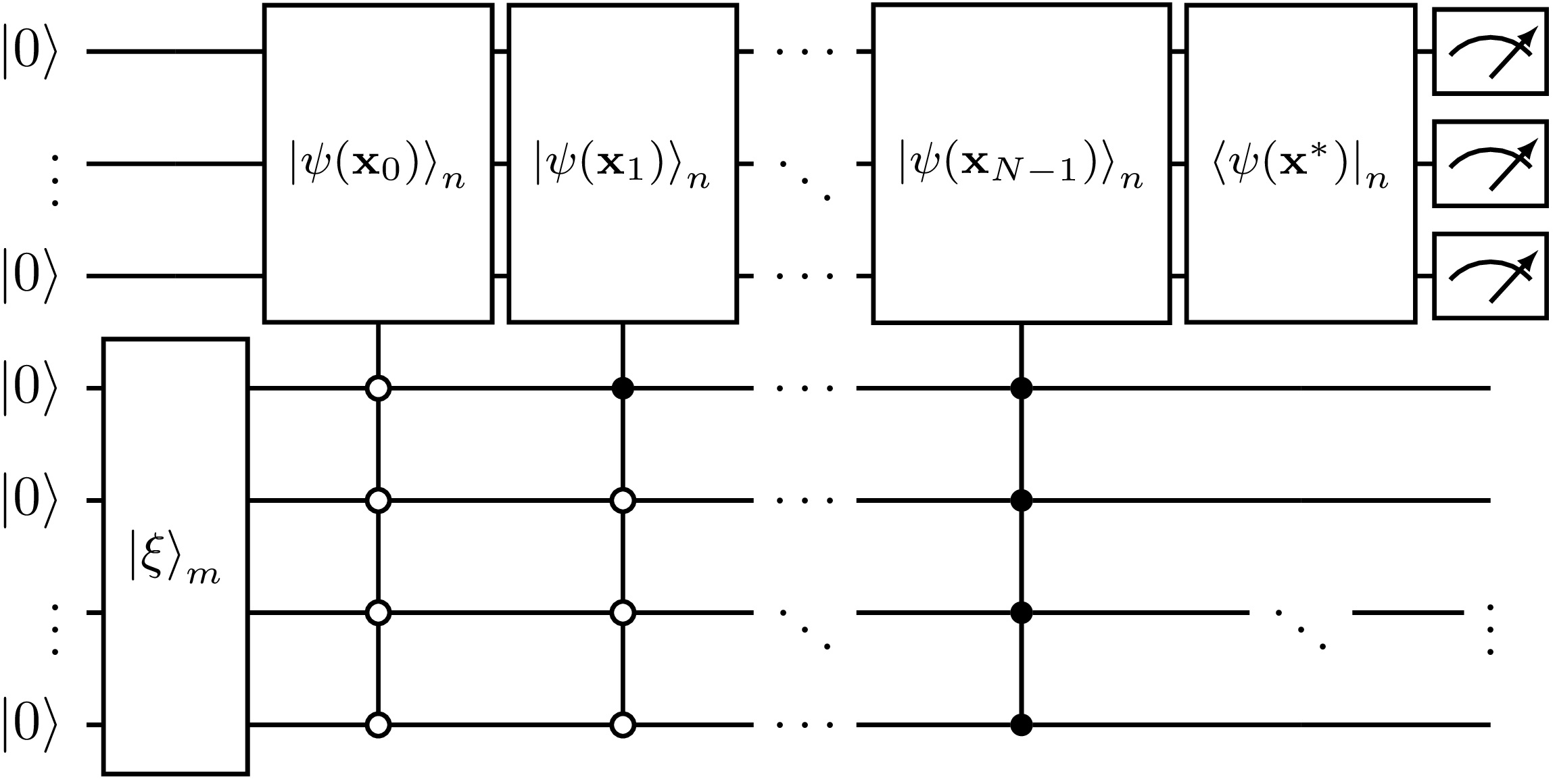}
	\caption{QNO-DMKDE quantum circuit to estimate the expected value of a density matrix as an ensemble of pure states, here $\ket{\xi}_m = \sum_{i=0}^{N-1} \sqrt{1/N} \ket{i}_m$, $\{\ket{\psi(\mathbf{x}_j)}\}$ are the training features and $\ket{\psi(\mathbf{x^*})}$ is the test feature whose density we aim to estimate.}
	\label{fig:QDE NO-DMKDE circuit}
\end{figure}

The quantum non-orthogonal DMKDE (QNO-DMKDE) estimates the expected value of the training density matrix as a mixture of non-orthogonal pure states $\bra{\psi(\mathbf{x^*})}\rho_{\text{train}}\ket{\psi(\mathbf{x^*})}$, where $\rho_{\text{train}} = (1/N)\sum_{j=0}^{N-1}\ket{\psi(\mathbf{x}_j)}\bra{\psi(\mathbf{x}_j)} $, and $\{\ket{\psi(\mathbf{x}_j)}\}$, $\ket{\psi(\mathbf{x^*})}$ $\in \mathbb{C}^d$ are the quantum features of the training data set and the test sample respectively. In contrast to the Q-DMKDE, this quantum algorithm does not require the spectral decomposition of $\rho_\text{train}$. The QNO-DMKDE algorithm requires $(m + n)$ qubits, such that $n = \lceil\log{d}\rceil$ and $m = \lceil\log{N}\rceil$. The method starts by building a purification of the training density matrix by initializing the pure state $\ket{\xi}_m = \sum_{i=0}^{N-1} \sqrt{1/N} \ket{i}_m$, on the last $m$ qubits, followed by a series of $N$ uniformly controlled isometries in the form of state preparations, with controls the last $m$ qubits and targets the first $n$ qubits; these controlled state preparations are applied following the structure of a multiplexer gate \cite{mottonen2004quantum, shende2006synthesis}; the $i^{\text{th}}$ controlled isometry sets the $\ket{i}_m$ state from the second half as the control and target the first $n$ qubits, with the isometry $U_n^i$, that corresponds to a quantum state preparation $U_n^i\ket{0}_n \rightarrow \ket{\psi(\mathbf{x}_i)}_n $, as shown in Fig. \ref{fig:QDE NO-DMKDE circuit}. The resulting state of these $N$ uniformly controlled isometries is the purification, 
\begin{equation}
	\sum_{i=0}^{N-1}\sqrt{\frac{1}{N}}\ket{\psi(\mathbf{x}_i)}_n \otimes \ket{i}_m.
\end{equation}

We then apply an inverse state preparation $\mathcal{U^\dag}_n$ in the first $n$ qubits, such that $\mathcal{U}_n\ket{0}_n = \ket{\psi(\mathbf{x^*})}_n$, is an isometry that prepares the quantum test sample, thus obtaining, 
\begin{equation}
	\big(\mathcal{U^\dag}_n\sum_{i=0}^{N-1}\sqrt{\frac{1}{N}}\ket{\psi(\mathbf{x}_i)}_n\big) \otimes \ket{i}_m,
\end{equation}
and we conclude by performing a measurement over the first $n$ qubits, obtaining that $P(\ket{0}_n) = (1/N)\sum_{i=0}^{N-1}\abs{\langle\psi(\mathbf{x^*})\vert\psi(\mathbf{x}_i)\rangle}^2 = \bra{\psi(\mathbf{x^*})}\rho_\text{train}\ket{\psi(\mathbf{x^*})}$.

The computational complexity of the QNO-DMKDE quantum algorithm is $O(dNS)$ since it takes $O(dN)$ to construct the purification of the training density matrix, $O(d)$ to build the isometry which prepares the test quantum state and it is required $S$ measurements of $\ket{0}_n$ basis state to estimate the probability density.

\subsubsection{Computational complexity of the QNO-DEMDE}\label{Sec: Complexity QNO-DMKDE quantum algorithm}

The computational complexity of both training and testing stages of the QNO-DEMDE with QAFF is $O(TD\max{(d, D)} + MDd + MNdS)$, assuming $T>N$, since it requires $O(TDd + Td^2)$ to train classically the QAFF, see Appendix. \ref{QDE sec: gradient classical QAFF}, $O(NDd)$ and $O(MDd)$ to predict the Fourier features of the training and testing data respectively, and $O(MNdS)$ to find an estimation of the probability densities of the testing data, because for each test sample it is necessary to build the QNO-DMKDE quantum algorithm which has complexity $O(dNS)$, as explained in the previous section \ref{NO-DMKDE quantum algorithm}. Furthermore, the complexity of QNO-DEMDE with QRFF is $O(NDd + MDd + MNdS)$, since it is not required to train the Fourier weights.

It should be noted that for most practical applications, it would be preferable to use the Q-DEMDE for quantum-classical density estimation, for instance, for both Q-DEMDE and QNO-DEMDE the computational complexity is dominated by the quantum density estimation protocols Q-DMKDE and QNO-DMKDE, which scale in the form $O(MdrS)$ and $O(MNdS)$, respectively. Indeed, in most cases, the number of training data $N$ would be larger than the rank $r$ of the training density matrix.

\section{Method evaluation}\label{QDE: density estimation results}

In this section, we present the results of the proposed quantum-classical density estimation method Q-DEMDE for one and two-dimensional density estimation with both QRFF and QAFF in a quantum simulator and a real quantum computer.

\subsection{One-dimensional quantum-classical density estimation}\label{QDE: method illustration 1D}

\begin{figure*}
\includegraphics[scale=0.51]{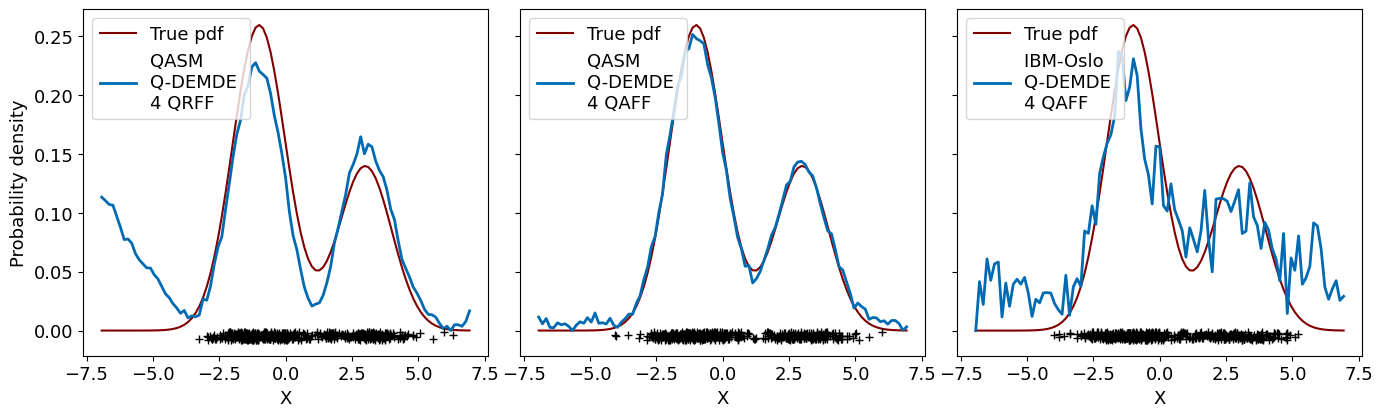}
\caption{Results of one-dimensional quantum-classical density estimation with 4 Fourier features: left, Q-DEMDE with QRFF in QASM simulator, center, Q-DEMDE with QAFF learned in a classical computer and predictions in the QASM quantum simulator, right, Q-DEMDE with QAFF learned in a classical computer and Q-DMKDE predictions in real IBM-Oslo quantum computer.}{\label{fig:QDE 1-Dimensional}}
\end{figure*}

We present a demonstration of one-dimensional density estimation in both a quantum simulator and a real quantum computer.

\subsubsection{Setup}

As mentioned in Sect. \ref{QDE: preliminaries DMKDE}, the DMKDE algorithm, along with quantum random or quantum adaptive Fourier features, encodes a probability distribution of the training data in a density matrix and estimates the probability density of new test samples by computing an expectation value.

To test the Q-DEMDE method with the quantum Fourier features, we constructed a one-dimensional probability density distribution corresponding to a mixture of two Gaussians. The training data set consisted of 1000 points sampled from the probability density function (pdf), whose probability density we wanted to estimate, and the test data set consisted of 250 equidistant points in $[-7, 7]$.

Two quantum feature maps, based on QAFF and QRFF, were applied to both the training and test data sets; we chose 4 Fourier components for each mapping. As in conventional RFF \cite{Rahimi2009RandomMachines}, the weights of the quantum random Fourier features were sampled from $\mathcal{N}(0, 1)$. To train in a classical computer the quantum adaptive Fourier features, we use the 1-dimensional Gaussian kernel training data set, see Sect. \ref{QAFF kernel traning data set}, i.e., we used a set 1-dimensional data points sampled from $\mathcal{N}(0, 1)$ with $\gamma = 0.5$ as kernel training data samples, which allowed to better approximate the weights of the quantum Fourier features; when doing this feature learning the obtained weights also resulted with a standard deviation close to 1. We set $\gamma=1$ when predicting the QRFF and QAFF mappings of the training and test data sets $\{\mathbf{x}_j\}$ and $\mathbf{x^*}$. We then constructed the training density matrix with dimensions $4\times4$, see Eq. \ref{QDE: rho train}, and computed its spectral decomposition with $4$ eigenvalues with a precision of 64 bytes on a classical computer.

For the prediction step, we computed the probability density estimator, see Eq. \ref{QDE: density estimator}, which corresponds to the expected value of the training density matrix, with our proposed Q-DMKDE quantum circuit, see Fig. \ref{fig:QDE DMKDE circuit}. We used the noisy QASM quantum simulator of the Qiskit IBM python library and the real IBM-Oslo quantum computer to estimate the expected value of the   $4\times4$ training density matrix using 4 qubits and 12000 circuit measurements (shots); in Appendix \ref{sec:oslo_description}, we describe the characteristics of the IBM-Oslo quantum computer. Since we were working with a small number of features, we also performed a normalization of the density values by subtracting the minimum value from the density estimates; although this is not necessary for a large number of Fourier features. We evaluated the Q-DEMDE with QRFF and QAFF for density estimation, by comparing their density estimates with the real probability distribution that generated the data. For this purpose, we selected as metrics the Kullback-Leibler (KL) divergence, mean average error (MAE) and Spearman correlation which are widely used to compare probability distributions.

\subsubsection{Results and discussion}

\begin{table}[h!]
\centering
\begin{tabular}{ | c | c | c |  c | c |}
\hline
Method & Feature Map & KL-Div  &  \ \ MAE \ \ & Spearman  \\
\hline

Q-DEMDE & \multirow{2}*{QRFF: 4} & \multirow{2}*{$0.218$} & \multirow{2}*{$0.026$}  & \multirow{2}*{$0.663$}  \\
 QASM & &  & &  \\
\hline
Q-DEMDE& \multirow{2}*{QAFF: 4} & \multirow{2}*{$\mathbf{0.025}$} & \multirow{2}*{$\mathbf{0.005}$} & \multirow{2}*{$\mathbf{0.971}$}   \\
QASM &  & &  & \\
\hline
Q-DEMDE & \multirow{2}*{QAFF: 4} & \multirow{2}*{$0.215$} & \multirow{2}*{$0.030$} & \multirow{2}*{$0.862$}  \\
IBM QC &  &  & &  \\

\hline
\end{tabular}
\caption{Results of one-dimensional quantum-classical density estimation using the Q-DEMDE method with four QRFF and four QAFF in the QASM quantum simulator and the real IBM-Oslo quantum computer. The best results are shown in bold.}
\label{table:15}
\end{table}

We report three demonstrations of density estimation, see Fig. \ref{fig:QDE 1-Dimensional} and Table. \ref{table:15}, two using the QASM quantum simulator for the Q-DMKDE and both QRFF and QAFF and one using QAFF and computing the density estimator on the real IBM-Oslo quantum computer.

In the subfigures of the left and center of Fig. \ref{fig:QDE 1-Dimensional}, we show the probability density estimation with both quantum random and quantum adaptive Fourier features with the QASM simulator. In contrast to QRFF, the QAFF produces a better approximation of the pdf, especially in low-density regions. Furthermore, the density estimation metrics shown in Table \ref{table:15} illustrate that QAFF is more suitable for 1-dimensional density estimation in the case of a low number of Fourier features. In the subfigure of the right of Fig. \ref{fig:QDE 1-Dimensional}, we present the results of QAFF learned in a classical computer and the density estimates predicted with the Q-DMKDE quantum algorithm in the real IBM-Oslo quantum computer. As expected, the results on the real quantum computer were much noisier than the results on the QASM simulator; however, the results and the density estimation metrics show that it is possible to perform 1-dimensional density estimation on real quantum computers with relatively good performance.

\subsection{Two-dimensional quantum-classical density estimation}\label{sec: QDE 2-dimensional DE experiments}

\begin{figure}
    \centering
    \includegraphics[scale = 0.6]{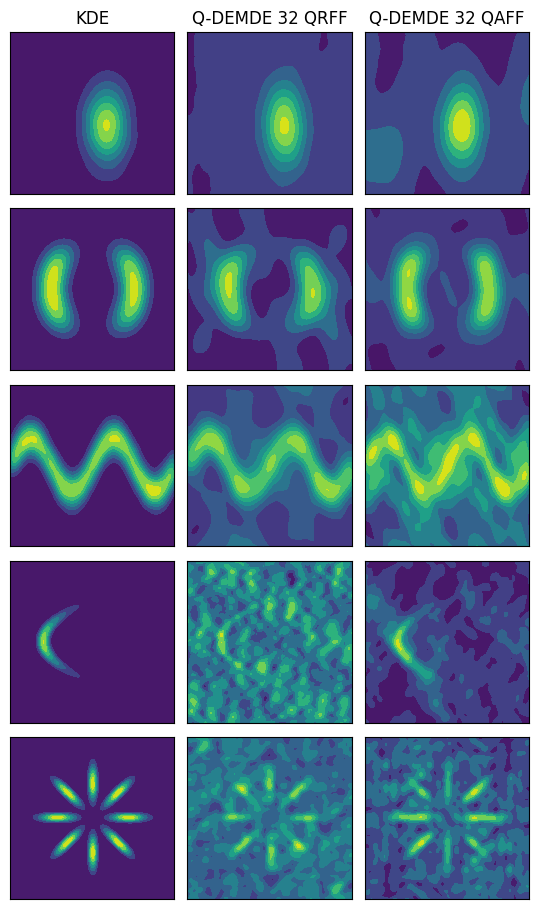}
    \caption{Two-dimensional quantum-classical density estimation with a classical computer and a quantum simulator from Pennylane. From top to bottom, the two-dimensional data sets correspond to Binomial,
Potential 1, Potential 2, Arc, and Star Eight.}
    \label{fig:QDEFig3}
\end{figure}

We evaluate the proposed hybrid density estimation method for two-dimensional density estimation.

\subsubsection{Setup}

We estimated the probability density functions of five two-dimensional data sets sampled from five two-dimensional distributions \cite{gallego2022quantum}, named from top to bottom: Binomial, Potential 1, Potential 2, Arc and Star Eight, see Fig. \ref{fig:QDEFig3}; the statistical characteristics of these distributions are described in the Appendix \ref{sec:aptwodimensional}. We performed density estimation with three models: kernel density estimation (KDE) as a reference model, and the Q-DEMDE with both QRFF and QAFF with 32 components. The training data set for the density estimation tasks corresponded to 10000 points sampled from the two-dimensional distributions, while the test data set was constructed by dividing the five two-dimensional spaces by grids of $120 \times 120$ points.

We use a classical computer to build the QRFF and QAFF mappings. For each data set, we constructed the quantum random Fourier features by sampling the weights from $\mathcal{N}(\mathbf{0}, \mathbf{I}_2)$, and we used these random weights as the initial parameters to train the QAFF mapping. To illustrate the DE improvement of QAFF over QRFF, we trained the QAFF independently for each 2D data set, although we could have learned the quantum adaptive Fourier weights only once for all the 2D data sets, considering that we used the Gaussian kernel training data set for the kernel learning, see Sect. \ref{QAFF kernel traning data set}, which is independent of the training and test density estimation data sets. After learning the optimal weights of the QAFF, for each of the five 2-dimensional data sets, we set the appropriate $\gamma$ scaling parameter to predict the QRFF and QAFF; we also used the same corresponding $\gamma$ parameter for the KDE method.

After establishing the QAFF and the QRFF maps for each 2-D data set, we built for each mapping the $32\times 32$ training density matrix and found its spectral decomposition with $32$ eigenvalues with a precision of 64 bytes in a classical computer. We then measured the expected value of the training density matrix with the quantum Fourier states of the test data set, using the 10-qubit Q-DMKDE quantum algorithm implemented in the Pennylane quantum library \cite{bergholm2018pennylane}, obtaining the density estimation of the test data set. To validate the density estimation method, we compared the obtained probability distributions of Q-DEMDE with quantum random and quantum adaptive Fourier features with the probability density obtained with kernel density estimation, using KL-divergence, MAE and Spearman correlation. 

\subsubsection{Results and discussion}

In Fig. \ref{fig:QDEFig3}, we show the density estimation of the test data set for all five 2-D data sets: in the left column, we show the density estimation using kernel density estimation (KDE) \cite{Rosenblatt1956RemarksFunction, Parzen1962OnMode}, in the middle figures, we plot the proposed quantum-classical DE method Q-DEMDE using 32 quantum random Fourier features, and in the right figures, we display the Q-DEMDE using 32 quantum adaptive Fourier features.

Since Q-DEMDE with QRFF and QAFF are both approximations of KDE, we aimed to replicate the 2-dimensional density estimation results of the KDE in all data sets. Despite the reduced number of quantum Fourier components,  Fig. \ref{fig:QDEFig3} and Table. \ref{table:2} show that both models with adaptive and random features obtained satisfactory results for two-dimensional quantum density estimation, albeit they could have been improved by increasing the number of Fourier components. Results also show that for most datasets there is an advantage of optimizing the Fourier weights with QAFF, illustrating that this adaptive mapping improves the kernel approximation. Out of the five data sets, the Arc and Star Eight were the most challenging data sets for density estimation considering that they required to approximate a kernel with a small bandwidth parameter.

It is worth mentioning that the proposed hybrid density estimation method is non-parametric, i.e., it does not require a prior probability distribution to fit the data, in contrast to parametric density estimation \cite{varanasi1989parametric, vapnik1999support}, and it is optimization-free. Furthermore, although we did not perform two-dimensional density estimation in real quantum hardware, the quantum simulation used a low number of qubits (only 10), showing that Q-DEMDE might be suitable for current quantum hardware.

\begin{table}[h!]
\centering
\begin{tabular}{ | c | c | c | c |}
\hline
\multirow{2}*{Data Set} & \multirow{2}*{Metric} & Q-DEMDE & Q-DEMDE \\
 &  & 32 QRFF  & 32 QAFF  \\
\hline

& KL-Div & $\mathbf{0.784}$ & $0.824$  \\
Binomial & MAE  & $0.025$ & $\mathbf{0.024}$ \\
 
  & Spearman & $\mathbf{0.750}$ & $0.425$ \\

\hline
& KL-Div & $0.571$ & $\mathbf{0.555}$ \\
Potential 1  & MAE  & $0.045$ & $\mathbf{0.043}$ \\
 
 & Spearman & $0.704$ & $\mathbf{0.782}$ \\

\hline

& KL-Div & $\mathbf{0.789}$ & $0.946$  \\
Potential 2 & MAE  & $0.171$ & $\mathbf{0.164}$ \\
 
 & Spearman & $0.726$ & $\mathbf{0.806}$  \\

\hline

& KL-Div & $2.713$  & $\mathbf{2.193}$ \\
Arc & MAE  & $0.159$ & $\mathbf{0.020}$  \\
 
 & Spearman & $0.050$ & $\mathbf{0.219}$  \\

\hline

& KL-Div & $1.643$ & $\mathbf{1.626}$  \\
Star Eight  & MAE  & $\mathbf{0.157}$ & $0.159$  \\
 
& Spearman & $0.230$ & $\mathbf{0.353}$\\

\hline

\hline
\end{tabular}
\caption{2D quantum-classical density estimation results using the Q-DEMDE algorithm with both QRFF and QAFF. These results illustrate the similarity between the probability densities of test data obtained with the quantum-classical Q-DEMDE method and the classical KDE algorithm. The best results are shown in bold.}
\label{table:2}
\end{table}

\subsubsection{Additional: Results of hybrid density estimation with quantum variational QAFF}\label{sec: additional hybrid DE experiments with quantum variational QAFF}

While the proposed Q-DEMDE strategy uses classical computers to optimize the quantum adaptive Fourier features, in Sect. \ref{QAFF Quantum implementation}, we describe an alternative method to train the QAFF in quantum hardware using a quantum variational algorithm. In Appendix \ref{sec:Hybrid DE experiments with quantum QAFF}, we include some additional quantum simulations that show that it is possible to perform 2-dimensional density estimation with the Q-DEMDE method using the QAFF optimized in a Pennylane noiseless quantum simulator. Although the Q-DEMDE method with classical QAFF has less computational complexity and it is therefore more optimal for hybrid density estimation, the feasibility of learning the QAFF mapping in quantum hardware may have applications to other hybrid quantum machine learning methods that, unlike the Q-DEMDE algorithm, do not require an intermediate classical computation.

\section{Method application: Quantum-classical anomaly detection}\label{QDE: anomaly detection results}

\begin{figure}
    \centering
    \includegraphics[scale=0.524]{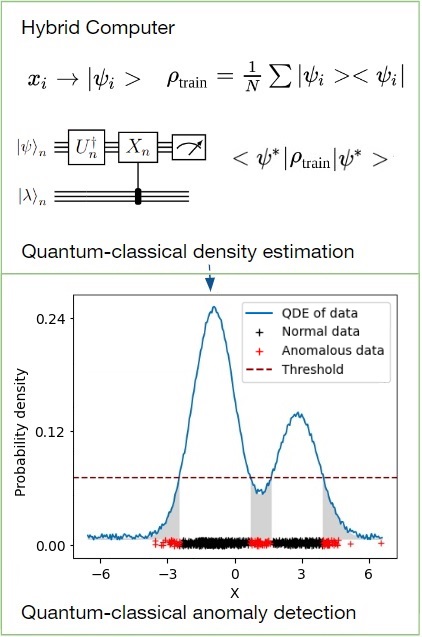} 
    \caption{Quantum-classical anomaly detection model for quantum computers based on the proposed quantum-classical density estimation strategy Q-DEMDE. Use Q-DEMDE to find the density values of the test data, and set a threshold to classify anomalous data, assuming that anomalous samples are found in low-density regions}
    \label{fig:QDE Anomaly Detection}
\end{figure}

In this section, we present a quantum-classical anomaly detection strategy as an application of our proposed hybrid density estimation method, see Fig. \ref{fig:QDE Anomaly Detection}.

Anomaly detection aims to determine which samples from a given data set are ``ordinary" or ``normal" (being the interpretation of ``normal" defined for each particular case) and which samples depart, or are deviated, from ``normal" data (commonly known as ``anomalies"). Some common applications of anomaly detection methods include fraud detection \cite{VanVlasselaer2015APATE:Extensions} and medical diagnosis \cite{Chen2011AnInspection}. This problem has been widely addressed in classical computers \cite{noble2003graph, gonzalez2003anomaly}, while quantum anomaly detection is a more recent field, from which some proposals have shown notable speed-ups in contrast to their classical
counterparts \cite{liang2019quantum, Liu2018QuantumDetection, Herr2021AnomalyNetworks}.

In order to detect anomalous data in a given data set, the proposed method starts by dividing the data samples into three partitions: train, validation, and test, maintaining the same proportion of ``normal" and ``anomalous" data in all three partitions. After this, the method performs the following steps: i) use the proposed quantum-classical density estimation method Q-DEMDE to estimate the probability density values of all samples in validation and test partitions with the training partition, ii) use the validation data set to select a percentile-based threshold to separate the samples: if a sample has a density lower than the threshold, then it is considered an ``anomalous" sample, and iii) finally, use this threshold to classify ``normal" and ``anomalous" data in the test partition. The main assumption on which we base this approach is that, unlike ``normal" samples, the ``anomalous" samples lie in regions with lower probability density values, as shown in Fig. \ref{fig:QDE Anomaly Detection}.

\subsection{QAD: Setup}

We used the quantum-classical density estimation method Q-DEMDE to detect anomalies in a previously labeled AD data set. We used a modified version, from Ref. \cite{Aggarwal2015TheoreticalEnsembles}, of the Cardiotocography data set from the UCI Machine Learning Repository related to fetal heart rate. It consists of 1831 samples, where each sample contains 21 attributes, with two classes: the \textit{normal} class for the inliers, and the \textit{pathologic} class for the outliers. This data set was divided into three partitions: training (60\%), validation (20\%), and test (20\%), all of which contained both normal samples and outliers in the same proportion (approximately 9.6\% of all samples were labeled as outliers).

As a first step, QRFF and QAFF were applied to the samples of all partitions as quantum feature mappings. We worked with 4 and 8 Fourier features corresponding to quantum states of 2 and 3 qubits respectively. The QAFF were trained using a classical computer. To train the QAFF with $\gamma=0.5$, we used the Gaussian kernel training data set, see Sect. \ref{QAFF kernel traning data set}, corresponding to $1\times10^5$ samples from $\mathcal{N}(\mathbf{0}, \frac{1}{21}\mathbf{I}_{21})$, the number of trainable parameters, which corresponds to the Fourier weights, were respectively $21\times 4 = 84$ and $21\times 8 = 168$ in the cases of 4 and 8 components. Once we learned the Fourier weights, we predicted the quantum Fourier features of the training, test and validation data sets by searching for the optimal value of $\gamma$ of the kernel (considering only powers of two ranging from $2^{-10}$ to $2^0$), then setting a value of $\gamma=2^{-7}$ for both QRFF and QAFF. We also used a classical computer to build the $4\times 4$ and $8\times8$ training density matrices, see Eq. \ref{QDE: rho train}, and to compute their spectral decompositions. We then performed the density estimations with the 4-qubit and 6-qubit Q-DMKDE quantum algorithm in a noiseless quantum simulator of the Pennylane quantum library \cite{bergholm2018pennylane}. As it is common in anomaly detection tasks, the labels were not used to train the algorithm.

After performing the density estimations of the validation and test partitions, we classified each sample as ``normal" or ``outlier" by comparing its density value with a threshold value $t$. We used the validation partition to search for the threshold by setting a percentile value $p$ such that $9.6\%$ of the samples had a density below $t$ and were thus considered anomalous samples. We then used the obtained boundary $t$ to classify the testing data.

We also performed a demonstration of the QAD algorithm on the real IBM-Oslo quantum computer, refer to Appendix \ref{sec:oslo_description} for its characteristics. To achieve this, we first applied the QRFF and QAFF mappings to all three data partitions using a classical computer with 4 Fourier features and $\gamma = 2^{-5}$. Due to satisfactory results, we directly used the anomaly detection data to train the QAFF, as described in Sect. \ref{QAFF kernel traning data set}, instead of the Gaussian synthetic kernel training data, although for a larger number of Fourier features and more complex data sets the synthetic Gaussian data set would be the preferable choice. Once the Fourier mappings were constructed, we used 4 qubits to prepare the training density matrix from its spectral decomposition using the Q-DMKDE quantum algorithm in the IBM-Oslo quantum processor, and estimated the probability densities of the test and validation data sets using $12000$ shots.

We report the F1 score of the ``anomalous" class, the accuracy, and the AUC (area under the characteristic curve). Since the random weights of the QRFF and the random initialization of the QAFF produce slightly different DE results in the case of a small number of Fourier features, we performed ten trials with different random seeds using the Pennylane quantum simulator. In contrast, in the real IBM-Oslo quantum computer, we performed the quantum demonstration only once for both quantum mappings. Furthermore, we initialized the weights of the QAFF from a uniform distribution $U[0, 1]$ in all cases, considering that the results were not strongly affected by the initial distribution of the Fourier weights.

\subsection{QAD: Results and discussion}

The results obtained by applying the proposed methods for hybrid anomaly detection with the Pennylane quantum simulator are reported in Tables \ref{table:10} and \ref{table:10b}. Compared to QRFF, the QAFF obtained better results in the three evaluated metrics, F1 Score, accuracy, and AUC, thanks to the optimization of the Fourier weights. Regarding the standard deviations, we observe more consistent results across different trials with QAFF, which shows that QRFF is more sensitive to randomness than QAFF in a low-dimensional setting.

\begin{table}[h!]
\centering
\begin{tabular}{ | c | c | c | c |}
\hline
Method & F1 Score & Accuracy & AUC \\
\hline
Q-DEMDE & \multirow{2}*{$0.422 \pm 0.073$} & \multirow{2}*{$0.896 \pm 0.013$} & \multirow{2}*{$0.776 \pm 0.061$} \\
4 QRFF &  &  & \\
\hline
Q-DEMDE & \multirow{2}*{$\mathbf{0.455 \pm 0.067}$} & \multirow{2}*{$\mathbf{0.899\pm0.010}$} & \multirow{2}*{$\mathbf{0.852\pm0.051}$} \\
4 QAFF &  &  & \\
\hline
\end{tabular}

\caption{Obtained metrics for 10 hybrid anomaly detection trials on Cardio data set using Q-DEMDE method with 4 Fourier features for both QRFF and QAFF with the Pennylane quantum simulator.}
\label{table:10}
\end{table}

\begin{table}[h!]
\centering
\begin{tabular}{ | c | c | c | c |}
\hline
Method & F1 Score & Accuracy & AUC \\
\hline
Q-DEMDE & \multirow{2}*{$0.516\pm0.126$} & \multirow{2}*{$0.911\pm0.020$} & \multirow{2}*{$\mathbf{0.920\pm0.039}$} \\
8 QRFF &  &  & \\
\hline
Q-DEMDE & \multirow{2}*{$\mathbf{0.573\pm0.083}$} & \multirow{2}*{$\mathbf{0.920\pm0.010}$} & \multirow{2}*{$0.915\pm0.028$} \\
8 QAFF &  &  & \\
\hline
\end{tabular}

\caption{Obtained metrics for 10 hybrid anomaly detection trials on Cardio data set using Q-DEMDE method with 8 Fourier features for both QRFF and QAFF with the Pennylane quantum simulator.}
\label{table:10b}
\end{table}

Furthermore, in Table \ref{table:20}, we present the results of the hybrid anomaly detection algorithm on the real IBM-Oslo quantum computer; we performed only one demonstration of the QAD strategy for each quantum mapping. The results follow a similar pattern as in the quantum simulator: QAFF leads to higher performance than QRFF. In contrast to the quantum simulator, the demonstration with QRFF on the real quantum computer presented inferior results, probably because the density predictions with random features had low variance, making it difficult to make a distinction between the two classes due to quantum noise. Nevertheless, the Q-DEMDE with QAFF in the real quantum computer obtained satisfactory results for the anomaly detection task, showing that this strategy is suitable for current quantum devices.

\begin{table}[h!]
\centering
\begin{tabular}{ | c | c | c | c |}
\hline
Method & F1 Score & Accuracy &\ \ AUC \ \ \\
\hline
Q-DEMDE 4 QRFF & \multirow{2}*{$0.170$} &  \multirow{2}*{$0.867$} & \multirow{2}*{$0.541$} \\
IBM-Oslo  &  &  & \\
\hline
Q-DEMDE 4 QAFF & \multirow{2}*{$\mathbf{0.648}$} & \multirow{2}*{$\mathbf{0.932}$} & \multirow{2}*{$\mathbf{0.933}$} \\
IBM-Oslo &  &  & \\
 \hline

\end{tabular}
\caption{Obtained metrics for hybrid anomaly detection on the real IBM-Oslo quantum computer using the Q-DEMDE density estimation strategy with both QRFF and QAFF with 4 Fourier components. We used a classical computer to learn the optimal QAFF, and the real quantum computer to predict the density of validation and test samples with the Q-DMKDE quantum circuit.}
\label{table:20}
\end{table}

\section{Conclusions}\label{QDE: conclusions}

In this article, we presented a quantum-classical density estimation method called Q-DEMDE for current quantum computers based on the expected values of density matrices and quantum Fourier features. Within our method, we proposed a quantum protocol for mixed-state preparation and expected value estimation, which was used as a subroutine to implement the classical density estimation algorithm DMKDE \cite{gonzalez2022learning} in quantum hardware. We also introduced two quantum feature maps called quantum random and quantum adaptive Fourier features, which allow to approximate a Gaussian kernel between data samples in both classical and quantum computers. We performed various demonstrations on different data sets for density estimation on a quantum simulator and on a real quantum computer.

The quantum-classical density estimation results show that it is possible to exploit the properties of mixed states and Fourier features to efficiently represent probability distributions of data in quantum hardware. A key feature of the proposed hybrid DE method is that it is non-parametric, i.e., it does not assume a prior probability distribution to fit the training data, in contrast to parametric quantum DE methods \cite{liang2019quantum, guo2022quantum}. Furthermore, the Q-DEMDE method with the QRFF map is optimization-free, i.e., it does not require any optimization, while the Q-DEMDE method with QAFF requires optimization only to construct the quantum map. Although the Q-DEMDE algorithm does not achieve a quantum advantage over the classical DMKDE, it shows that it is possible to perform quantum-classical density estimation with a reduced number of quantum Fourier features and relatively small quantum circuits, illustrating its viability for present-day quantum computers.

We also presented an application of the hybrid density estimation method for quantum-classical anomaly detection. Despite the current limitations of quantum hardware, the Q-DEMDE strategy showed satisfactory results for anomaly detection in a real quantum computer. Considering the numerous other applications of density estimation \cite{fraley2002model, anderson2009kernel, PhysRevD.101.075042, hu2018anomaly, bigdeli2020learning}, the results indicate that the proposed hybrid density estimation method can be used as a building block for other quantum machine learning algorithms, and could provide novel avenues to solve new problems with quantum computers.

Future work of the proposed quantum-classical density estimation technique includes comparing our method with other classical and quantum density estimation methods, exploring other quantum density estimation algorithms that could possibly achieve quantum advantage, and integrating the proposed Q-DMKDE quantum circuit with variational quantum algorithms to learn mixed quantum states in quantum hardware.

\section*{Code availability}\label{Code_section}

The source code necessary to replicate the demonstrations presented in this paper is available in the following GitHub repository \href{https://github.com/diegour1/QDEMDE}{https://github.com/diegour1/QDEMDE}.

\section*{Acknowledgments}\label{qanomalyacknowledgments_section}

We acknowledge the use of IBM Quantum services for this work. The views expressed in this work are those of the authors and do not reflect the official policy or position of IBM or the IBM Quantum team. D.H.U. acknowledges Google Research for the support through the Google Ph.D. Fellowship.

\bibliography{mainref}

\begin{thebibliography}{77}%
\makeatletter
\providecommand \@ifxundefined [1]{%
 \@ifx{#1\undefined}
}%
\providecommand \@ifnum [1]{%
 \ifnum #1\expandafter \@firstoftwo
 \else \expandafter \@secondoftwo
 \fi
}%
\providecommand \@ifx [1]{%
 \ifx #1\expandafter \@firstoftwo
 \else \expandafter \@secondoftwo
 \fi
}%
\providecommand \natexlab [1]{#1}%
\providecommand \enquote  [1]{``#1''}%
\providecommand \bibnamefont  [1]{#1}%
\providecommand \bibfnamefont [1]{#1}%
\providecommand \citenamefont [1]{#1}%
\providecommand \href@noop [0]{\@secondoftwo}%
\providecommand \href [0]{\begingroup \@sanitize@url \@href}%
\providecommand \@href[1]{\@@startlink{#1}\@@href}%
\providecommand \@@href[1]{\endgroup#1\@@endlink}%
\providecommand \@sanitize@url [0]{\catcode `\\12\catcode `\$12\catcode
  `\&12\catcode `\#12\catcode `\^12\catcode `\_12\catcode `\%12\relax}%
\providecommand \@@startlink[1]{}%
\providecommand \@@endlink[0]{}%
\providecommand \url  [0]{\begingroup\@sanitize@url \@url }%
\providecommand \@url [1]{\endgroup\@href {#1}{\urlprefix }}%
\providecommand \urlprefix  [0]{URL }%
\providecommand \Eprint [0]{\href }%
\providecommand \doibase [0]{https://doi.org/}%
\providecommand \selectlanguage [0]{\@gobble}%
\providecommand \bibinfo  [0]{\@secondoftwo}%
\providecommand \bibfield  [0]{\@secondoftwo}%
\providecommand \translation [1]{[#1]}%
\providecommand \BibitemOpen [0]{}%
\providecommand \bibitemStop [0]{}%
\providecommand \bibitemNoStop [0]{.\EOS\space}%
\providecommand \EOS [0]{\spacefactor3000\relax}%
\providecommand \BibitemShut  [1]{\csname bibitem#1\endcsname}%
\let\auto@bib@innerbib\@empty
\bibitem [{\citenamefont {Bigdeli}\ \emph {et~al.}(2023)\citenamefont
  {Bigdeli}, \citenamefont {Lin}, \citenamefont {Dunbar}, \citenamefont
  {Portenier},\ and\ \citenamefont {Zwicker}}]{bigdeli2020learning}%
  \BibitemOpen
  \bibfield  {author} {\bibinfo {author} {\bibfnamefont {S.~A.}\ \bibnamefont
  {Bigdeli}}, \bibinfo {author} {\bibfnamefont {G.}~\bibnamefont {Lin}},
  \bibinfo {author} {\bibfnamefont {L.~A.}\ \bibnamefont {Dunbar}}, \bibinfo
  {author} {\bibfnamefont {T.}~\bibnamefont {Portenier}},\ and\ \bibinfo
  {author} {\bibfnamefont {M.}~\bibnamefont {Zwicker}},\ }\bibfield  {title}
  {\bibinfo {title} {Learning generative models using denoising density
  estimators},\ }\href@noop {} {\bibfield  {journal} {\bibinfo  {journal} {IEEE
  Transactions on Neural Networks and Learning Systems}\ } (\bibinfo {year}
  {2023})}\BibitemShut {NoStop}%
\bibitem [{\citenamefont {Nachman}\ and\ \citenamefont
  {Shih}(2020)}]{PhysRevD.101.075042}%
  \BibitemOpen
  \bibfield  {author} {\bibinfo {author} {\bibfnamefont {B.}~\bibnamefont
  {Nachman}}\ and\ \bibinfo {author} {\bibfnamefont {D.}~\bibnamefont {Shih}},\
  }\bibfield  {title} {\bibinfo {title} {Anomaly detection with density
  estimation},\ }\href {https://doi.org/10.1103/PhysRevD.101.075042} {\bibfield
   {journal} {\bibinfo  {journal} {Phys. Rev. D}\ }\textbf {\bibinfo {volume}
  {101}},\ \bibinfo {pages} {075042} (\bibinfo {year} {2020})}\BibitemShut
  {NoStop}%
\bibitem [{\citenamefont {Hu}\ \emph {et~al.}(2018)\citenamefont {Hu},
  \citenamefont {Gao}, \citenamefont {Li}, \citenamefont {Wu}, \citenamefont
  {Du},\ and\ \citenamefont {Maybank}}]{hu2018anomaly}%
  \BibitemOpen
  \bibfield  {author} {\bibinfo {author} {\bibfnamefont {W.}~\bibnamefont
  {Hu}}, \bibinfo {author} {\bibfnamefont {J.}~\bibnamefont {Gao}}, \bibinfo
  {author} {\bibfnamefont {B.}~\bibnamefont {Li}}, \bibinfo {author}
  {\bibfnamefont {O.}~\bibnamefont {Wu}}, \bibinfo {author} {\bibfnamefont
  {J.}~\bibnamefont {Du}},\ and\ \bibinfo {author} {\bibfnamefont
  {S.}~\bibnamefont {Maybank}},\ }\bibfield  {title} {\bibinfo {title} {Anomaly
  detection using local kernel density estimation and context-based
  regression},\ }\href@noop {} {\bibfield  {journal} {\bibinfo  {journal} {IEEE
  Transactions on Knowledge and Data Engineering}\ }\textbf {\bibinfo {volume}
  {32}},\ \bibinfo {pages} {218} (\bibinfo {year} {2018})}\BibitemShut
  {NoStop}%
\bibitem [{\citenamefont {Bortoloti}\ \emph {et~al.}(2021)\citenamefont
  {Bortoloti}, \citenamefont {de~Oliveira},\ and\ \citenamefont
  {Ciarelli}}]{bortoloti2021supervised}%
  \BibitemOpen
  \bibfield  {author} {\bibinfo {author} {\bibfnamefont {F.~D.}\ \bibnamefont
  {Bortoloti}}, \bibinfo {author} {\bibfnamefont {E.}~\bibnamefont
  {de~Oliveira}},\ and\ \bibinfo {author} {\bibfnamefont {P.~M.}\ \bibnamefont
  {Ciarelli}},\ }\bibfield  {title} {\bibinfo {title} {Supervised kernel
  density estimation k-means},\ }\href@noop {} {\bibfield  {journal} {\bibinfo
  {journal} {Expert Systems with Applications}\ }\textbf {\bibinfo {volume}
  {168}},\ \bibinfo {pages} {114350} (\bibinfo {year} {2021})}\BibitemShut
  {NoStop}%
\bibitem [{\citenamefont {Fraley}\ and\ \citenamefont
  {Raftery}(2002)}]{fraley2002model}%
  \BibitemOpen
  \bibfield  {author} {\bibinfo {author} {\bibfnamefont {C.}~\bibnamefont
  {Fraley}}\ and\ \bibinfo {author} {\bibfnamefont {A.~E.}\ \bibnamefont
  {Raftery}},\ }\bibfield  {title} {\bibinfo {title} {Model-based clustering,
  discriminant analysis, and density estimation},\ }\href@noop {} {\bibfield
  {journal} {\bibinfo  {journal} {Journal of the American statistical
  Association}\ }\textbf {\bibinfo {volume} {97}},\ \bibinfo {pages} {611}
  (\bibinfo {year} {2002})}\BibitemShut {NoStop}%
\bibitem [{\citenamefont {Anderson}(2009)}]{anderson2009kernel}%
  \BibitemOpen
  \bibfield  {author} {\bibinfo {author} {\bibfnamefont {T.~K.}\ \bibnamefont
  {Anderson}},\ }\bibfield  {title} {\bibinfo {title} {Kernel density
  estimation and k-means clustering to profile road accident hotspots},\
  }\href@noop {} {\bibfield  {journal} {\bibinfo  {journal} {Accident Analysis
  \& Prevention}\ }\textbf {\bibinfo {volume} {41}},\ \bibinfo {pages} {359}
  (\bibinfo {year} {2009})}\BibitemShut {NoStop}%
\bibitem [{\citenamefont {Gudhe}\ \emph {et~al.}(2022)\citenamefont {Gudhe},
  \citenamefont {Behravan}, \citenamefont {Sudah}, \citenamefont {Okuma},
  \citenamefont {Vanninen}, \citenamefont {Kosma},\ and\ \citenamefont
  {Mannermaa}}]{gudhe2022area}%
  \BibitemOpen
  \bibfield  {author} {\bibinfo {author} {\bibfnamefont {N.~R.}\ \bibnamefont
  {Gudhe}}, \bibinfo {author} {\bibfnamefont {H.}~\bibnamefont {Behravan}},
  \bibinfo {author} {\bibfnamefont {M.}~\bibnamefont {Sudah}}, \bibinfo
  {author} {\bibfnamefont {H.}~\bibnamefont {Okuma}}, \bibinfo {author}
  {\bibfnamefont {R.}~\bibnamefont {Vanninen}}, \bibinfo {author}
  {\bibfnamefont {V.-M.}\ \bibnamefont {Kosma}},\ and\ \bibinfo {author}
  {\bibfnamefont {A.}~\bibnamefont {Mannermaa}},\ }\bibfield  {title} {\bibinfo
  {title} {Area-based breast percentage density estimation in mammograms using
  weight-adaptive multitask learning},\ }\href@noop {} {\bibfield  {journal}
  {\bibinfo  {journal} {Scientific reports}\ }\textbf {\bibinfo {volume}
  {12}},\ \bibinfo {pages} {12060} (\bibinfo {year} {2022})}\BibitemShut
  {NoStop}%
\bibitem [{\citenamefont {Kato}(2021)}]{kato2021development}%
  \BibitemOpen
  \bibfield  {author} {\bibinfo {author} {\bibfnamefont {H.}~\bibnamefont
  {Kato}},\ }\bibfield  {title} {\bibinfo {title} {Development of a
  spatio-temporal analysis method to support the prevention of covid-19
  infection: space-time kernel density estimation using gps location history
  data},\ }\href@noop {} {\bibfield  {journal} {\bibinfo  {journal} {Urban
  Informatics and Future Cities}\ ,\ \bibinfo {pages} {51}} (\bibinfo {year}
  {2021})}\BibitemShut {NoStop}%
\bibitem [{\citenamefont {Srikanth}\ and\ \citenamefont
  {Srikanth}(2020)}]{srikanth2020case}%
  \BibitemOpen
  \bibfield  {author} {\bibinfo {author} {\bibfnamefont {L.}~\bibnamefont
  {Srikanth}}\ and\ \bibinfo {author} {\bibfnamefont {I.}~\bibnamefont
  {Srikanth}},\ }\bibfield  {title} {\bibinfo {title} {A case study on kernel
  density estimation and hotspot analysis methods in traffic safety
  management},\ }in\ \href@noop {} {\emph {\bibinfo {booktitle} {2020
  International Conference on COMmunication Systems \& NETworkS (COMSNETS)}}}\
  (\bibinfo {organization} {IEEE},\ \bibinfo {year} {2020})\ pp.\ \bibinfo
  {pages} {99--104}\BibitemShut {NoStop}%
\bibitem [{\citenamefont {Van~Vlasselaer}\ \emph {et~al.}(2015)\citenamefont
  {Van~Vlasselaer}, \citenamefont {Bravo}, \citenamefont {Caelen},
  \citenamefont {Eliassi-Rad}, \citenamefont {Akoglu}, \citenamefont {Snoeck},\
  and\ \citenamefont {Baesens}}]{VanVlasselaer2015APATE:Extensions}%
  \BibitemOpen
  \bibfield  {author} {\bibinfo {author} {\bibfnamefont {V.}~\bibnamefont
  {Van~Vlasselaer}}, \bibinfo {author} {\bibfnamefont {C.}~\bibnamefont
  {Bravo}}, \bibinfo {author} {\bibfnamefont {O.}~\bibnamefont {Caelen}},
  \bibinfo {author} {\bibfnamefont {T.}~\bibnamefont {Eliassi-Rad}}, \bibinfo
  {author} {\bibfnamefont {L.}~\bibnamefont {Akoglu}}, \bibinfo {author}
  {\bibfnamefont {M.}~\bibnamefont {Snoeck}},\ and\ \bibinfo {author}
  {\bibfnamefont {B.}~\bibnamefont {Baesens}},\ }\bibfield  {title} {\bibinfo
  {title} {{APATE: A novel approach for automated credit card transaction fraud
  detection using network-based extensions}},\ }\href
  {https://doi.org/10.1016/J.DSS.2015.04.013} {\bibfield  {journal} {\bibinfo
  {journal} {Decision Support Systems}\ }\textbf {\bibinfo {volume} {75}},\
  \bibinfo {pages} {38} (\bibinfo {year} {2015})}\BibitemShut {NoStop}%
\bibitem [{\citenamefont {Papamakarios}\ \emph {et~al.}(2017)\citenamefont
  {Papamakarios}, \citenamefont {Pavlakou},\ and\ \citenamefont
  {Murray}}]{papamakarios2017masked}%
  \BibitemOpen
  \bibfield  {author} {\bibinfo {author} {\bibfnamefont {G.}~\bibnamefont
  {Papamakarios}}, \bibinfo {author} {\bibfnamefont {T.}~\bibnamefont
  {Pavlakou}},\ and\ \bibinfo {author} {\bibfnamefont {I.}~\bibnamefont
  {Murray}},\ }\bibfield  {title} {\bibinfo {title} {Masked autoregressive flow
  for density estimation},\ }in\ \href
  {https://proceedings.neurips.cc/paper_files/paper/2017/file/6c1da886822c67822bcf3679d04369fa-Paper.pdf}
  {\emph {\bibinfo {booktitle} {Advances in Neural Information Processing
  Systems}}},\ Vol.~\bibinfo {volume} {30},\ \bibinfo {editor} {edited by\
  \bibinfo {editor} {\bibfnamefont {I.}~\bibnamefont {Guyon}}, \bibinfo
  {editor} {\bibfnamefont {U.~V.}\ \bibnamefont {Luxburg}}, \bibinfo {editor}
  {\bibfnamefont {S.}~\bibnamefont {Bengio}}, \bibinfo {editor} {\bibfnamefont
  {H.}~\bibnamefont {Wallach}}, \bibinfo {editor} {\bibfnamefont
  {R.}~\bibnamefont {Fergus}}, \bibinfo {editor} {\bibfnamefont
  {S.}~\bibnamefont {Vishwanathan}},\ and\ \bibinfo {editor} {\bibfnamefont
  {R.}~\bibnamefont {Garnett}}}\ (\bibinfo  {publisher} {Curran Associates,
  Inc.},\ \bibinfo {year} {2017})\BibitemShut {NoStop}%
\bibitem [{\citenamefont {Varanasi}\ and\ \citenamefont
  {Aazhang}(1989)}]{varanasi1989parametric}%
  \BibitemOpen
  \bibfield  {author} {\bibinfo {author} {\bibfnamefont {M.~K.}\ \bibnamefont
  {Varanasi}}\ and\ \bibinfo {author} {\bibfnamefont {B.}~\bibnamefont
  {Aazhang}},\ }\bibfield  {title} {\bibinfo {title} {Parametric generalized
  gaussian density estimation},\ }\href@noop {} {\bibfield  {journal} {\bibinfo
   {journal} {The Journal of the Acoustical Society of America}\ }\textbf
  {\bibinfo {volume} {86}},\ \bibinfo {pages} {1404} (\bibinfo {year}
  {1989})}\BibitemShut {NoStop}%
\bibitem [{\citenamefont {Vapnik}\ and\ \citenamefont
  {Mukherjee}(1999)}]{vapnik1999support}%
  \BibitemOpen
  \bibfield  {author} {\bibinfo {author} {\bibfnamefont {V.}~\bibnamefont
  {Vapnik}}\ and\ \bibinfo {author} {\bibfnamefont {S.}~\bibnamefont
  {Mukherjee}},\ }\bibfield  {title} {\bibinfo {title} {Support vector method
  for multivariate density estimation},\ }in\ \href
  {https://proceedings.neurips.cc/paper_files/paper/1999/file/207f88018f72237565570f8a9e5ca240-Paper.pdf}
  {\emph {\bibinfo {booktitle} {Advances in Neural Information Processing
  Systems}}},\ Vol.~\bibinfo {volume} {12},\ \bibinfo {editor} {edited by\
  \bibinfo {editor} {\bibfnamefont {S.}~\bibnamefont {Solla}}, \bibinfo
  {editor} {\bibfnamefont {T.}~\bibnamefont {Leen}},\ and\ \bibinfo {editor}
  {\bibfnamefont {K.}~\bibnamefont {M\"{u}ller}}}\ (\bibinfo  {publisher} {MIT
  Press},\ \bibinfo {year} {1999})\BibitemShut {NoStop}%
\bibitem [{\citenamefont {Rosenblatt}(1956)}]{Rosenblatt1956RemarksFunction}%
  \BibitemOpen
  \bibfield  {author} {\bibinfo {author} {\bibfnamefont {M.}~\bibnamefont
  {Rosenblatt}},\ }\bibfield  {title} {\bibinfo {title} {{Remarks on Some
  Nonparametric Estimates of a Density Function}},\ }\href
  {https://doi.org/10.1214/AOMS/1177728190} {\bibfield  {journal} {\bibinfo
  {journal} {https://doi.org/10.1214/aoms/1177728190}\ }\textbf {\bibinfo
  {volume} {27}},\ \bibinfo {pages} {832} (\bibinfo {year} {1956})}\BibitemShut
  {NoStop}%
\bibitem [{\citenamefont {Parzen}(1962)}]{Parzen1962OnMode}%
  \BibitemOpen
  \bibfield  {author} {\bibinfo {author} {\bibfnamefont {E.}~\bibnamefont
  {Parzen}},\ }\bibfield  {title} {\bibinfo {title} {{On Estimation of a
  Probability Density Function and Mode}},\ }\href
  {https://doi.org/10.1214/aoms/1177704472} {\bibfield  {journal} {\bibinfo
  {journal} {The Annals of Mathematical Statistics}\ }\textbf {\bibinfo
  {volume} {33}},\ \bibinfo {pages} {1065 } (\bibinfo {year}
  {1962})}\BibitemShut {NoStop}%
\bibitem [{\citenamefont {Wang}\ and\ \citenamefont
  {Scott}(2019)}]{wang2019nonparametric}%
  \BibitemOpen
  \bibfield  {author} {\bibinfo {author} {\bibfnamefont {Z.}~\bibnamefont
  {Wang}}\ and\ \bibinfo {author} {\bibfnamefont {D.~W.}\ \bibnamefont
  {Scott}},\ }\bibfield  {title} {\bibinfo {title} {Nonparametric density
  estimation for high-dimensional data—algorithms and applications},\
  }\href@noop {} {\bibfield  {journal} {\bibinfo  {journal} {Wiley
  Interdisciplinary Reviews: Computational Statistics}\ }\textbf {\bibinfo
  {volume} {11}},\ \bibinfo {pages} {e1461} (\bibinfo {year}
  {2019})}\BibitemShut {NoStop}%
\bibitem [{\citenamefont {Liang}\ \emph {et~al.}(2019)\citenamefont {Liang},
  \citenamefont {Shen}, \citenamefont {Li},\ and\ \citenamefont
  {Li}}]{liang2019quantum}%
  \BibitemOpen
  \bibfield  {author} {\bibinfo {author} {\bibfnamefont {J.-M.}\ \bibnamefont
  {Liang}}, \bibinfo {author} {\bibfnamefont {S.-Q.}\ \bibnamefont {Shen}},
  \bibinfo {author} {\bibfnamefont {M.}~\bibnamefont {Li}},\ and\ \bibinfo
  {author} {\bibfnamefont {L.}~\bibnamefont {Li}},\ }\bibfield  {title}
  {\bibinfo {title} {Quantum anomaly detection with density estimation and
  multivariate gaussian distribution},\ }\href@noop {} {\bibfield  {journal}
  {\bibinfo  {journal} {Physical Review A}\ }\textbf {\bibinfo {volume} {99}},\
  \bibinfo {pages} {052310} (\bibinfo {year} {2019})}\BibitemShut {NoStop}%
\bibitem [{\citenamefont {Guo}\ \emph {et~al.}(2022)\citenamefont {Guo},
  \citenamefont {Liu}, \citenamefont {Li}, \citenamefont {Li}, \citenamefont
  {Gao}, \citenamefont {Qin},\ and\ \citenamefont {Wen}}]{guo2022quantum}%
  \BibitemOpen
  \bibfield  {author} {\bibinfo {author} {\bibfnamefont {M.}~\bibnamefont
  {Guo}}, \bibinfo {author} {\bibfnamefont {H.}~\bibnamefont {Liu}}, \bibinfo
  {author} {\bibfnamefont {Y.}~\bibnamefont {Li}}, \bibinfo {author}
  {\bibfnamefont {W.}~\bibnamefont {Li}}, \bibinfo {author} {\bibfnamefont
  {F.}~\bibnamefont {Gao}}, \bibinfo {author} {\bibfnamefont {S.}~\bibnamefont
  {Qin}},\ and\ \bibinfo {author} {\bibfnamefont {Q.}~\bibnamefont {Wen}},\
  }\bibfield  {title} {\bibinfo {title} {Quantum algorithms for anomaly
  detection using amplitude estimation},\ }\href@noop {} {\bibfield  {journal}
  {\bibinfo  {journal} {Physica A: Statistical Mechanics and its Applications}\
  }\textbf {\bibinfo {volume} {604}},\ \bibinfo {pages} {127936} (\bibinfo
  {year} {2022})}\BibitemShut {NoStop}%
\bibitem [{\citenamefont {Useche}\ \emph {et~al.}(2021)\citenamefont {Useche},
  \citenamefont {Giraldo-Carvajal}, \citenamefont {Zuluaga-Bucheli},
  \citenamefont {Jaramillo-Villegas},\ and\ \citenamefont
  {Gonz{\'{a}}lez}}]{Useche2021QuantumQudits}%
  \BibitemOpen
  \bibfield  {author} {\bibinfo {author} {\bibfnamefont {D.~H.}\ \bibnamefont
  {Useche}}, \bibinfo {author} {\bibfnamefont {A.}~\bibnamefont
  {Giraldo-Carvajal}}, \bibinfo {author} {\bibfnamefont {H.~M.}\ \bibnamefont
  {Zuluaga-Bucheli}}, \bibinfo {author} {\bibfnamefont {J.~A.}\ \bibnamefont
  {Jaramillo-Villegas}},\ and\ \bibinfo {author} {\bibfnamefont {F.~A.}\
  \bibnamefont {Gonz{\'{a}}lez}},\ }\bibfield  {title} {\bibinfo {title}
  {{Quantum measurement classification with qudits}},\ }\href
  {https://doi.org/10.1007/S11128-021-03363-Y} {\bibfield  {journal} {\bibinfo
  {journal} {Quantum Information Processing 2021 21:1}\ }\textbf {\bibinfo
  {volume} {21}},\ \bibinfo {pages} {1} (\bibinfo {year} {2021})}\BibitemShut
  {NoStop}%
\bibitem [{\citenamefont {Vargas-Calder{\'o}n}\ \emph
  {et~al.}(2022)\citenamefont {Vargas-Calder{\'o}n}, \citenamefont
  {Gonz{\'a}lez},\ and\ \citenamefont {Vinck-Posada}}]{vargas2022optimisation}%
  \BibitemOpen
  \bibfield  {author} {\bibinfo {author} {\bibfnamefont {V.}~\bibnamefont
  {Vargas-Calder{\'o}n}}, \bibinfo {author} {\bibfnamefont {F.~A.}\
  \bibnamefont {Gonz{\'a}lez}},\ and\ \bibinfo {author} {\bibfnamefont
  {H.}~\bibnamefont {Vinck-Posada}},\ }\bibfield  {title} {\bibinfo {title}
  {Optimisation-free density estimation and classification with quantum
  circuits},\ }\href@noop {} {\bibfield  {journal} {\bibinfo  {journal}
  {Quantum Machine Intelligence}\ }\textbf {\bibinfo {volume} {4}},\ \bibinfo
  {pages} {1} (\bibinfo {year} {2022})}\BibitemShut {NoStop}%
\bibitem [{\citenamefont {Lloyd}\ \emph {et~al.}(2014)\citenamefont {Lloyd},
  \citenamefont {Mohseni},\ and\ \citenamefont
  {Rebentrost}}]{lloyd2014quantum}%
  \BibitemOpen
  \bibfield  {author} {\bibinfo {author} {\bibfnamefont {S.}~\bibnamefont
  {Lloyd}}, \bibinfo {author} {\bibfnamefont {M.}~\bibnamefont {Mohseni}},\
  and\ \bibinfo {author} {\bibfnamefont {P.}~\bibnamefont {Rebentrost}},\
  }\bibfield  {title} {\bibinfo {title} {Quantum principal component
  analysis},\ }\href@noop {} {\bibfield  {journal} {\bibinfo  {journal} {Nature
  Physics}\ }\textbf {\bibinfo {volume} {10}},\ \bibinfo {pages} {631}
  (\bibinfo {year} {2014})}\BibitemShut {NoStop}%
\bibitem [{\citenamefont {Shor}(1994)}]{Shor1994AlgorithmsFactoring}%
  \BibitemOpen
  \bibfield  {author} {\bibinfo {author} {\bibfnamefont {P.~W.}\ \bibnamefont
  {Shor}},\ }\bibfield  {title} {\bibinfo {title} {{Algorithms for quantum
  computation: Discrete logarithms and factoring}},\ }\href
  {https://doi.org/10.1109/SFCS.1994.365700} {\bibfield  {journal} {\bibinfo
  {journal} {Proceedings - Annual IEEE Symposium on Foundations of Computer
  Science, FOCS}\ ,\ \bibinfo {pages} {124}} (\bibinfo {year}
  {1994})}\BibitemShut {NoStop}%
\bibitem [{\citenamefont {Haroche}(1998)}]{haroche1998entanglement}%
  \BibitemOpen
  \bibfield  {author} {\bibinfo {author} {\bibfnamefont {S.}~\bibnamefont
  {Haroche}},\ }\bibfield  {title} {\bibinfo {title} {Entanglement, decoherence
  and the quantum/classical boundary},\ }\href@noop {} {\bibfield  {journal}
  {\bibinfo  {journal} {Physics today}\ }\textbf {\bibinfo {volume} {51}},\
  \bibinfo {pages} {36} (\bibinfo {year} {1998})}\BibitemShut {NoStop}%
\bibitem [{\citenamefont {Preskill}(2018)}]{preskill2018quantum}%
  \BibitemOpen
  \bibfield  {author} {\bibinfo {author} {\bibfnamefont {J.}~\bibnamefont
  {Preskill}},\ }\bibfield  {title} {\bibinfo {title} {Quantum computing in the
  nisq era and beyond},\ }\href@noop {} {\bibfield  {journal} {\bibinfo
  {journal} {Quantum}\ }\textbf {\bibinfo {volume} {2}},\ \bibinfo {pages} {79}
  (\bibinfo {year} {2018})}\BibitemShut {NoStop}%
\bibitem [{\citenamefont {Lee}\ \emph {et~al.}(2021)\citenamefont {Lee},
  \citenamefont {Jun}, \citenamefont {Kim}, \citenamefont {Woo}, \citenamefont
  {Rao}, \citenamefont {Castillo},\ and\ \citenamefont {Yu}}]{lee2021quantum}%
  \BibitemOpen
  \bibfield  {author} {\bibinfo {author} {\bibfnamefont {H.}~\bibnamefont
  {Lee}}, \bibinfo {author} {\bibfnamefont {K.}~\bibnamefont {Jun}}, \bibinfo
  {author} {\bibfnamefont {B.~C.}\ \bibnamefont {Kim}}, \bibinfo {author}
  {\bibfnamefont {M.~M.}\ \bibnamefont {Woo}}, \bibinfo {author} {\bibfnamefont
  {P.}~\bibnamefont {Rao}}, \bibinfo {author} {\bibfnamefont {P.}~\bibnamefont
  {Castillo}},\ and\ \bibinfo {author} {\bibfnamefont {K.}~\bibnamefont {Yu}},\
  }\bibfield  {title} {\bibinfo {title} {{Quantum convolutional neural networks
  on NISQ processors}},\ }in\ \href {https://doi.org/10.1117/12.2594982} {\emph
  {\bibinfo {booktitle} {Quantum Communications and Quantum Imaging XIX}}},\
  Vol.\ \bibinfo {volume} {11835},\ \bibinfo {editor} {edited by\ \bibinfo
  {editor} {\bibfnamefont {K.~S.}\ \bibnamefont {Deacon}}\ and\ \bibinfo
  {editor} {\bibfnamefont {R.~E.}\ \bibnamefont {Meyers}}},\ \bibinfo
  {organization} {International Society for Optics and Photonics}\ (\bibinfo
  {publisher} {SPIE},\ \bibinfo {address} {Bellingham, Washington},\ \bibinfo
  {year} {2021})\ p.\ \bibinfo {pages} {1183509}\BibitemShut {NoStop}%
\bibitem [{\citenamefont {Banchi}(2022)}]{banchi2022robust}%
  \BibitemOpen
  \bibfield  {author} {\bibinfo {author} {\bibfnamefont {L.}~\bibnamefont
  {Banchi}},\ }\bibfield  {title} {\bibinfo {title} {Robust quantum classifiers
  via nisq adversarial learning},\ }\href@noop {} {\bibfield  {journal}
  {\bibinfo  {journal} {Nature Computational Science}\ }\textbf {\bibinfo
  {volume} {2}},\ \bibinfo {pages} {699} (\bibinfo {year} {2022})}\BibitemShut
  {NoStop}%
\bibitem [{\citenamefont {Gonz{\'a}lez}\ \emph {et~al.}(2022)\citenamefont
  {Gonz{\'a}lez}, \citenamefont {Gallego}, \citenamefont {Toledo-Cort{\'e}s},\
  and\ \citenamefont {Vargas-Calder{\'o}n}}]{gonzalez2022learning}%
  \BibitemOpen
  \bibfield  {author} {\bibinfo {author} {\bibfnamefont {F.~A.}\ \bibnamefont
  {Gonz{\'a}lez}}, \bibinfo {author} {\bibfnamefont {A.}~\bibnamefont
  {Gallego}}, \bibinfo {author} {\bibfnamefont {S.}~\bibnamefont
  {Toledo-Cort{\'e}s}},\ and\ \bibinfo {author} {\bibfnamefont
  {V.}~\bibnamefont {Vargas-Calder{\'o}n}},\ }\bibfield  {title} {\bibinfo
  {title} {Learning with density matrices and random features},\ }\href@noop {}
  {\bibfield  {journal} {\bibinfo  {journal} {Quantum Machine Intelligence}\
  }\textbf {\bibinfo {volume} {4}},\ \bibinfo {pages} {1} (\bibinfo {year}
  {2022})}\BibitemShut {NoStop}%
\bibitem [{\citenamefont {Rahimi}\ and\ \citenamefont
  {Recht}(2009)}]{Rahimi2009RandomMachines}%
  \BibitemOpen
  \bibfield  {author} {\bibinfo {author} {\bibfnamefont {A.}~\bibnamefont
  {Rahimi}}\ and\ \bibinfo {author} {\bibfnamefont {B.}~\bibnamefont {Recht}},\
  }\bibfield  {title} {\bibinfo {title} {{Random features for large-scale
  kernel machines}},\ }in\ \href@noop {} {\emph {\bibinfo {booktitle} {Advances
  in Neural Information Processing Systems 20 - Proceedings of the 2007
  Conference}}}\ (\bibinfo {year} {2009})\BibitemShut {NoStop}%
\bibitem [{\citenamefont {Blank}\ \emph {et~al.}(2022)\citenamefont {Blank},
  \citenamefont {da~Silva}, \citenamefont {de~Albuquerque}, \citenamefont
  {Petruccione},\ and\ \citenamefont {Park}}]{blank2022compact}%
  \BibitemOpen
  \bibfield  {author} {\bibinfo {author} {\bibfnamefont {C.}~\bibnamefont
  {Blank}}, \bibinfo {author} {\bibfnamefont {A.~J.}\ \bibnamefont {da~Silva}},
  \bibinfo {author} {\bibfnamefont {L.~P.}\ \bibnamefont {de~Albuquerque}},
  \bibinfo {author} {\bibfnamefont {F.}~\bibnamefont {Petruccione}},\ and\
  \bibinfo {author} {\bibfnamefont {D.~K.}\ \bibnamefont {Park}},\ }\bibfield
  {title} {\bibinfo {title} {Compact quantum distance-based binary
  classifier},\ }\href@noop {} {\bibfield  {journal} {\bibinfo  {journal}
  {arXiv preprint arXiv:2202.02151}\ } (\bibinfo {year} {2022})}\BibitemShut
  {NoStop}%
\bibitem [{\citenamefont {Nghiem}\ \emph {et~al.}(2021)\citenamefont {Nghiem},
  \citenamefont {Chen},\ and\ \citenamefont {Wei}}]{nghiem2021unified}%
  \BibitemOpen
  \bibfield  {author} {\bibinfo {author} {\bibfnamefont {N.~A.}\ \bibnamefont
  {Nghiem}}, \bibinfo {author} {\bibfnamefont {S.~Y.-C.}\ \bibnamefont
  {Chen}},\ and\ \bibinfo {author} {\bibfnamefont {T.-C.}\ \bibnamefont
  {Wei}},\ }\bibfield  {title} {\bibinfo {title} {Unified framework for quantum
  classification},\ }\href {https://doi.org/10.1103/PhysRevResearch.3.033056}
  {\bibfield  {journal} {\bibinfo  {journal} {Phys. Rev. Res.}\ }\textbf
  {\bibinfo {volume} {3}},\ \bibinfo {pages} {033056} (\bibinfo {year}
  {2021})}\BibitemShut {NoStop}%
\bibitem [{\citenamefont {James}\ \emph {et~al.}(2001)\citenamefont {James},
  \citenamefont {Kwiat}, \citenamefont {Munro},\ and\ \citenamefont
  {White}}]{James2005OnQubits}%
  \BibitemOpen
  \bibfield  {author} {\bibinfo {author} {\bibfnamefont {D.~F.~V.}\
  \bibnamefont {James}}, \bibinfo {author} {\bibfnamefont {P.~G.}\ \bibnamefont
  {Kwiat}}, \bibinfo {author} {\bibfnamefont {W.~J.}\ \bibnamefont {Munro}},\
  and\ \bibinfo {author} {\bibfnamefont {A.~G.}\ \bibnamefont {White}},\
  }\bibfield  {title} {\bibinfo {title} {Measurement of qubits},\ }\href
  {https://doi.org/10.1103/PhysRevA.64.052312} {\bibfield  {journal} {\bibinfo
  {journal} {Phys. Rev. A}\ }\textbf {\bibinfo {volume} {64}},\ \bibinfo
  {pages} {052312} (\bibinfo {year} {2001})}\BibitemShut {NoStop}%
\bibitem [{\citenamefont {Li}\ \emph {et~al.}(2019)\citenamefont {Li},
  \citenamefont {Zhang}, \citenamefont {Wang},\ and\ \citenamefont
  {Kumar}}]{li2019learning}%
  \BibitemOpen
  \bibfield  {author} {\bibinfo {author} {\bibfnamefont {Y.}~\bibnamefont
  {Li}}, \bibinfo {author} {\bibfnamefont {K.}~\bibnamefont {Zhang}}, \bibinfo
  {author} {\bibfnamefont {J.}~\bibnamefont {Wang}},\ and\ \bibinfo {author}
  {\bibfnamefont {S.}~\bibnamefont {Kumar}},\ }\bibfield  {title} {\bibinfo
  {title} {Learning adaptive random features},\ }\href
  {https://doi.org/10.1609/aaai.v33i01.33014229} {\bibfield  {journal}
  {\bibinfo  {journal} {Proceedings of the AAAI Conference on Artificial
  Intelligence}\ }\textbf {\bibinfo {volume} {33}},\ \bibinfo {pages} {4229}
  (\bibinfo {year} {2019})}\BibitemShut {NoStop}%
\bibitem [{\citenamefont {González}\ \emph {et~al.}(2023)\citenamefont
  {González}, \citenamefont {Ramos-Pollán},\ and\ \citenamefont
  {Gallego-Mejia}}]{gonzalez2023quantum}%
  \BibitemOpen
  \bibfield  {author} {\bibinfo {author} {\bibfnamefont {F.~A.}\ \bibnamefont
  {González}}, \bibinfo {author} {\bibfnamefont {R.}~\bibnamefont
  {Ramos-Pollán}},\ and\ \bibinfo {author} {\bibfnamefont {J.~A.}\
  \bibnamefont {Gallego-Mejia}},\ }\href@noop {} {\bibinfo {title} {Kernel
  density matrices for probabilistic deep learning}} (\bibinfo {year} {2023}),\
  \Eprint {https://arxiv.org/abs/2305.18204} {arXiv:2305.18204 [cs.LG]}
  \BibitemShut {NoStop}%
\bibitem [{\citenamefont {Rebentrost}\ \emph {et~al.}(2014)\citenamefont
  {Rebentrost}, \citenamefont {Mohseni},\ and\ \citenamefont
  {Lloyd}}]{rebentrost2014quantum}%
  \BibitemOpen
  \bibfield  {author} {\bibinfo {author} {\bibfnamefont {P.}~\bibnamefont
  {Rebentrost}}, \bibinfo {author} {\bibfnamefont {M.}~\bibnamefont
  {Mohseni}},\ and\ \bibinfo {author} {\bibfnamefont {S.}~\bibnamefont
  {Lloyd}},\ }\bibfield  {title} {\bibinfo {title} {Quantum support vector
  machine for big data classification},\ }\href@noop {} {\bibfield  {journal}
  {\bibinfo  {journal} {Physical review letters}\ }\textbf {\bibinfo {volume}
  {113}},\ \bibinfo {pages} {130503} (\bibinfo {year} {2014})}\BibitemShut
  {NoStop}%
\bibitem [{\citenamefont {Blank}\ \emph {et~al.}(2020)\citenamefont {Blank},
  \citenamefont {Park}, \citenamefont {Rhee},\ and\ \citenamefont
  {Petruccione}}]{blank2020quantum}%
  \BibitemOpen
  \bibfield  {author} {\bibinfo {author} {\bibfnamefont {C.}~\bibnamefont
  {Blank}}, \bibinfo {author} {\bibfnamefont {D.~K.}\ \bibnamefont {Park}},
  \bibinfo {author} {\bibfnamefont {J.-K.~K.}\ \bibnamefont {Rhee}},\ and\
  \bibinfo {author} {\bibfnamefont {F.}~\bibnamefont {Petruccione}},\
  }\bibfield  {title} {\bibinfo {title} {Quantum classifier with tailored
  quantum kernel},\ }\href@noop {} {\bibfield  {journal} {\bibinfo  {journal}
  {npj Quantum Information}\ }\textbf {\bibinfo {volume} {6}},\ \bibinfo
  {pages} {1} (\bibinfo {year} {2020})}\BibitemShut {NoStop}%
\bibitem [{\citenamefont {Sergioli}\ \emph {et~al.}(2019)\citenamefont
  {Sergioli}, \citenamefont {Giuntini},\ and\ \citenamefont
  {Freytes}}]{sergioli2019new}%
  \BibitemOpen
  \bibfield  {author} {\bibinfo {author} {\bibfnamefont {G.}~\bibnamefont
  {Sergioli}}, \bibinfo {author} {\bibfnamefont {R.}~\bibnamefont {Giuntini}},\
  and\ \bibinfo {author} {\bibfnamefont {H.}~\bibnamefont {Freytes}},\
  }\bibfield  {title} {\bibinfo {title} {A new quantum approach to binary
  classification},\ }\href@noop {} {\bibfield  {journal} {\bibinfo  {journal}
  {PloS one}\ }\textbf {\bibinfo {volume} {14}},\ \bibinfo {pages} {e0216224}
  (\bibinfo {year} {2019})}\BibitemShut {NoStop}%
\bibitem [{\citenamefont {Bennett}\ \emph {et~al.}(1996)\citenamefont
  {Bennett}, \citenamefont {Brassard}, \citenamefont {Popescu}, \citenamefont
  {Schumacher}, \citenamefont {Smolin},\ and\ \citenamefont
  {Wootters}}]{bennett1996purification}%
  \BibitemOpen
  \bibfield  {author} {\bibinfo {author} {\bibfnamefont {C.~H.}\ \bibnamefont
  {Bennett}}, \bibinfo {author} {\bibfnamefont {G.}~\bibnamefont {Brassard}},
  \bibinfo {author} {\bibfnamefont {S.}~\bibnamefont {Popescu}}, \bibinfo
  {author} {\bibfnamefont {B.}~\bibnamefont {Schumacher}}, \bibinfo {author}
  {\bibfnamefont {J.~A.}\ \bibnamefont {Smolin}},\ and\ \bibinfo {author}
  {\bibfnamefont {W.~K.}\ \bibnamefont {Wootters}},\ }\bibfield  {title}
  {\bibinfo {title} {Purification of noisy entanglement and faithful
  teleportation via noisy channels},\ }\href@noop {} {\bibfield  {journal}
  {\bibinfo  {journal} {Physical review letters}\ }\textbf {\bibinfo {volume}
  {76}},\ \bibinfo {pages} {722} (\bibinfo {year} {1996})}\BibitemShut
  {NoStop}%
\bibitem [{\citenamefont {Peruzzo}\ \emph {et~al.}(2014)\citenamefont
  {Peruzzo}, \citenamefont {McClean}, \citenamefont {Shadbolt}, \citenamefont
  {Yung}, \citenamefont {Zhou}, \citenamefont {Love}, \citenamefont
  {Aspuru-Guzik},\ and\ \citenamefont {O’brien}}]{peruzzo2014variational}%
  \BibitemOpen
  \bibfield  {author} {\bibinfo {author} {\bibfnamefont {A.}~\bibnamefont
  {Peruzzo}}, \bibinfo {author} {\bibfnamefont {J.}~\bibnamefont {McClean}},
  \bibinfo {author} {\bibfnamefont {P.}~\bibnamefont {Shadbolt}}, \bibinfo
  {author} {\bibfnamefont {M.-H.}\ \bibnamefont {Yung}}, \bibinfo {author}
  {\bibfnamefont {X.-Q.}\ \bibnamefont {Zhou}}, \bibinfo {author}
  {\bibfnamefont {P.~J.}\ \bibnamefont {Love}}, \bibinfo {author}
  {\bibfnamefont {A.}~\bibnamefont {Aspuru-Guzik}},\ and\ \bibinfo {author}
  {\bibfnamefont {J.~L.}\ \bibnamefont {O’brien}},\ }\bibfield  {title}
  {\bibinfo {title} {A variational eigenvalue solver on a photonic quantum
  processor},\ }\href@noop {} {\bibfield  {journal} {\bibinfo  {journal}
  {Nature communications}\ }\textbf {\bibinfo {volume} {5}},\ \bibinfo {pages}
  {1} (\bibinfo {year} {2014})}\BibitemShut {NoStop}%
\bibitem [{\citenamefont {Nielsen}\ and\ \citenamefont
  {Chuang}(2010)}]{Nielsen2010QuantumEdition}%
  \BibitemOpen
  \bibfield  {author} {\bibinfo {author} {\bibfnamefont {M.~A.}\ \bibnamefont
  {Nielsen}}\ and\ \bibinfo {author} {\bibfnamefont {I.~L.}\ \bibnamefont
  {Chuang}},\ }\bibfield  {title} {\bibinfo {title} {{Quantum Computation and
  Quantum Information: 10th Anniversary Edition}},\ }\bibfield  {journal}
  {\bibinfo  {journal} {Quantum Computation and Quantum Information}\ }\href
  {https://doi.org/10.1017/CBO9780511976667} {10.1017/CBO9780511976667}
  (\bibinfo {year} {2010})\BibitemShut {NoStop}%
\bibitem [{\citenamefont {Cramer}\ \emph {et~al.}(2010)\citenamefont {Cramer},
  \citenamefont {Plenio}, \citenamefont {Flammia}, \citenamefont {Somma},
  \citenamefont {Gross}, \citenamefont {Bartlett}, \citenamefont
  {Landon-Cardinal}, \citenamefont {Poulin},\ and\ \citenamefont
  {Liu}}]{Cramer2010EfficientTomography}%
  \BibitemOpen
  \bibfield  {author} {\bibinfo {author} {\bibfnamefont {M.}~\bibnamefont
  {Cramer}}, \bibinfo {author} {\bibfnamefont {M.~B.}\ \bibnamefont {Plenio}},
  \bibinfo {author} {\bibfnamefont {S.~T.}\ \bibnamefont {Flammia}}, \bibinfo
  {author} {\bibfnamefont {R.}~\bibnamefont {Somma}}, \bibinfo {author}
  {\bibfnamefont {D.}~\bibnamefont {Gross}}, \bibinfo {author} {\bibfnamefont
  {S.~D.}\ \bibnamefont {Bartlett}}, \bibinfo {author} {\bibfnamefont
  {O.}~\bibnamefont {Landon-Cardinal}}, \bibinfo {author} {\bibfnamefont
  {D.}~\bibnamefont {Poulin}},\ and\ \bibinfo {author} {\bibfnamefont {Y.~K.}\
  \bibnamefont {Liu}},\ }\bibfield  {title} {\bibinfo {title} {{Efficient
  quantum state tomography}},\ }\href {https://doi.org/10.1038/ncomms1147}
  {\bibfield  {journal} {\bibinfo  {journal} {Nature Communications 2010 1:1}\
  }\textbf {\bibinfo {volume} {1}},\ \bibinfo {pages} {1} (\bibinfo {year}
  {2010})}\BibitemShut {NoStop}%
\bibitem [{\citenamefont {González}\ \emph {et~al.}(2021)\citenamefont
  {González}, \citenamefont {Vargas-Calderón},\ and\ \citenamefont
  {Vinck-Posada}}]{Gonzalez2021ClassificationMeasurements}%
  \BibitemOpen
  \bibfield  {author} {\bibinfo {author} {\bibfnamefont {F.~A.}\ \bibnamefont
  {González}}, \bibinfo {author} {\bibfnamefont {V.}~\bibnamefont
  {Vargas-Calderón}},\ and\ \bibinfo {author} {\bibfnamefont {H.}~\bibnamefont
  {Vinck-Posada}},\ }\bibfield  {title} {\bibinfo {title} {Classification with
  quantum measurements},\ }\href {https://doi.org/10.7566/JPSJ.90.044002}
  {\bibfield  {journal} {\bibinfo  {journal} {Journal of the Physical Society
  of Japan}\ }\textbf {\bibinfo {volume} {90}},\ \bibinfo {pages} {044002}
  (\bibinfo {year} {2021})},\ \Eprint
  {https://arxiv.org/abs/https://doi.org/10.7566/JPSJ.90.044002}
  {https://doi.org/10.7566/JPSJ.90.044002} \BibitemShut {NoStop}%
\bibitem [{\citenamefont {Kammonen}\ \emph {et~al.}(2020)\citenamefont
  {Kammonen}, \citenamefont {Kiessling}, \citenamefont {Plecháč},
  \citenamefont {Sandberg},\ and\ \citenamefont
  {Szepessy}}]{2639-8001_2020_3_309}%
  \BibitemOpen
  \bibfield  {author} {\bibinfo {author} {\bibfnamefont {A.}~\bibnamefont
  {Kammonen}}, \bibinfo {author} {\bibfnamefont {J.}~\bibnamefont {Kiessling}},
  \bibinfo {author} {\bibfnamefont {P.}~\bibnamefont {Plecháč}}, \bibinfo
  {author} {\bibfnamefont {M.}~\bibnamefont {Sandberg}},\ and\ \bibinfo
  {author} {\bibfnamefont {A.}~\bibnamefont {Szepessy}},\ }\bibfield  {title}
  {\bibinfo {title} {Adaptive random fourier features with metropolis
  sampling},\ }\href@noop {} {\bibfield  {journal} {\bibinfo  {journal}
  {Foundations of Data Science}\ }\textbf {\bibinfo {volume} {2}},\ \bibinfo
  {pages} {309} (\bibinfo {year} {2020})}\BibitemShut {NoStop}%
\bibitem [{\citenamefont {Liu}\ \emph {et~al.}(2019)\citenamefont {Liu},
  \citenamefont {Xu}, \citenamefont {Yang},\ and\ \citenamefont
  {Jiang}}]{Liu2019AAlgorithm}%
  \BibitemOpen
  \bibfield  {author} {\bibinfo {author} {\bibfnamefont {Y.}~\bibnamefont
  {Liu}}, \bibinfo {author} {\bibfnamefont {Y.}~\bibnamefont {Xu}}, \bibinfo
  {author} {\bibfnamefont {J.}~\bibnamefont {Yang}},\ and\ \bibinfo {author}
  {\bibfnamefont {S.}~\bibnamefont {Jiang}},\ }\bibfield  {title} {\bibinfo
  {title} {{A Polarized Random Fourier Feature Kernel Least-Mean-Square
  Algorithm}},\ }\href {https://doi.org/10.1109/ACCESS.2019.2909304} {\bibfield
   {journal} {\bibinfo  {journal} {IEEE Access}\ }\textbf {\bibinfo {volume}
  {7}},\ \bibinfo {pages} {50833} (\bibinfo {year} {2019})}\BibitemShut
  {NoStop}%
\bibitem [{\citenamefont {Agrawal}\ \emph {et~al.}(2019)\citenamefont
  {Agrawal}, \citenamefont {Campbell}, \citenamefont {Huggins},\ and\
  \citenamefont {Broderick}}]{Agrawal2018Data-dependentApproximation}%
  \BibitemOpen
  \bibfield  {author} {\bibinfo {author} {\bibfnamefont {R.}~\bibnamefont
  {Agrawal}}, \bibinfo {author} {\bibfnamefont {T.}~\bibnamefont {Campbell}},
  \bibinfo {author} {\bibfnamefont {J.}~\bibnamefont {Huggins}},\ and\ \bibinfo
  {author} {\bibfnamefont {T.}~\bibnamefont {Broderick}},\ }\bibfield  {title}
  {\bibinfo {title} {Data-dependent compression of random features for
  large-scale kernel approximation},\ }in\ \href
  {https://proceedings.mlr.press/v89/agrawal19a.html} {\emph {\bibinfo
  {booktitle} {Proceedings of the Twenty-Second International Conference on
  Artificial Intelligence and Statistics}}},\ \bibinfo {series} {Proceedings of
  Machine Learning Research}, Vol.~\bibinfo {volume} {89},\ \bibinfo {editor}
  {edited by\ \bibinfo {editor} {\bibfnamefont {K.}~\bibnamefont {Chaudhuri}}\
  and\ \bibinfo {editor} {\bibfnamefont {M.}~\bibnamefont {Sugiyama}}}\
  (\bibinfo  {publisher} {PMLR},\ \bibinfo {year} {2019})\ pp.\ \bibinfo
  {pages} {1822--1831}\BibitemShut {NoStop}%
\bibitem [{\citenamefont {Yamasaki}\ \emph {et~al.}(2020)\citenamefont
  {Yamasaki}, \citenamefont {Subramanian}, \citenamefont {Sonoda},\ and\
  \citenamefont {Koashi}}]{yamasaki2020learning}%
  \BibitemOpen
  \bibfield  {author} {\bibinfo {author} {\bibfnamefont {H.}~\bibnamefont
  {Yamasaki}}, \bibinfo {author} {\bibfnamefont {S.}~\bibnamefont
  {Subramanian}}, \bibinfo {author} {\bibfnamefont {S.}~\bibnamefont
  {Sonoda}},\ and\ \bibinfo {author} {\bibfnamefont {M.}~\bibnamefont
  {Koashi}},\ }\bibfield  {title} {\bibinfo {title} {Learning with optimized
  random features: Exponential speedup by quantum machine learning without
  sparsity and low-rank assumptions},\ }\href@noop {} {\bibfield  {journal}
  {\bibinfo  {journal} {Advances in neural information processing systems}\
  }\textbf {\bibinfo {volume} {33}},\ \bibinfo {pages} {13674} (\bibinfo {year}
  {2020})}\BibitemShut {NoStop}%
\bibitem [{\citenamefont {Xie}\ \emph {et~al.}(2019)\citenamefont {Xie},
  \citenamefont {Liu}, \citenamefont {Wang},\ and\ \citenamefont
  {Huang}}]{xie2019deep}%
  \BibitemOpen
  \bibfield  {author} {\bibinfo {author} {\bibfnamefont {J.}~\bibnamefont
  {Xie}}, \bibinfo {author} {\bibfnamefont {F.}~\bibnamefont {Liu}}, \bibinfo
  {author} {\bibfnamefont {K.}~\bibnamefont {Wang}},\ and\ \bibinfo {author}
  {\bibfnamefont {X.}~\bibnamefont {Huang}},\ }\bibfield  {title} {\bibinfo
  {title} {Deep kernel learning via random fourier features},\ }\href@noop {}
  {\bibfield  {journal} {\bibinfo  {journal} {arXiv preprint arXiv:1910.02660}\
  } (\bibinfo {year} {2019})}\BibitemShut {NoStop}%
\bibitem [{\citenamefont {Bǎzǎvan}\ \emph {et~al.}(2012)\citenamefont
  {Bǎzǎvan}, \citenamefont {Li},\ and\ \citenamefont
  {Sminchisescu}}]{Bazavan2012FourierLearning}%
  \BibitemOpen
  \bibfield  {author} {\bibinfo {author} {\bibfnamefont {E.~G.}\ \bibnamefont
  {Bǎzǎvan}}, \bibinfo {author} {\bibfnamefont {F.}~\bibnamefont {Li}},\ and\
  \bibinfo {author} {\bibfnamefont {C.}~\bibnamefont {Sminchisescu}},\
  }\bibfield  {title} {\bibinfo {title} {{Fourier Kernel Learning}},\ }\href
  {https://doi.org/10.1007/978-3-642-33709-3{\_}33} {\bibfield  {journal}
  {\bibinfo  {journal} {Lecture Notes in Computer Science (including subseries
  Lecture Notes in Artificial Intelligence and Lecture Notes in
  Bioinformatics)}\ }\textbf {\bibinfo {volume} {7573 LNCS}},\ \bibinfo {pages}
  {459} (\bibinfo {year} {2012})}\BibitemShut {NoStop}%
\bibitem [{\citenamefont {Schuld}\ \emph
  {et~al.}(2021{\natexlab{a}})\citenamefont {Schuld}, \citenamefont {Sweke},\
  and\ \citenamefont {Meyer}}]{schuld2021effect}%
  \BibitemOpen
  \bibfield  {author} {\bibinfo {author} {\bibfnamefont {M.}~\bibnamefont
  {Schuld}}, \bibinfo {author} {\bibfnamefont {R.}~\bibnamefont {Sweke}},\ and\
  \bibinfo {author} {\bibfnamefont {J.~J.}\ \bibnamefont {Meyer}},\ }\bibfield
  {title} {\bibinfo {title} {Effect of data encoding on the expressive power of
  variational quantum-machine-learning models},\ }\href@noop {} {\bibfield
  {journal} {\bibinfo  {journal} {Physical Review A}\ }\textbf {\bibinfo
  {volume} {103}},\ \bibinfo {pages} {032430} (\bibinfo {year}
  {2021}{\natexlab{a}})}\BibitemShut {NoStop}%
\bibitem [{\citenamefont {Havl{\'\i}{\v{c}}ek}\ \emph
  {et~al.}(2019)\citenamefont {Havl{\'\i}{\v{c}}ek}, \citenamefont
  {C{\'o}rcoles}, \citenamefont {Temme}, \citenamefont {Harrow}, \citenamefont
  {Kandala}, \citenamefont {Chow},\ and\ \citenamefont
  {Gambetta}}]{havlivcek2019supervised}%
  \BibitemOpen
  \bibfield  {author} {\bibinfo {author} {\bibfnamefont {V.}~\bibnamefont
  {Havl{\'\i}{\v{c}}ek}}, \bibinfo {author} {\bibfnamefont {A.~D.}\
  \bibnamefont {C{\'o}rcoles}}, \bibinfo {author} {\bibfnamefont
  {K.}~\bibnamefont {Temme}}, \bibinfo {author} {\bibfnamefont {A.~W.}\
  \bibnamefont {Harrow}}, \bibinfo {author} {\bibfnamefont {A.}~\bibnamefont
  {Kandala}}, \bibinfo {author} {\bibfnamefont {J.~M.}\ \bibnamefont {Chow}},\
  and\ \bibinfo {author} {\bibfnamefont {J.~M.}\ \bibnamefont {Gambetta}},\
  }\bibfield  {title} {\bibinfo {title} {Supervised learning with
  quantum-enhanced feature spaces},\ }\href@noop {} {\bibfield  {journal}
  {\bibinfo  {journal} {Nature}\ }\textbf {\bibinfo {volume} {567}},\ \bibinfo
  {pages} {209} (\bibinfo {year} {2019})}\BibitemShut {NoStop}%
\bibitem [{\citenamefont {Chatterjee}\ and\ \citenamefont
  {Yu}(2016)}]{chatterjee2016generalized}%
  \BibitemOpen
  \bibfield  {author} {\bibinfo {author} {\bibfnamefont {R.}~\bibnamefont
  {Chatterjee}}\ and\ \bibinfo {author} {\bibfnamefont {T.}~\bibnamefont
  {Yu}},\ }\bibfield  {title} {\bibinfo {title} {Generalized coherent states,
  reproducing kernels, and quantum support vector machines},\ }\href@noop {}
  {\bibfield  {journal} {\bibinfo  {journal} {arXiv preprint arXiv:1612.03713}\
  } (\bibinfo {year} {2016})}\BibitemShut {NoStop}%
\bibitem [{\citenamefont {Schuld}\ \emph
  {et~al.}(2021{\natexlab{b}})\citenamefont {Schuld}, \citenamefont
  {Petruccione}, \citenamefont {Schuld},\ and\ \citenamefont
  {Petruccione}}]{schuld2021quantum}%
  \BibitemOpen
  \bibfield  {author} {\bibinfo {author} {\bibfnamefont {M.}~\bibnamefont
  {Schuld}}, \bibinfo {author} {\bibfnamefont {F.}~\bibnamefont {Petruccione}},
  \bibinfo {author} {\bibfnamefont {M.}~\bibnamefont {Schuld}},\ and\ \bibinfo
  {author} {\bibfnamefont {F.}~\bibnamefont {Petruccione}},\ }\bibfield
  {title} {\bibinfo {title} {Quantum models as kernel methods},\ }\href@noop {}
  {\bibfield  {journal} {\bibinfo  {journal} {Machine Learning with Quantum
  Computers}\ ,\ \bibinfo {pages} {217}} (\bibinfo {year}
  {2021}{\natexlab{b}})}\BibitemShut {NoStop}%
\bibitem [{\citenamefont {Stoudenmire}\ and\ \citenamefont
  {Schwab}(2016)}]{stoudenmire2016supervised}%
  \BibitemOpen
  \bibfield  {author} {\bibinfo {author} {\bibfnamefont {E.}~\bibnamefont
  {Stoudenmire}}\ and\ \bibinfo {author} {\bibfnamefont {D.~J.}\ \bibnamefont
  {Schwab}},\ }\bibfield  {title} {\bibinfo {title} {Supervised learning with
  tensor networks},\ }\href@noop {} {\bibfield  {journal} {\bibinfo  {journal}
  {Advances in neural information processing systems}\ }\textbf {\bibinfo
  {volume} {29}} (\bibinfo {year} {2016})}\BibitemShut {NoStop}%
\bibitem [{\citenamefont {Rudin}(2017)}]{rudin2017fourier}%
  \BibitemOpen
  \bibfield  {author} {\bibinfo {author} {\bibfnamefont {W.}~\bibnamefont
  {Rudin}},\ }\href@noop {} {\emph {\bibinfo {title} {Fourier analysis on
  groups}}}\ (\bibinfo  {publisher} {Courier Dover Publications},\ \bibinfo
  {address} {Mineola, New York},\ \bibinfo {year} {2017})\BibitemShut {NoStop}%
\bibitem [{\citenamefont {Raykar}\ \emph {et~al.}(2010)\citenamefont {Raykar},
  \citenamefont {Duraiswami},\ and\ \citenamefont {Zhao}}]{raykar2010fast}%
  \BibitemOpen
  \bibfield  {author} {\bibinfo {author} {\bibfnamefont {V.~C.}\ \bibnamefont
  {Raykar}}, \bibinfo {author} {\bibfnamefont {R.}~\bibnamefont {Duraiswami}},\
  and\ \bibinfo {author} {\bibfnamefont {L.~H.}\ \bibnamefont {Zhao}},\
  }\bibfield  {title} {\bibinfo {title} {Fast computation of kernel
  estimators},\ }\href@noop {} {\bibfield  {journal} {\bibinfo  {journal}
  {Journal of Computational and Graphical Statistics}\ }\textbf {\bibinfo
  {volume} {19}},\ \bibinfo {pages} {205} (\bibinfo {year} {2010})}\BibitemShut
  {NoStop}%
\bibitem [{\citenamefont {Yang}\ \emph {et~al.}(2004)\citenamefont {Yang},
  \citenamefont {Duraiswami},\ and\ \citenamefont {Davis}}]{yang2004efficient}%
  \BibitemOpen
  \bibfield  {author} {\bibinfo {author} {\bibfnamefont {C.}~\bibnamefont
  {Yang}}, \bibinfo {author} {\bibfnamefont {R.}~\bibnamefont {Duraiswami}},\
  and\ \bibinfo {author} {\bibfnamefont {L.~S.}\ \bibnamefont {Davis}},\
  }\bibfield  {title} {\bibinfo {title} {Efficient kernel machines using the
  improved fast gauss transform},\ }\href@noop {} {\bibfield  {journal}
  {\bibinfo  {journal} {Advances in neural information processing systems}\
  }\textbf {\bibinfo {volume} {17}} (\bibinfo {year} {2004})}\BibitemShut
  {NoStop}%
\bibitem [{\citenamefont {Yang}\ \emph {et~al.}(2003)\citenamefont {Yang},
  \citenamefont {Duraiswami},\ and\ \citenamefont
  {Gumerov}}]{yang2003improved}%
  \BibitemOpen
  \bibfield  {author} {\bibinfo {author} {\bibnamefont {Yang}}, \bibinfo
  {author} {\bibnamefont {Duraiswami}},\ and\ \bibinfo {author} {\bibnamefont
  {Gumerov}},\ }\bibfield  {title} {\bibinfo {title} {Improved fast gauss
  transform and efficient kernel density estimation},\ }in\ \href@noop {}
  {\emph {\bibinfo {booktitle} {Proceedings ninth IEEE international conference
  on computer vision}}}\ (\bibinfo {organization} {IEEE},\ \bibinfo {year}
  {2003})\ pp.\ \bibinfo {pages} {664--671}\BibitemShut {NoStop}%
\bibitem [{\citenamefont {Iten}\ \emph {et~al.}(2016)\citenamefont {Iten},
  \citenamefont {Colbeck}, \citenamefont {Kukuljan}, \citenamefont {Home},\
  and\ \citenamefont {Christandl}}]{Iten2016QuantumIsometries}%
  \BibitemOpen
  \bibfield  {author} {\bibinfo {author} {\bibfnamefont {R.}~\bibnamefont
  {Iten}}, \bibinfo {author} {\bibfnamefont {R.}~\bibnamefont {Colbeck}},
  \bibinfo {author} {\bibfnamefont {I.}~\bibnamefont {Kukuljan}}, \bibinfo
  {author} {\bibfnamefont {J.}~\bibnamefont {Home}},\ and\ \bibinfo {author}
  {\bibfnamefont {M.}~\bibnamefont {Christandl}},\ }\bibfield  {title}
  {\bibinfo {title} {Quantum circuits for isometries},\ }\href
  {https://doi.org/10.1103/PhysRevA.93.032318} {\bibfield  {journal} {\bibinfo
  {journal} {Phys. Rev. A}\ }\textbf {\bibinfo {volume} {93}},\ \bibinfo
  {pages} {032318} (\bibinfo {year} {2016})}\BibitemShut {NoStop}%
\bibitem [{\citenamefont {Shende}\ \emph {et~al.}(2006)\citenamefont {Shende},
  \citenamefont {Bullock},\ and\ \citenamefont {Markov}}]{shende2006synthesis}%
  \BibitemOpen
  \bibfield  {author} {\bibinfo {author} {\bibfnamefont {V.~V.}\ \bibnamefont
  {Shende}}, \bibinfo {author} {\bibfnamefont {S.~S.}\ \bibnamefont
  {Bullock}},\ and\ \bibinfo {author} {\bibfnamefont {I.~L.}\ \bibnamefont
  {Markov}},\ }\bibfield  {title} {\bibinfo {title} {Synthesis of quantum-logic
  circuits},\ }\href@noop {} {\bibfield  {journal} {\bibinfo  {journal} {IEEE
  Transactions on Computer-Aided Design of Integrated Circuits and Systems}\
  }\textbf {\bibinfo {volume} {25}},\ \bibinfo {pages} {1000} (\bibinfo {year}
  {2006})}\BibitemShut {NoStop}%
\bibitem [{\citenamefont {M{\"o}tt{\"o}nen}\ \emph {et~al.}(2005)\citenamefont
  {M{\"o}tt{\"o}nen}, \citenamefont {Vartiainen}, \citenamefont {Bergholm},\
  and\ \citenamefont {Salomaa}}]{mottonen2005transformation}%
  \BibitemOpen
  \bibfield  {author} {\bibinfo {author} {\bibfnamefont {M.}~\bibnamefont
  {M{\"o}tt{\"o}nen}}, \bibinfo {author} {\bibfnamefont {J.~J.}\ \bibnamefont
  {Vartiainen}}, \bibinfo {author} {\bibfnamefont {V.}~\bibnamefont
  {Bergholm}},\ and\ \bibinfo {author} {\bibfnamefont {M.~M.}\ \bibnamefont
  {Salomaa}},\ }\bibfield  {title} {\bibinfo {title} {Transformation of quantum
  states using uniformly controlled rotations},\ }\href@noop {} {\bibfield
  {journal} {\bibinfo  {journal} {Quantum Information \& Computation}\ }\textbf
  {\bibinfo {volume} {5}},\ \bibinfo {pages} {467} (\bibinfo {year}
  {2005})}\BibitemShut {NoStop}%
\bibitem [{\citenamefont {Liu}\ \emph {et~al.}(2021)\citenamefont {Liu},
  \citenamefont {Arunachalam},\ and\ \citenamefont {Temme}}]{liu2021rigorous}%
  \BibitemOpen
  \bibfield  {author} {\bibinfo {author} {\bibfnamefont {Y.}~\bibnamefont
  {Liu}}, \bibinfo {author} {\bibfnamefont {S.}~\bibnamefont {Arunachalam}},\
  and\ \bibinfo {author} {\bibfnamefont {K.}~\bibnamefont {Temme}},\ }\bibfield
   {title} {\bibinfo {title} {A rigorous and robust quantum speed-up in
  supervised machine learning},\ }\href@noop {} {\bibfield  {journal} {\bibinfo
   {journal} {Nature Physics}\ }\textbf {\bibinfo {volume} {17}},\ \bibinfo
  {pages} {1013} (\bibinfo {year} {2021})}\BibitemShut {NoStop}%
\bibitem [{\citenamefont {Mitarai}\ \emph {et~al.}(2018)\citenamefont
  {Mitarai}, \citenamefont {Negoro}, \citenamefont {Kitagawa},\ and\
  \citenamefont {Fujii}}]{mitarai2018quantum}%
  \BibitemOpen
  \bibfield  {author} {\bibinfo {author} {\bibfnamefont {K.}~\bibnamefont
  {Mitarai}}, \bibinfo {author} {\bibfnamefont {M.}~\bibnamefont {Negoro}},
  \bibinfo {author} {\bibfnamefont {M.}~\bibnamefont {Kitagawa}},\ and\
  \bibinfo {author} {\bibfnamefont {K.}~\bibnamefont {Fujii}},\ }\bibfield
  {title} {\bibinfo {title} {Quantum circuit learning},\ }\href@noop {}
  {\bibfield  {journal} {\bibinfo  {journal} {Physical Review A}\ }\textbf
  {\bibinfo {volume} {98}},\ \bibinfo {pages} {032309} (\bibinfo {year}
  {2018})}\BibitemShut {NoStop}%
\bibitem [{\citenamefont {Cincio}\ \emph {et~al.}(2018)\citenamefont {Cincio},
  \citenamefont {Suba{\c{s}}{\i}}, \citenamefont {Sornborger},\ and\
  \citenamefont {Coles}}]{cincio2018learning}%
  \BibitemOpen
  \bibfield  {author} {\bibinfo {author} {\bibfnamefont {L.}~\bibnamefont
  {Cincio}}, \bibinfo {author} {\bibfnamefont {Y.}~\bibnamefont
  {Suba{\c{s}}{\i}}}, \bibinfo {author} {\bibfnamefont {A.~T.}\ \bibnamefont
  {Sornborger}},\ and\ \bibinfo {author} {\bibfnamefont {P.~J.}\ \bibnamefont
  {Coles}},\ }\bibfield  {title} {\bibinfo {title} {Learning the quantum
  algorithm for state overlap},\ }\href@noop {} {\bibfield  {journal} {\bibinfo
   {journal} {New Journal of Physics}\ }\textbf {\bibinfo {volume} {20}},\
  \bibinfo {pages} {113022} (\bibinfo {year} {2018})}\BibitemShut {NoStop}%
\bibitem [{\citenamefont {Schuld}\ \emph {et~al.}(2020)\citenamefont {Schuld},
  \citenamefont {Bocharov}, \citenamefont {Svore},\ and\ \citenamefont
  {Wiebe}}]{schuld2020circuit}%
  \BibitemOpen
  \bibfield  {author} {\bibinfo {author} {\bibfnamefont {M.}~\bibnamefont
  {Schuld}}, \bibinfo {author} {\bibfnamefont {A.}~\bibnamefont {Bocharov}},
  \bibinfo {author} {\bibfnamefont {K.~M.}\ \bibnamefont {Svore}},\ and\
  \bibinfo {author} {\bibfnamefont {N.}~\bibnamefont {Wiebe}},\ }\bibfield
  {title} {\bibinfo {title} {Circuit-centric quantum classifiers},\ }\href@noop
  {} {\bibfield  {journal} {\bibinfo  {journal} {Physical Review A}\ }\textbf
  {\bibinfo {volume} {101}},\ \bibinfo {pages} {032308} (\bibinfo {year}
  {2020})}\BibitemShut {NoStop}%
\bibitem [{\citenamefont {Watkins}\ \emph {et~al.}(2023)\citenamefont
  {Watkins}, \citenamefont {Chen},\ and\ \citenamefont
  {Yoo}}]{watkins2023quantum}%
  \BibitemOpen
  \bibfield  {author} {\bibinfo {author} {\bibfnamefont {W.~M.}\ \bibnamefont
  {Watkins}}, \bibinfo {author} {\bibfnamefont {S.~Y.-C.}\ \bibnamefont
  {Chen}},\ and\ \bibinfo {author} {\bibfnamefont {S.}~\bibnamefont {Yoo}},\
  }\bibfield  {title} {\bibinfo {title} {Quantum machine learning with
  differential privacy},\ }\href@noop {} {\bibfield  {journal} {\bibinfo
  {journal} {Scientific Reports}\ }\textbf {\bibinfo {volume} {13}},\ \bibinfo
  {pages} {2453} (\bibinfo {year} {2023})}\BibitemShut {NoStop}%
\bibitem [{\citenamefont {Cardoso}\ \emph {et~al.}(2021)\citenamefont
  {Cardoso}, \citenamefont {Akamatsu}, \citenamefont {Campo~Junior},
  \citenamefont {Duzzioni}, \citenamefont {Jaramillo},\ and\ \citenamefont
  {Villas-Boas}}]{cardoso2021detailed}%
  \BibitemOpen
  \bibfield  {author} {\bibinfo {author} {\bibfnamefont {F.~R.}\ \bibnamefont
  {Cardoso}}, \bibinfo {author} {\bibfnamefont {D.~Y.}\ \bibnamefont
  {Akamatsu}}, \bibinfo {author} {\bibfnamefont {V.~L.}\ \bibnamefont
  {Campo~Junior}}, \bibinfo {author} {\bibfnamefont {E.~I.}\ \bibnamefont
  {Duzzioni}}, \bibinfo {author} {\bibfnamefont {A.}~\bibnamefont
  {Jaramillo}},\ and\ \bibinfo {author} {\bibfnamefont {C.~J.}\ \bibnamefont
  {Villas-Boas}},\ }\bibfield  {title} {\bibinfo {title} {Detailed account of
  complexity for implementation of circuit-based quantum algorithms},\
  }\href@noop {} {\bibfield  {journal} {\bibinfo  {journal} {Frontiers in
  Physics}\ ,\ \bibinfo {pages} {582}} (\bibinfo {year} {2021})}\BibitemShut
  {NoStop}%
\bibitem [{\citenamefont {M{\"o}tt{\"o}nen}\ \emph {et~al.}(2004)\citenamefont
  {M{\"o}tt{\"o}nen}, \citenamefont {Vartiainen}, \citenamefont {Bergholm},\
  and\ \citenamefont {Salomaa}}]{mottonen2004quantum}%
  \BibitemOpen
  \bibfield  {author} {\bibinfo {author} {\bibfnamefont {M.}~\bibnamefont
  {M{\"o}tt{\"o}nen}}, \bibinfo {author} {\bibfnamefont {J.~J.}\ \bibnamefont
  {Vartiainen}}, \bibinfo {author} {\bibfnamefont {V.}~\bibnamefont
  {Bergholm}},\ and\ \bibinfo {author} {\bibfnamefont {M.~M.}\ \bibnamefont
  {Salomaa}},\ }\bibfield  {title} {\bibinfo {title} {Quantum circuits for
  general multiqubit gates},\ }\href@noop {} {\bibfield  {journal} {\bibinfo
  {journal} {Physical review letters}\ }\textbf {\bibinfo {volume} {93}},\
  \bibinfo {pages} {130502} (\bibinfo {year} {2004})}\BibitemShut {NoStop}%
\bibitem [{\citenamefont {Gallego-Mejia}\ and\ \citenamefont
  {González}(2023)}]{gallego2022quantum}%
  \BibitemOpen
  \bibfield  {author} {\bibinfo {author} {\bibfnamefont {J.~A.}\ \bibnamefont
  {Gallego-Mejia}}\ and\ \bibinfo {author} {\bibfnamefont {F.~A.}\ \bibnamefont
  {González}},\ }\bibfield  {title} {\bibinfo {title} {Demande: Density matrix
  neural density estimation},\ }\href
  {https://doi.org/10.1109/ACCESS.2023.3279123} {\bibfield  {journal} {\bibinfo
   {journal} {IEEE Access}\ }\textbf {\bibinfo {volume} {11}},\ \bibinfo
  {pages} {53062} (\bibinfo {year} {2023})}\BibitemShut {NoStop}%
\bibitem [{\citenamefont {Bergholm}\ \emph {et~al.}(2018)\citenamefont
  {Bergholm}, \citenamefont {Izaac}, \citenamefont {Schuld}, \citenamefont
  {Gogolin}, \citenamefont {Alam}, \citenamefont {Ahmed}, \citenamefont
  {Arrazola}, \citenamefont {Blank}, \citenamefont {Delgado}, \citenamefont
  {Jahangiri} \emph {et~al.}}]{bergholm2018pennylane}%
  \BibitemOpen
  \bibfield  {author} {\bibinfo {author} {\bibfnamefont {V.}~\bibnamefont
  {Bergholm}}, \bibinfo {author} {\bibfnamefont {J.}~\bibnamefont {Izaac}},
  \bibinfo {author} {\bibfnamefont {M.}~\bibnamefont {Schuld}}, \bibinfo
  {author} {\bibfnamefont {C.}~\bibnamefont {Gogolin}}, \bibinfo {author}
  {\bibfnamefont {M.~S.}\ \bibnamefont {Alam}}, \bibinfo {author}
  {\bibfnamefont {S.}~\bibnamefont {Ahmed}}, \bibinfo {author} {\bibfnamefont
  {J.~M.}\ \bibnamefont {Arrazola}}, \bibinfo {author} {\bibfnamefont
  {C.}~\bibnamefont {Blank}}, \bibinfo {author} {\bibfnamefont
  {A.}~\bibnamefont {Delgado}}, \bibinfo {author} {\bibfnamefont
  {S.}~\bibnamefont {Jahangiri}}, \emph {et~al.},\ }\bibfield  {title}
  {\bibinfo {title} {Pennylane: Automatic differentiation of hybrid
  quantum-classical computations},\ }\href@noop {} {\bibfield  {journal}
  {\bibinfo  {journal} {arXiv preprint arXiv:1811.04968}\ } (\bibinfo {year}
  {2018})}\BibitemShut {NoStop}%
\bibitem [{\citenamefont {Chen}(2011)}]{Chen2011AnInspection}%
  \BibitemOpen
  \bibfield  {author} {\bibinfo {author} {\bibfnamefont {L.~F.}\ \bibnamefont
  {Chen}},\ }\bibfield  {title} {\bibinfo {title} {{An improved negative
  selection approach for anomaly detection: with applications in medical
  diagnosis and quality inspection}},\ }\href
  {https://doi.org/10.1007/S00521-011-0781-5} {\bibfield  {journal} {\bibinfo
  {journal} {Neural Computing and Applications 2011 22:5}\ }\textbf {\bibinfo
  {volume} {22}},\ \bibinfo {pages} {901} (\bibinfo {year} {2011})}\BibitemShut
  {NoStop}%
\bibitem [{\citenamefont {Noble}\ and\ \citenamefont
  {Cook}(2003)}]{noble2003graph}%
  \BibitemOpen
  \bibfield  {author} {\bibinfo {author} {\bibfnamefont {C.~C.}\ \bibnamefont
  {Noble}}\ and\ \bibinfo {author} {\bibfnamefont {D.~J.}\ \bibnamefont
  {Cook}},\ }\bibfield  {title} {\bibinfo {title} {Graph-based anomaly
  detection},\ }in\ \href {https://doi.org/10.1145/956750.956831} {\emph
  {\bibinfo {booktitle} {Proceedings of the Ninth ACM SIGKDD International
  Conference on Knowledge Discovery and Data Mining}}},\ \bibinfo {series and
  number} {KDD '03}\ (\bibinfo  {publisher} {Association for Computing
  Machinery},\ \bibinfo {address} {New York, NY, USA},\ \bibinfo {year}
  {2003})\ p.\ \bibinfo {pages} {631–636}\BibitemShut {NoStop}%
\bibitem [{\citenamefont {Gonz{\'a}lez}\ and\ \citenamefont
  {Dasgupta}(2003)}]{gonzalez2003anomaly}%
  \BibitemOpen
  \bibfield  {author} {\bibinfo {author} {\bibfnamefont {F.~A.}\ \bibnamefont
  {Gonz{\'a}lez}}\ and\ \bibinfo {author} {\bibfnamefont {D.}~\bibnamefont
  {Dasgupta}},\ }\bibfield  {title} {\bibinfo {title} {Anomaly detection using
  real-valued negative selection},\ }\href@noop {} {\bibfield  {journal}
  {\bibinfo  {journal} {Genetic Programming and Evolvable Machines}\ }\textbf
  {\bibinfo {volume} {4}},\ \bibinfo {pages} {383} (\bibinfo {year}
  {2003})}\BibitemShut {NoStop}%
\bibitem [{\citenamefont {Liu}\ and\ \citenamefont
  {Rebentrost}(2018)}]{Liu2018QuantumDetection}%
  \BibitemOpen
  \bibfield  {author} {\bibinfo {author} {\bibfnamefont {N.}~\bibnamefont
  {Liu}}\ and\ \bibinfo {author} {\bibfnamefont {P.}~\bibnamefont
  {Rebentrost}},\ }\bibfield  {title} {\bibinfo {title} {{Quantum machine
  learning for quantum anomaly detection}},\ }\href
  {https://doi.org/10.1103/PHYSREVA.97.042315/FIGURES/1/MEDIUM} {\bibfield
  {journal} {\bibinfo  {journal} {Physical Review A}\ }\textbf {\bibinfo
  {volume} {97}},\ \bibinfo {pages} {042315} (\bibinfo {year}
  {2018})}\BibitemShut {NoStop}%
\bibitem [{\citenamefont {Herr}\ \emph {et~al.}(2021)\citenamefont {Herr},
  \citenamefont {Obert},\ and\ \citenamefont
  {Rosenkranz}}]{Herr2021AnomalyNetworks}%
  \BibitemOpen
  \bibfield  {author} {\bibinfo {author} {\bibfnamefont {D.}~\bibnamefont
  {Herr}}, \bibinfo {author} {\bibfnamefont {B.}~\bibnamefont {Obert}},\ and\
  \bibinfo {author} {\bibfnamefont {M.}~\bibnamefont {Rosenkranz}},\ }\bibfield
   {title} {\bibinfo {title} {{Anomaly detection with variational quantum
  generative adversarial networks}},\ }\href
  {https://doi.org/10.1088/2058-9565/AC0D4D} {\bibfield  {journal} {\bibinfo
  {journal} {Quantum Science and Technology}\ }\textbf {\bibinfo {volume}
  {6}},\ \bibinfo {pages} {045004} (\bibinfo {year} {2021})}\BibitemShut
  {NoStop}%
\bibitem [{\citenamefont {Aggarwal}\ and\ \citenamefont
  {Sathe}(2015)}]{Aggarwal2015TheoreticalEnsembles}%
  \BibitemOpen
  \bibfield  {author} {\bibinfo {author} {\bibfnamefont {C.~C.}\ \bibnamefont
  {Aggarwal}}\ and\ \bibinfo {author} {\bibfnamefont {S.}~\bibnamefont
  {Sathe}},\ }\bibfield  {title} {\bibinfo {title} {{Theoretical Foundations
  and Algorithms for Outlier Ensembles}},\ }\href
  {https://doi.org/10.1145/2830544.2830549} {\bibfield  {journal} {\bibinfo
  {journal} {ACM SIGKDD Explorations Newsletter}\ }\textbf {\bibinfo {volume}
  {17}},\ \bibinfo {pages} {24} (\bibinfo {year} {2015})}\BibitemShut {NoStop}%
\bibitem [{\citenamefont {Perdomo}\ \emph {et~al.}(2021)\citenamefont
  {Perdomo}, \citenamefont {Leyton-Ortega},\ and\ \citenamefont
  {Perdomo-Ortiz}}]{perdomo2021entanglement}%
  \BibitemOpen
  \bibfield  {author} {\bibinfo {author} {\bibfnamefont {O.}~\bibnamefont
  {Perdomo}}, \bibinfo {author} {\bibfnamefont {V.}~\bibnamefont
  {Leyton-Ortega}},\ and\ \bibinfo {author} {\bibfnamefont {A.}~\bibnamefont
  {Perdomo-Ortiz}},\ }\bibfield  {title} {\bibinfo {title} {Entanglement types
  for two-qubit states with real amplitudes},\ }\href@noop {} {\bibfield
  {journal} {\bibinfo  {journal} {Quantum Information Processing}\ }\textbf
  {\bibinfo {volume} {20}},\ \bibinfo {pages} {1} (\bibinfo {year}
  {2021})}\BibitemShut {NoStop}%
\bibitem [{\citenamefont {Ezzell}\ \emph {et~al.}(2022)\citenamefont {Ezzell},
  \citenamefont {Ball}, \citenamefont {Siddiqui}, \citenamefont {Wilde},
  \citenamefont {Sornborger}, \citenamefont {Coles},\ and\ \citenamefont
  {Holmes}}]{ezzell2022quantum}%
  \BibitemOpen
  \bibfield  {author} {\bibinfo {author} {\bibfnamefont {N.}~\bibnamefont
  {Ezzell}}, \bibinfo {author} {\bibfnamefont {E.~M.}\ \bibnamefont {Ball}},
  \bibinfo {author} {\bibfnamefont {A.~U.}\ \bibnamefont {Siddiqui}}, \bibinfo
  {author} {\bibfnamefont {M.~M.}\ \bibnamefont {Wilde}}, \bibinfo {author}
  {\bibfnamefont {A.~T.}\ \bibnamefont {Sornborger}}, \bibinfo {author}
  {\bibfnamefont {P.~J.}\ \bibnamefont {Coles}},\ and\ \bibinfo {author}
  {\bibfnamefont {Z.}~\bibnamefont {Holmes}},\ }\bibfield  {title} {\bibinfo
  {title} {Quantum mixed state compiling},\ }\href@noop {} {\bibfield
  {journal} {\bibinfo  {journal} {arXiv preprint arXiv:2209.00528}\ } (\bibinfo
  {year} {2022})}\BibitemShut {NoStop}%
\bibitem [{\citenamefont {Di~Bartolomeo}\ \emph {et~al.}(2023)\citenamefont
  {Di~Bartolomeo}, \citenamefont {Vischi}, \citenamefont {Cesa}, \citenamefont
  {Wixinger}, \citenamefont {Grossi}, \citenamefont {Donadi},\ and\
  \citenamefont {Bassi}}]{di2023noisy}%
  \BibitemOpen
  \bibfield  {author} {\bibinfo {author} {\bibfnamefont {G.}~\bibnamefont
  {Di~Bartolomeo}}, \bibinfo {author} {\bibfnamefont {M.}~\bibnamefont
  {Vischi}}, \bibinfo {author} {\bibfnamefont {F.}~\bibnamefont {Cesa}},
  \bibinfo {author} {\bibfnamefont {R.}~\bibnamefont {Wixinger}}, \bibinfo
  {author} {\bibfnamefont {M.}~\bibnamefont {Grossi}}, \bibinfo {author}
  {\bibfnamefont {S.}~\bibnamefont {Donadi}},\ and\ \bibinfo {author}
  {\bibfnamefont {A.}~\bibnamefont {Bassi}},\ }\bibfield  {title} {\bibinfo
  {title} {Noisy gates for simulating quantum computers},\ }\href@noop {}
  {\bibfield  {journal} {\bibinfo  {journal} {Physical Review Research}\
  }\textbf {\bibinfo {volume} {5}},\ \bibinfo {pages} {043210} (\bibinfo {year}
  {2023})}\BibitemShut {NoStop}%
\end{thebibliography}%

\onecolumngrid

\appendix

\section{Quantum adaptive Fourier features in a quantum computer with one and two qubit gates}\label{QDE: QAFF 1-2 gates}

We present the quantum circuit implementation of the quantum adaptive Fourier features mapping,
\begin{equation}
	\ket{\psi(\boldsymbol{x})}_n = \sqrt{\frac{1}{d}}\begin{bmatrix}1, e^{i\sqrt{\gamma}\mathbf{w}_1\cdot\boldsymbol{x}}, e^{i\sqrt{\gamma}\mathbf{w}_2\cdot\boldsymbol{x}}, \cdots , e^{i\sqrt{\gamma}\mathbf{w}_{d-1}\cdot\boldsymbol{x}}  \end{bmatrix}',
\end{equation}\label{QDE:Eq QAFF Appendix}
with the use of single qubit gates and CNOT gates, instead of uniformly controlled rotations, see Sect. \ref{QAFF Quantum implementation}. The proposed method is based on the following reparametrization $\{\sqrt{\gamma}\mathbf{w}_j\cdot\boldsymbol{x}\}\rightarrow\{\sqrt{\gamma}\boldsymbol{\theta}_j\cdot\boldsymbol{x}\}$, where  $\{\mathbf{w}_j\}, \{\boldsymbol{\theta}_j\}_{j=1,\cdots, d-1} \in \mathbb{R}^D$.

\begin{figure}
	\centering
	\includegraphics[scale = 0.645]{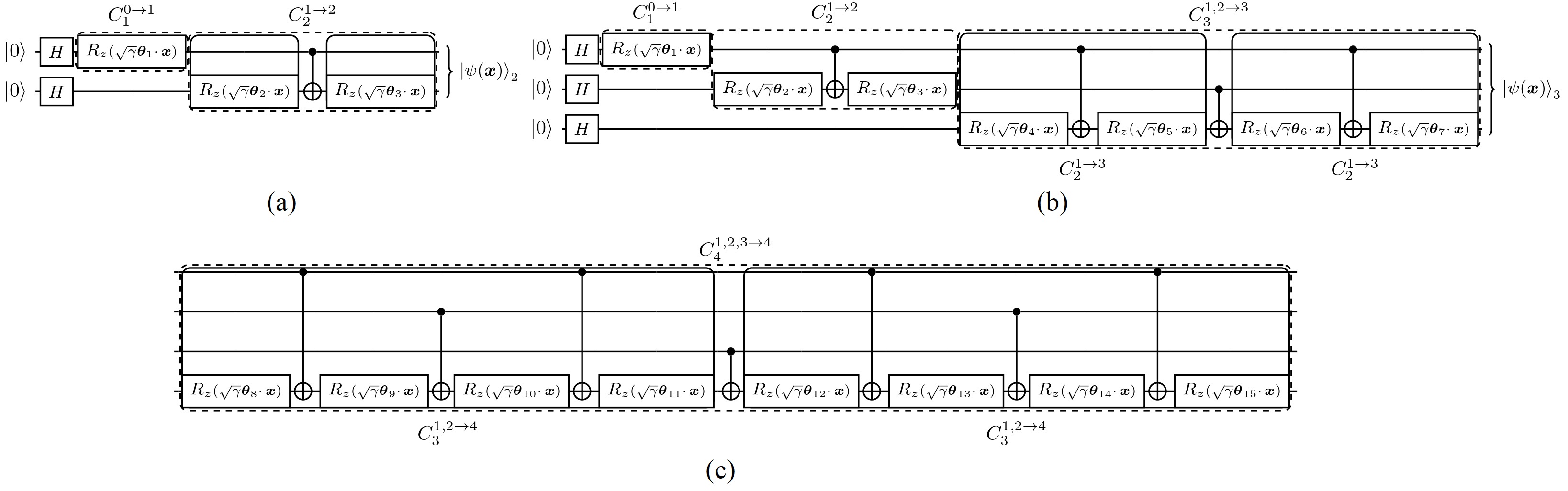}
	\caption{Proposed quantum circuits for the Quantum adaptive Fourier features operation, (a) shows the circuit for two qubits and a size of 4 QAFF, and (b) shows the circuit for three qubits and 8 QAFF. As an example of the recursive construction of these operations, (c) illustrates the operation $C_4^{1,2,3\rightarrow 4}$ of the circuit for 16 QAFF, formed by two copies of the block $C_3^{1, 2\rightarrow 3}$ in (b) and a CNOT gate.}
	\label{fig:QAFF2Qubits}
\end{figure}

We illustrate the proposed method starting from base cases starting with the method to prepare the state $\ket{\psi(\boldsymbol{x})}_2 = \sqrt{\frac{1}{4}}\begin{bmatrix} 1, e^{i\sqrt{\gamma} \mathbf{w}_1\cdot\boldsymbol{x}}, e^{i\sqrt{\gamma}\mathbf{w}_2\cdot\boldsymbol{x}} , e^{i\sqrt{\gamma}\mathbf{w}_3\cdot\boldsymbol{x}} \end{bmatrix}$ that corresponds to 2 qubits, 4 QAFF and the classical feature $\boldsymbol{x} \in \mathbb{R}^D$. The method starts by applying a single Hadamard gate to each qubit, followed by  rotations $R_z(\sqrt{\gamma}\boldsymbol{\theta}_1\cdot\boldsymbol{x})$ in the first qubit and $R_z(\sqrt{\gamma}\boldsymbol{\theta}_2\cdot\boldsymbol{x})$ in the second qubit. We then apply a CNOT gate which controls the first qubit and targets the second qubit followed by a last rotation $R_z(\sqrt{\gamma}\boldsymbol{\theta}_3\cdot\boldsymbol{x})$ on the second qubit as show in the Fig. \ref{fig:QAFF2Qubits}a. These operations create the state $\ket{\psi(\boldsymbol{x})}_2=\sqrt{\frac{1}{4}}\begin{bmatrix}1, e^{i\sqrt{\gamma}(\boldsymbol{\theta}_1+\boldsymbol{\theta}_2)\cdot\boldsymbol{x}}, e^{i\sqrt{\gamma}(\boldsymbol{\theta}_2+\boldsymbol{\theta}_3)\cdot\boldsymbol{x}}, e^{i\sqrt{\gamma}(\boldsymbol{\theta}_1+\boldsymbol{\theta}_3)\cdot\boldsymbol{x}} \end{bmatrix}'$. The structure of this quantum state preparation for two qubits has been explored in Ref. \cite{perdomo2021entanglement} in the context of amplitude encoding. The desired quantum Fourier map is obtained by setting the parameters $\{\boldsymbol{\theta}_j\}$ such that $\boldsymbol{\theta}_1 + \boldsymbol{\theta}_2 = \mathbf{w}_1$, $\boldsymbol{\theta}_2 + \boldsymbol{\theta}_3 = \mathbf{w}_2$, and $\boldsymbol{\theta}_1 + \boldsymbol{\theta}_3 = \mathbf{w}_3$.

It is also possible to rewrite the previous state preparation by defining the operations $C_1^{0\rightarrow1}$ and $C^{1\rightarrow 2}_2$ as controlled rotations in the $z$ axis, where the subindex indicates the number of qubits involved in the operation, and the superindex indicates the controlled qubits followed by an arrow to the single target qubit, with the top qubit of the circuit considered as the qubit 1, and the cero index indicating no controlled qubit. The two-qubit state preparation can be realized by applying a $H$ gate to each qubit followed by the $C_1^{0\rightarrow1}$ operation in the first qubit corresponding to a $R_z$ rotation, followed by the $C^{1\rightarrow 2}_2$, which is built by grouping the two single $R_z$ rotations in the second qubit with the CNOT, as shown in Fig. \ref{fig:QAFF2Qubits}a.

We now illustrate the quantum state preparation for eight quantum Fourier features and three qubits with single $R_z$ qubit rotations with angles $\{\sqrt{\gamma}\boldsymbol{\theta}_j\cdot\boldsymbol{x}\}_{j=1,\cdots,7}$. First, we apply a Hadamard gate to each of the three qubits, and then the preparation of the first two qubits corresponds to the previous two-qubit state preparation, i.e., we apply the operations $C_1^{0\rightarrow1}$, and $C^{1\rightarrow 2}_2$, and then we apply twice the operation $C^{1\rightarrow 3}_2$ that shares the same structure of $C^{1\rightarrow 2}_2$; the CNOT has the same control qubit (first qubit), but it changes the $R_z$ rotations and the target of the CNOT from qubit 2 to qubit 3. Hence, we apply twice $C^{1\rightarrow 3}_2$  with different angular rotations in the $z$ basis as shown in Fig. \ref{fig:QAFF2Qubits}b and then we connect these operations with a CNOT which controls the second qubit and targets the third qubit, creating the state,
\begin{equation}
	\ket{\psi(\boldsymbol{x})}_3 = \sqrt{\frac{1}{8}}\exp \Big( i\sqrt{\gamma}\begin{bmatrix}0, \boldsymbol{\Theta}_{1,2,5,6}\cdot\boldsymbol{x}, \boldsymbol{\Theta}_{2,3,4,5}\cdot\boldsymbol{x}, \boldsymbol{\Theta}_{1,3,4,6}\cdot\boldsymbol{x} , \boldsymbol{\Theta}_{4,5,6,7}\cdot\boldsymbol{x}, \boldsymbol{\Theta}_{1,2,4,7}\cdot\boldsymbol{x}, \boldsymbol{\Theta}_{2,3,6,7}\cdot\boldsymbol{x}, \boldsymbol{\Theta}_{1,3,5,7}\cdot\boldsymbol{x} \end{bmatrix}'\Big),
\end{equation}\label{QDE:Eq QAFF 2 qubits}
where we define $\boldsymbol{\Theta}_{r,s,t,u} = (\boldsymbol{\theta}_r + \boldsymbol{\theta}_s + \boldsymbol{\theta}_t + \boldsymbol{\theta}_u)$, and we set the corresponding parameters equal to the corresponding weights $\{\mathbf{w}_j\}_{j=1,\cdots,7}$ that build the QAFF with three qubits.

\begin{figure}
	\centering
	\includegraphics[scale = 0.197]{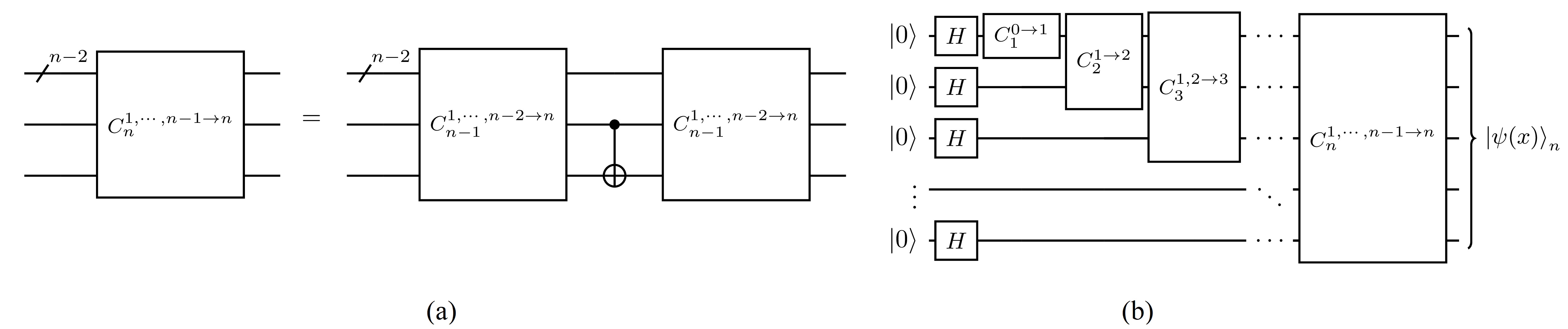}
	\caption{Recursive construction of the quantum adaptive Fourier features on $n$ qubits, (a) shows the construction of the operator $C_n^{{1,\cdots,{n-1}} \rightarrow n}$ as the concatenation of two operators $C_{n-1}^{{1,\cdots,{n-2}} \rightarrow n}$ and a CNOT gate in between which controls the $(n-1)$th qubit  and targets the $(n)$th qubit, and (b) shows the construction of the total QAFF circuit for an arbitrary number of qubits $n$ in terms of these recursive operators.}
	\label{fig:QAFF2REcursion Formula}
\end{figure}

We may generalize these previous state preparation strategies by defining recursively the operation $C_n^{{1,\cdots,{n-1}} \rightarrow n}$, which corresponds to a transformation on $n$ qubits with controls the first $n-1$ qubits and target the $n^{\text{th}}$ qubit. This operation is defined by considering the previous case $C_{n-1}^{{1,\cdots,{n-2}} \rightarrow {n-1}}$ whose target is the $(n-1)^{\text{th}}$ qubit. We may construct an equivalent operation $C_{n-1}^{{1,\cdots,{n-2}} \rightarrow {n}}$ that shares the same sequence of gates as the $C_{n-1}^{{1,\cdots,{n-2}} \rightarrow {n-1}}$, namely, both share the same control qubits for the CNOTs, but the $R_z$ rotations and targets of the CNOTs move from the $(n-1)^{\text{th}}$ qubit to the $n^{\text{th}}$ qubit. Then, we construct recursively $C_n^{{1,\cdots,{n-1}} \rightarrow n}$, by applying twice $C_{n-1}^{{1,\cdots,{n-2}} \rightarrow {n}}$ on the $n$ qubits, each duplicate with distinct $R_z$ rotation angles and connecting these two duplicates with a CNOT which controls the $(n-1)$th qubit and targets the $n$th qubit as shown in Fig. \ref{fig:QAFF2REcursion Formula}a. To further illustrate the method, Fig. \ref{fig:QAFF2Qubits}c shows the $C^{1,2,3\rightarrow 4}_4$ case  which is defined from $C^{1,2\rightarrow 3}_3$ (Fig. \ref{fig:QAFF2Qubits}b).

Based on the transformations $\{C_i^{{1,\cdots,{i-1}} \rightarrow i}\}$, we may construct an ansatz for QAFF that also corresponds to an arbitrary quantum state preparation in the $z$ basis on $n$ qubits, by first applying a $H$ gate to each of the $n$ qubits followed by the $C_1^{0\rightarrow1}$, $C^{1\rightarrow 2}_2$, $C^{1,2\rightarrow 3}_3$, $C^{1,2,3\rightarrow 4}_4$, $\cdots$, $C_n^{{1,\cdots,{n-1}} \rightarrow n}$. The final quantum state $\ket{\psi(\boldsymbol{x})}_n$ will be given by,

\begin{equation}
	\ket{\psi(\boldsymbol{x})}_n = \Big[\prod_{i=1}^n{(C_{n-i+1}^{{1,\cdots,{n-i}} \rightarrow n-i+1}\otimes I_{i-1}})\Big]H_{n}\ket{0}_n,
\end{equation}

where we define $I_{i-1} = I^{\otimes i-1}$ as the identity matrix for $i-1$ qubits, and $H_n = H^{\otimes n}$ to the $n$ initial Hadamard gates applied to each qubit. Also, each $C_i^{{1,\cdots,{i-1}} \rightarrow i}$ is composed of $2^{i-1}$ rotations from the set $\{R_z(\sqrt{\gamma}{\boldsymbol{\theta}_j}\cdot \boldsymbol{x})\}$ and $2^{i-1}-1$ CNOT gates. The depth of the QAFF ansatz (total number of CNOTs) is given by $\sum_{i=1}^n(2^{i-1}-1)=2^{n}-(n+1)$, and it requires $\sum_{i=1}^n2^{i-1}=2^{n}-1$ $R_z$ rotations, which match the number of independent phase variables of an arbitrary pure state on $n$ qubits.

As stated before, this recursive method can also be used to build an ansatz for an arbitrary pure quantum state on $n$ qubits. To build an arbitrary state in the $z$ basis, we apply the same ansatz as the QAFF (which starts by applying $H_n$ to the $\ket{0}_n$ state), but replace the single-qubit rotations $\{R_z(\sqrt{\gamma}{\boldsymbol{\theta}_j}\cdot\boldsymbol{x})\}_{j=1, \cdots d-1}$ with the rotations $\{R_z(\delta_j)\}_{j=1, \cdots d-1}$, where the $\{\delta_j\}$ depend on the phases of the pure state to be learned or to be prepared and $d = 2^n$. We can also build a pure quantum state with only real components by ignoring the $H_n$ operation, and applying the sequence of $\{C_i^{{1,\cdots,{i-1}} \rightarrow i}\}$ to the $\ket{0}_n$ state with $R_y$ rotations instead of $R_z$ rotations. Furthermore, the QAFF ansatz can be used to prepare an arbitrary complex-valued quantum state of $n$ qubits by first building a pure state with real components with the $R_y$ rotations followed by the $R_z$ rotations to account for the phases of the complex components. Compared to a common mechanism for building quantum states such as the presented in Ref. \cite{shende2006synthesis}, our method uses fewer CNOT gates for the preparation of arbitrary quantum states in either the $y$ basis or in the $z$ basis, but has the same depth and structure when combining both $y$ and $z$ rotations to build arbitrary quantum states.

Considering that the depth of the QAFF map $\ket{\psi(\boldsymbol{x})}_n$ (number of CNOTs) is $2^{n}-(n+1) = d - (\log{d}-1)$, which is independent on the size $D$ of the original data, this mapping can represent high-dimensional data in quantum computers with a low number of qubits. Furthermore, the complexity of the QAFF method for estimating the Gaussian kernel with the projection $\abs{\langle\psi(\boldsymbol{x}_l)\vert\psi(\boldsymbol{x}_m)\rangle}^2$ is $O(TDd + TdR)$ where $T$ is the number of data pairs $\{(\boldsymbol{x}_l, \boldsymbol{x}_m)\}$, and $R$ is the total number of $\ket{0}_n$ measurements required to estimate $\hat{k}_{\{\mathbf{w}_j\}}(\boldsymbol{x}_l, \boldsymbol{x}_m) = \abs{\langle\psi(\boldsymbol{x}_l)\vert\psi(\boldsymbol{x}_m)\rangle}^2$ with the quantum kernel estimation algorithm \cite{liu2021rigorous}; this result follows from the fact that for each data pair $(\boldsymbol{x}_l, \boldsymbol{x}_m)$ it is required $O(Dd)$ steps to find the angles $\{\sqrt{\gamma} \boldsymbol{\theta}_j \cdot \boldsymbol{x}\}$, $O(d) $ to prepare the QAFF states and $R$ measurement shots to estimate the projection. In Appendix \ref{QDE sec: gradient classical QAFF}, we illustrate the overall quantum complexity of training QAFF by taking into account the gradients of the method.

\section{Gradients and complexity of the classical and quantum QAFF}\label{QDE sec: gradient classical QAFF}

In this section, we illustrate how to find the gradients of the loss function $\mathcal{L}_{\{\mathbf{w}_k\}}$ that depends on the operation $\hat{k}_{\{\mathbf{w}_k\}}(\boldsymbol{x}_l, \boldsymbol{x}_m) = \abs{\langle\psi(\boldsymbol{x}_l)\vert\psi(\boldsymbol{x}_m)\rangle}^2$ for the QAFF, see Eqs. \ref{QDE: QAFF map} and \ref{QDE: QAFF loss}, in both classical and quantum computers. 

The gradient of the loss function $\mathcal{L}_{\{\mathbf{w}_k\}}$ with respect to some $\mathbf{w}_j$ may be written as,
\begin{equation}
    \nabla_{\mathbf{w}_j}\mathcal{L}_{\{\mathbf{w}_k\}} = \frac{1}{N^2}\sum_{l=1}^{N}\sum_{m=1}^{N}\Big[-2\big(k(\boldsymbol{x}_l, \boldsymbol{x}_m) - \abs{\langle\psi(\boldsymbol{x}_l)\vert\psi(\boldsymbol{x}_m)\rangle}^2\big)\nabla_{\mathbf{w}_j} \big( \abs{\langle\psi(\boldsymbol{x}_l)\vert\psi(\boldsymbol{x}_m)\rangle}^2\big)\Big],\label{QDE:eq gradients QAFF}
\end{equation}

where $\nabla_{\mathbf{w}_j} \big( \abs{\langle\psi(\boldsymbol{x}_l)\vert\psi(\boldsymbol{x}_m)\rangle}^2\big)$ may be written explicitly as, 
\begin{eqnarray}
    \nabla_{\mathbf{w}_j} \Big[\big(\sum_{k=0}^{d-1}\frac{1}{d}e^{i\sqrt{\gamma}\mathbf{w}_k\cdot(\boldsymbol{x}_l-\boldsymbol{x}_m)}\big)\big(\sum_{k=0}^{d-1}\frac{1}{d}e^{-i\sqrt{\gamma}\mathbf{w}_k\cdot(\boldsymbol{x}_l-\boldsymbol{x}_m)}\big)\Big] =\notag\\
     \frac{1}{d}(i\sqrt{\gamma}(\boldsymbol{x}_l-\boldsymbol{x}_m))\Big[e^{i\sqrt{\gamma}\mathbf{w}_j\cdot(\boldsymbol{x}_l-\boldsymbol{x}_m)}\langle\psi(\boldsymbol{x}_l)\vert\psi(\boldsymbol{x}_m)\rangle- e^{-i\sqrt{\gamma}\mathbf{w}_j\cdot(\boldsymbol{x}_l-\boldsymbol{x}_m)}\langle\psi(\boldsymbol{x}_m)\vert\psi(\boldsymbol{x}_l)\rangle\Big].\label{QDE: eq: gradient loss function}
\end{eqnarray}

Based on the previous equation, for each data pair, these $d$ gradients can be calculated on a classical computer with complexity $O(Dd+d^2)$, therefore the total complexity of training classically QAFF is $O(TDd+Td^2)$. 

To build this operation in a quantum computer we may use parameter-shift rule \cite{mitarai2018quantum}, by shifting the parameter $(\sqrt{\gamma}\mathbf{w}_j\cdot(\boldsymbol{x}_l-\boldsymbol{x}_m))$ of $\abs{\langle\psi(\boldsymbol{x}_l)\vert\psi(\boldsymbol{x}_m)\rangle}^2$ to $(\sqrt{\gamma}\mathbf{w}_j\cdot(\boldsymbol{x}_l-\boldsymbol{x}_m) + \pi/2)$ and $(\sqrt{\gamma}\mathbf{w}_j\cdot(\boldsymbol{x}_l-\boldsymbol{x}_m) - \pi/2)$, since $\abs{\langle\psi(\boldsymbol{x}_l)\vert\psi(\boldsymbol{x}_m)\rangle}^2$ might be written as $\bra{x}U^\dag(\sqrt{\gamma}\mathbf{w}_j\cdot(\boldsymbol{x}_l-\boldsymbol{x}_m))\hat{B}U(\sqrt{\gamma}\mathbf{w}_j\cdot(\boldsymbol{x}_l-\boldsymbol{x}_m))\ket{x}$, for some state $\ket{x}$ and some operator $\hat{B}$ on $n$ qubits that do not depend on $\mathbf{w}_j$. Hence from parameter-shift rule the gradient can also be computed by, 
\begin{eqnarray}
    \nabla_{\mathbf{w}_j} \big( \abs{\langle\psi(\boldsymbol{x}_l)\vert\psi(\boldsymbol{x}_m)\rangle}^2\big) = \frac{1}{2}\sqrt{\gamma}(\boldsymbol{x}_l-\boldsymbol{x}_m)\Big[\bra{x}U^\dag(\sqrt{\gamma}\mathbf{w}_j\cdot(\boldsymbol{x}_l-\boldsymbol{x}_m)+\pi/2)\hat{B}U(\sqrt{\gamma}\mathbf{w}_j\cdot(\boldsymbol{x}_l-\boldsymbol{x}_m)+\pi/2)\ket{x} \notag \\ - \bra{x}U^\dag(\sqrt{\gamma}\mathbf{w}_j\cdot(\boldsymbol{x}_l-\boldsymbol{x}_m)-\pi/2)\hat{B}U(\sqrt{\gamma}\mathbf{w}_j\cdot(\boldsymbol{x}_l-\boldsymbol{x}_m)-\pi/2)\ket{x}\Big] = \notag \\
     \frac{1}{2}\sqrt{\gamma}(\boldsymbol{x}_l-\boldsymbol{x}_m)\Big[ \abs{\langle\psi(\boldsymbol{x}_l)\vert\psi(\boldsymbol{x}_m)\rangle}^2_{\sqrt{\gamma}\mathbf{w}_j\cdot(\boldsymbol{x}_l-\boldsymbol{x}_m)\rightarrow\sqrt{\gamma}\mathbf{w}_j\cdot(\boldsymbol{x}_l-\boldsymbol{x}_m)+\pi/2} - \abs{\langle\psi(\boldsymbol{x}_l)\vert\psi(\boldsymbol{x}_m)\rangle}^2_{\sqrt{\gamma}\mathbf{w}_j\cdot(\boldsymbol{x}_l-\boldsymbol{x}_m)\rightarrow\sqrt{\gamma}\mathbf{w}_j\cdot(\boldsymbol{x}_l-\boldsymbol{x}_m)-\pi/2} \Big],\notag
\end{eqnarray}

where in the last line the arrow in the subindex indicates the applied parameter-shift rule. Hence, we may use the QAFF mapping along with the kernel estimation algorithm \cite{liu2021rigorous} to estimate the gradients of the QAFF. Although, in this argument we parameterized the QAFF by $\{\sqrt{\gamma}\mathbf{w}_j\cdot\boldsymbol{x}\}$ which correspond to uniformly controlled $R_z$ rotations as explained in Sect. \ref{QAFF Quantum implementation}, we can also apply an equivalent parameter-shift rule to the QAFF reparameterized with single $R_z$ rotations with angles $\{\sqrt{\gamma}\boldsymbol{\theta}_j\cdot\boldsymbol{x}\}$, see Appendix \ref{QDE: QAFF 1-2 gates}.

Knowing the gradients of the loss function is sufficient to optimize the weights of the QAFF hence the total quantum complexity of QAFF lies on estimating these gradients, see Eq. \ref{QDE:eq gradients QAFF}. In Appendix \ref{QDE: QAFF 1-2 gates}, we showed that to estimate $\abs{\langle\psi(\boldsymbol{x}_l)\vert\psi(\boldsymbol{x}_m)\rangle}^2$ it is required $O(TDd + TdR)$ where $D$ is the size of the classical data, $d$ is the number of Fourier features, $T$ is the number of data pairs $\{(\boldsymbol{x}_l, \boldsymbol{x}_m)\}$, and $R$ is the number of $\ket{0}_n$ measurements of the quantum kernel estimation circuit. The estimation of the gradients $\nabla_{\mathbf{w}_j} ( \abs{\langle\psi(\boldsymbol{x}_l)\vert\psi(\boldsymbol{x}_m)\rangle}^2)$ accounts for an additional quantum complexity of $O(Td^2R)$ considering that for each of the $d$ Fourier vector weights $\{\mathbf{w}_j\}$ (or $\{\boldsymbol{\theta}_j\}$) and each data pair $(\boldsymbol{x}_l, \boldsymbol{x}_m)$ we need to find its gradient by estimating the projection $\abs{\langle\psi(\boldsymbol{x}_l)\vert\psi(\boldsymbol{x}_m)\rangle}^2$ with shifted parameter $\mathbf{w}_j\cdot(\boldsymbol{x}_l-\boldsymbol{x}_m)$ with complexity $O(dR)$, because we only recalculate the angles $\{\sqrt{\gamma}\mathbf{w}_j\cdot\boldsymbol{x}\}$ when doing the gradient estimation of a different data pair. Therefore, the overall quantum complexity to train the QAFF corresponds to $O(TDd + Td^2R)$. Furthermore, the complexity of extracting the quantum Fourier features of a data set with $N$ samples is $O(NDd + Nd\log{d}R)$, because for each data sample, it is required $O(Dd)$ to find the Fourier angles of the QAFF, and $O(d\log{d}R)$ to prepare the $d$ Fourier components and to measure $R$ times the $n = \log{d}$ bit strings that form the basis. 

\section{SDM ansatz, density matrix preparation from eigenbasis}\label{QDE sec:mixedstateinitialization2}

\begin{figure}
    \centering
    \includegraphics[width=0.35\columnwidth]{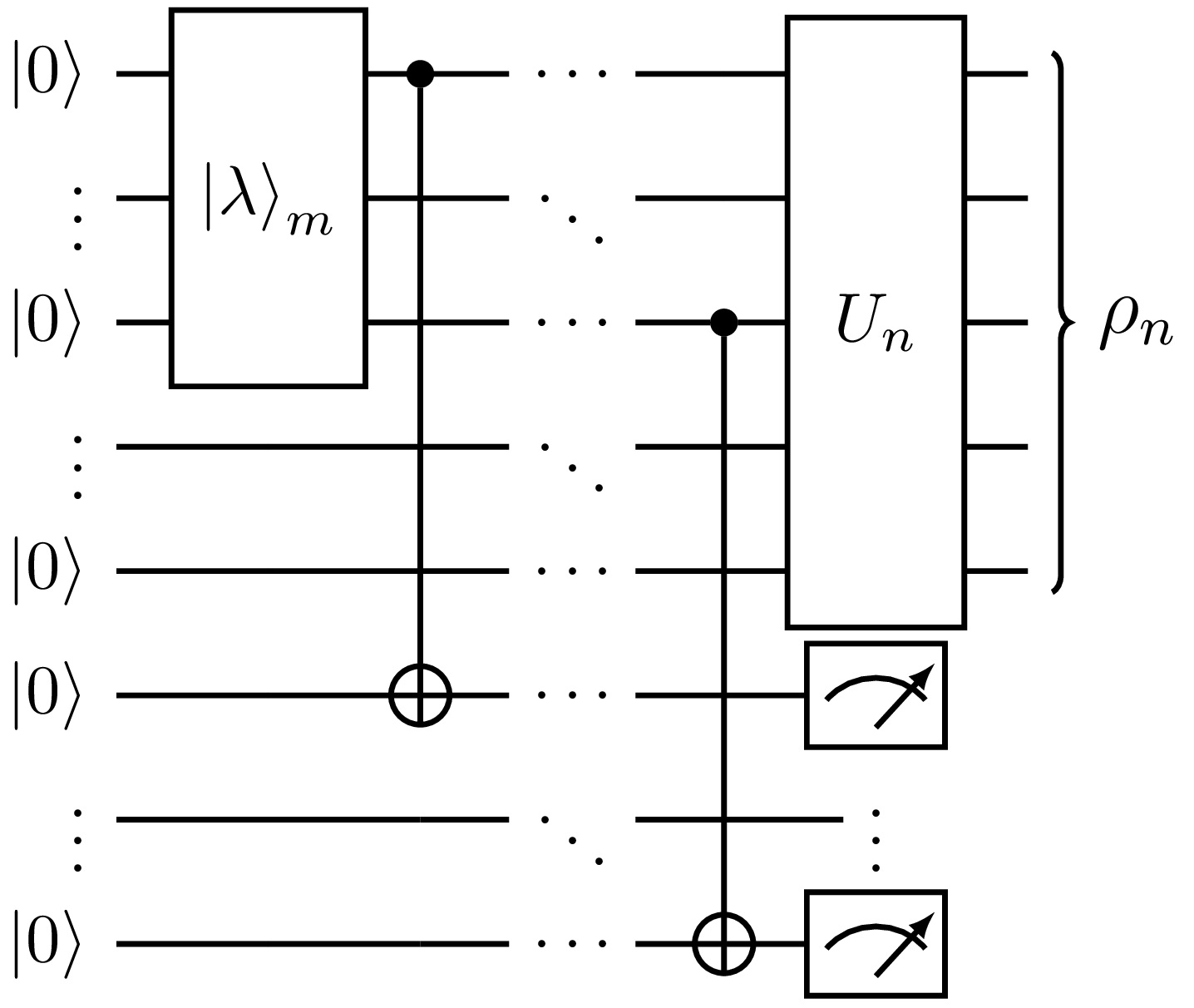}
    \caption{SDM ansatz: mixed state preparation from its spectral decomposition. Considering the eigendecomposion of the density matrix $\rho = U \Lambda U^\dag$, we prepare $\Lambda$ with an entangled state between the first $m$ qubits and the last $m$ qubits and $U$ with an isometry in the first $n$ qubits.}
    \label{fig:QMCFig1}
\end{figure}

The SDM (spectral density matrix) ansatz to initialize a density matrix $\rho \in \mathbb{C}^{d \times d}$ with rank $r$ departing from its spectral decomposition requires  $n + m$ qubits, such that $n = \lceil \log{(d)} \rceil$, and $m = \lceil \log{(r)} \rceil$. Let $\rho = U \Lambda U^\dag$ be the spectral decomposition of the density matrix, where $U$ is a unitary matrix whose first $r$ columns are the eigenvectors of $\rho$ and $\Lambda = \big[\sum_{i=0}^{r-1} \lambda_i \ket{i}\bra{i}+\sum_{i=r}^{d-1} 0 \ket{i}\bra{i}\big]$ is the diagonal matrix of eigenvalues. The protocol begins by initializing the first $m$ qubits with the state $\ket{\lambda}_m = \sum_{i=0}^{r-1}\sqrt{\lambda_i}\ket{i}_m$, thanks to amplitude encoding \cite{shende2006synthesis, mottonen2005transformation}. Therefore, the circuit starts with the state, 
\begin{equation}
	\sum_{i=0}^{r-1}\sqrt{\lambda_i}{\textstyle\ket{i}_m} \otimes \ket{0}_{(n-m)} \otimes \ket{0}_m, 
\end{equation}
then, a cascade of $m$ CNOT gates between the first $m$ qubits and the last $m$ qubits are applied to the circuit, as shown in Fig.  \ref{fig:QMCFig1}. The $j^{\text{th}}$ CNOT controls the $j$th qubit and targets the $(n + j)$th qubit for $j \in \{0, \ldots m - 1\}$. The resulting quantum state of this operation is, 
\begin{equation}
	\sum_{i=0}^{r-1}\sqrt{\lambda_i}\ket{i}_m\otimes\ket{0}_{(n-m)}\otimes\ket{i}_m,
\end{equation}
followed by a partial trace on the purification by performing a measurement on the last $m$ qubits, which results in a mixed state in the first $m$ qubits with only classical probability, 
\begin{equation}
\Lambda_n = (\sum_{i=0}^{r-1} \lambda_i \ket{i}\bra{i}_m) \otimes \ket{0}\bra{0}_{n-m}.
\end{equation}
The desired density matrix is obtained by applying to the state $\Lambda_n$ an isometry \cite{Iten2016QuantumIsometries} $U_n$ from $m$ qubits to $n$ qubits, given by,
\begin{equation}
U_n = 
\left( \begin{array}{c|c}
   U & 0 \\
   \hline
   0 & I \\
\end{array}\right),
\end{equation}
which prepares the $r$ eigenvectors of $\rho$ in the first $r$ columns of $U_n$ and $I$ is the identity matrix with rank $2^n-d$, the process results in the desired mixed state, 
\begin{equation}
     \rho_n = U_n\Lambda_n U^\dag_n.
\end{equation}

as shown in Fig \ref{fig:QMCFig1}. Note that measuring the last $m$ qubits after applying the isometry $U_{n}$ leads to the same result. 

The mixed state preparation protocol from its spectral decomposition was first proposed in qudit-based quantum computers \cite{Useche2021QuantumQudits}, here we extend the method to quantum computers based on qubits. After the initial version of this manuscript, an independent work on the preparation of mixed states with qubits was released \cite{ezzell2022quantum}, applied to the problem of mixed state learning.

\section{Additional simulations for hybrid 2-D density estimation  with quantum variational QAFF}\label{sec:Hybrid DE experiments with quantum QAFF}

Although the hybrid density estimation method Q-DEMDE uses classical computers to train the quantum adaptive Fourier features, in this section, we demonstrate the feasibility of using a quantum simulator to learn the quantum adaptive Fourier features for density estimation via the quantum variational algorithm described in Sect. \ref{QAFF Quantum implementation}.

\begin{figure}
    \centering
    \includegraphics[scale = 0.6]{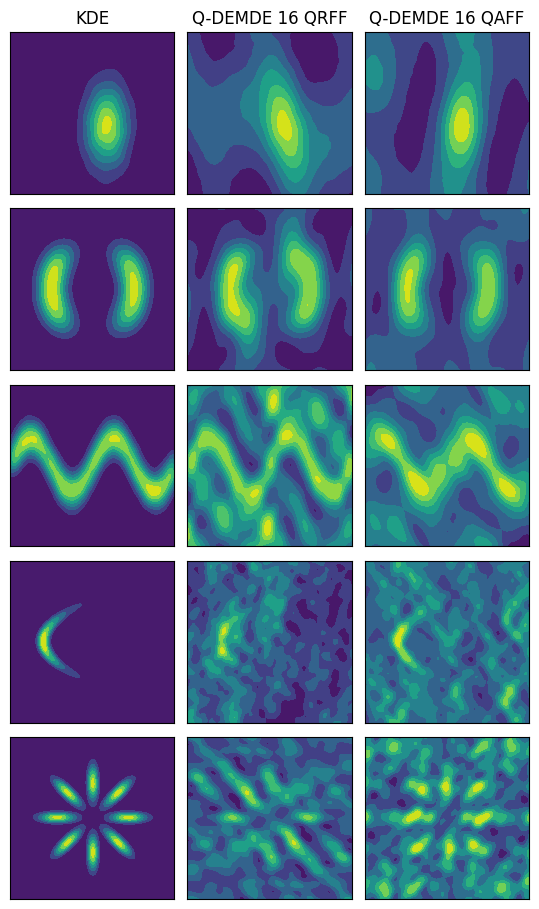}
    \caption{Two-dimensional quantum-classical density estimation with the proposed Q-DEMDE algorithm with 16 Fourier components for both QRFF and QAFF. In contrast to the results of Sect. \ref{sec: QDE 2-dimensional DE experiments}, which use classical hardware to train the QAFF, here, we present the results of using a Pennylane noiseless quantum simulator to optimize the Fourier features. From top to bottom, the two-dimensional data sets correspond to Binomial, Potential 1, Potential 2, Arc, and Star Eight.}
    \label{fig:QDE with QAFF in quantum simulator}
\end{figure}

\subsection{Setup}

We present the results for 2-dimensional density estimation using the Q-DEMDE algorithm with the QAFF learned in a Pennylane noiseless quantum simulator. Similar to the simulations of Sect. \ref{sec: QDE 2-dimensional DE experiments}, we tested the algorithm on five 2-dimensional density estimation data sets: Binomial, Potential 1, Potential 2, Arc, and Star, shown from top to bottom in Fig. \ref{fig:QDE with QAFF in quantum simulator}. The training points corresponded to 10000 points sampled from these distributions, and the test samples, whose probability density we wanted to estimate, were the result of dividing the space of each 2D data set into a grid of $120 \times 120$ points.

We used the quantum variational strategy described in Sect. \ref{QAFF Quantum implementation} and Appendix \ref{QDE: QAFF 1-2 gates} to train the quantum adaptive Fourier features in a noiseless quantum simulator of the Pennylane python library \cite{bergholm2018pennylane}. In total, we trained $16$ adaptive Fourier features, requiring a quantum circuit of $4$ qubits and $15 \times 2 = 30$ variational parameters, corresponding to $15$ Fourier vector weights $\{\boldsymbol{\theta}_j\}\in \mathbb{R}^2$, using the QAFF mapping described in the Appendix \ref{QDE: QAFF 1-2 gates}. For each 2D density estimation data set, we initialized the adaptive Fourier weights from $U[0, 1]$, and used these 2-dimensional data sets as the kernel training data sets, see Sect. \ref{QAFF kernel traning data set}. After learning the optimal Fourier weights, $\{\boldsymbol{\theta}_j\}$, we prepared the QAFF mappings of the training and test density estimation data samples by making predictions from the QAFF quantum circuit. Once the QAFF were prepared, we followed the steps of the main Q-DEMDE method, i.e., we constructed the $16\times 16$ training density matrix, computed its spectral decomposition with $16$ eigenvalues, and estimated the density estimation of the test samples using the 8-qubit Q-DMKDE quantum algorithm with the Pennylane quantum simulator.

We evaluated the hybrid Q-DEMDE with quantum variational QAFF by comparing the obtained probability density values with the probability densities obtained with KDE \cite{Parzen1962OnMode, Rosenblatt1956RemarksFunction}, using the density estimation metrics KL-Div, MAE, and Spearman correlation. As a baseline, we also performed hybrid density estimation using the Q-DEMDE method with 16 QRFF. 

\subsection{Results and discussion}

We present the results of the Q-DEMDE density estimation method with quantum variational QAFF in Fig. \ref{fig:QDE with QAFF in quantum simulator} and Table \ref{table:QDE with QAFF in quantum simulator}. The results of the simulations show that it is possible to perform quantum-classical density estimation by learning the QAFF feature map in a quantum simulator, for instance, the QDE method achieved satisfactory results for density estimation, especially on the Potential 2 data set; it is worth highlighting that the adaptive Fourier weights were randomly initialized  from $U[0, 1]$, thus the method was able to capture the optimal Fourier parameters. However, the results also show better performance of the Q-DEMDE method with QRFF on the Binomial, Potential 1, Arc and Star data sets, indicating that the results of Q-DEMDE method could have been improved by increasing the number of training iterations and using the Gaussian kernel training data set, see Sect. \ref{QAFF kernel traning data set}, which is to be explored in future work.

Although the primary Q-DEMDE strategy uses classical computers to prepare the QAFF, it is worth reemphasizing, as in Sect. \ref{sec: additional hybrid DE experiments with quantum variational QAFF}, that learning the Gaussian kernel with QAFF in quantum hardware could have applications to other quantum machine learning algorithms that do not require performing intermediate classical computations.

\begin{table}[h]
\centering
\begin{tabular}{ | c | c | c | c |}
\hline
\multirow{2}*{Data Set} & \multirow{2}*{Metric} & Q-DEMDE & Q-DEMDE \\
 &  & 16 QRFF  & 16 QAFF  \\
\hline

& KL-Div & $\mathbf{0.795}$ & $1.053$ \\
Binomial & MAE  & $\mathbf{0.029}$ & $0.073$ \\
 
  & Spearman & $\mathbf{0.573}$ & $0.283$ \\

\hline
& KL-Div & $0.639$ & $\mathbf{0.631}$ \\
Potential 1  & MAE  & $0.079$ & $0.079$ \\
 
 & Spearman & $\mathbf{0.803}$ & $0.539$ \\

\hline

& KL-Div & $1.050$ & $\mathbf{0.929}$ \\
Potential 2 & MAE  & $\mathbf{0.316}$ & $0.330$   \\
 
 & Spearman & $0.389$ & $\mathbf{0.538}$ \\

\hline

& KL-Div & $\mathbf{2.040}$ & $2.431$ \\
Arc & MAE  & $\mathbf{0.017}$ & $0.040$ \\
 
 & Spearman & $\mathbf{0.218}$ & $-0.027$ \\

\hline

& KL-Div & $\mathbf{1.383}$ & $1.433$ \\
Star Eight  & MAE  & $0.162$ & $\mathbf{0.156}$ \\
 
& Spearman & $\mathbf{0.325}$ & $0.119$ \\

\hline

\hline
\end{tabular}
\caption{2D quantum-classical density estimation results with Q-DEMDE algorithm with 16 Fourier components for both QRFF and QAFF. The QAFF were trained using a Pennylane noiseless quantum simulator. The best results are shown in bold.}
\label{table:QDE with QAFF in quantum simulator}
\end{table}

\section{Description of the IBM-Oslo quantum device}
\label{sec:oslo_description}

\begin{figure}[h]
    \centering
    \includegraphics[scale=0.45]{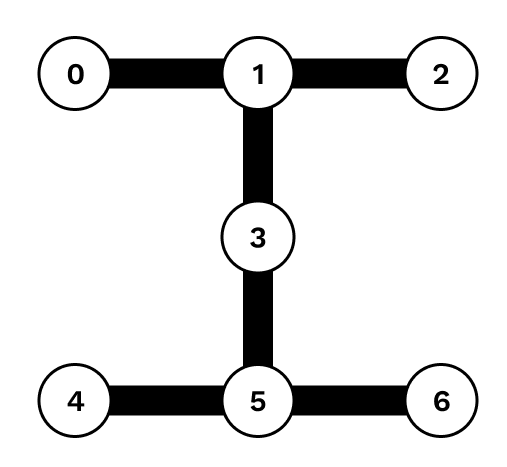}
    \caption{Topological map of the IBM-Oslo quantum processor.}
    \label{fig:oslo_layout}
\end{figure}

All the demonstrations related to applications of the methods presented in this paper on real quantum systems were performed on the IBM-Oslo quantum computer, one of the quantum processors made publicly available through the IBM Quantum Platform. The layout of the quantum processor is shown in Fig. \ref{fig:oslo_layout}, and other relevant features of the processor are listed in Table \ref{tab:oslo_features}. 

\begin{table}[h]
    \centering
    \begin{tabular}{|l|c|}
    \hline
        CLOPS (Circuit layer operations per second) & 2.6K \\ \hline
        Median fundamental transition frequency & 5.046 GHz \\ \hline
        Median anharmonicity & -0.3429 GHz \\ \hline
        Median qubit lifetime $T_1$ & 128 $\mu$s \\ \hline
        Median coherence time $T_2$ & 79.89 $\mu$s \\ \hline
        Single qubit gate error & $1.65\times10^{-4}$ to $6.698\times10^{-4}$ \\ \hline
        CNOT gate error & $6.47\times10^{-3}$ to $2.067\times10^{-2}$ \\ \hline
        Average readout error & 0.0216 \\ \hline
    \end{tabular}
    \caption{Main features of the IBM-Oslo quantum processor. The IBM-Oslo parameters were taken from Ref. \cite{di2023noisy}.}
    \label{tab:oslo_features}
\end{table}

\section{Two-dimensional data sets}\label{sec:aptwodimensional}

We used five synthetic two-dimensional data sets from Ref. \cite{gallego2022quantum} to evaluate our proposed quantum density estimation method. The data sets are characterized as follows:

\begin{itemize}
    \item \textit{Binomial} data set corresponds to a two-dimensional random sample drawn from the random vector $\bm{X}=(X_1, X_2)$ with a probability density function given by multinomial distribution $\mathcal{N}(\bm{x}|\bm{\mu}_1,\bm{\Sigma_1})$ with mean $\bm{\mu}_1 = [1, -1]^T$ and covariance matrix $\bm{\Sigma_1} = \left[\begin{array}{cc} 1 & 0 \\ 0 & 2 \end{array}\right]$.
 
   \item \textit{Potential 1} data set corresponds to a two-dimensional random sample drawn from a random vector $\bm{X}=(X_1,X_2)$ with probability density function given by \begin{scriptsize}
    $$f(\bm{x}_1,\bm{x}_2)=\frac{1}{2}\left(\frac{||\bm{x}||-2}{0.4}\right)^2  - \ln{\left(\exp\left\{-\frac{1}{2}\left[\frac{\bm{x}_1-2}{0.6}\right]^2\right\}+\exp\left\{-\frac{1}{2}\left[\frac{\bm{x}_2+2}{0.6}\right]^2\right\}\right)}$$
    \end{scriptsize}
    with a normalizing constant of approximately 6.52 calculated by Monte Carlo integration.
    \item \textit{Potential 2} data set corresponds to a two-dimensional random sample drawn from a random vector $\bm{X}=(X_1,X_2)$ with probability density function given by  $$f(\bm{x}_1,\bm{x}_2)=\frac{1}{2}\left[ \frac{\bm{x}_2-w_1(\bm{x})}{0.4}\right]^2$$ 
  where $w_1(\bm{x})=\sin{(\frac{2\pi x_1}{4})}$ with a normalizing constant of approximately 8 calculated by Monte Carlo integration.
      \item \textit{Arc} data set corresponds to a two-dimensional random sample drawn from a random vector $\bm{X}=(X_1,X_2)$ with probability density function given by $$f(\bm{x}_1,\bm{x}_2)=\mathcal{N}(\bm{x}_2|0,4)\mathcal{N}(\bm{x}_1|0.25\bm{x}_2^2,1)$$
              where $\mathcal{N}(u|\mu,\sigma^2)$ denotes the density function of a normal distribution with mean $\mu$ and variance $\sigma^2$.
    \item \textit{Star Eight} data set corresponds to a two-dimensional random sample drawn from a random vector $\bm{X}=(X_1,X_2)$. The bivariate probability distribution function is generated from a Gaussian bivariate random variable $\bm{z}$ as  $$f(\bm{x}_1, \bm{x}_2)=-\ln(e^{-\frac{1}{2}\left[\frac{\bm{z}_2-w_1(\bm{z})}{0.35}\right]^2}+e^{-\frac{1}{2}\left[\frac{\bm{z}_2-w_1(\bm{z})+w_2(\bm{z})}{0.35}\right]^2})$$ where   $w_1(\bm{z}) = \sin{(\frac{2\pi\bm{z}}{4})}$ and   $w_2(\bm{z})=3e^{-\frac{1}{2}\left[\frac{\bm{z}-1}{0.6}\right]^2}$.
\end{itemize}

\end{document}